\documentclass[10pt]{article}
\pdfoutput=1
\usepackage{jheppub}
\usepackage{physics}
\usepackage{relsize}
\usepackage{tikz}
\usepackage{pgfplots}
\usepackage{comment}
\usepackage{amsmath}
\usepackage{mathtools}
\usepackage[english]{babel}
\usepackage{ragged2e}
\usepackage{xspace}
\usepackage{cancel}
\usepackage[normalem]{ulem}
\usepackage{upgreek}
\usepackage[bottom]{footmisc}
\makeatletter
\def\@fpheader{\relax}
\makeatother

\usepackage[utf8]{inputenc}

\definecolor{blue3}{RGB}{31,119,180}
\definecolor{red3}{RGB}{214,39,40}
\definecolor{orange3}{RGB}{255,127,14}
\definecolor{green3}{RGB}{44,160,44}

\newcommand{\bfk}{{\mathbf{k}}}

\newcommand{\CRT}{{$\bf{CRT}$}\xspace}

\def\RR{$\bf{RR}$\xspace}
\def\CRT{$\bf{CRT}$\xspace}

\def\bfk{\textbf{k}}

\numberwithin{equation}{section}

\newcommand*\diff{\mathop{}\!\mathrm{d}}

\usepackage{pgfplots}
\pgfplotsset{width=10cm,compat=1.9}
\usepgfplotslibrary{fillbetween}

\allowdisplaybreaks[1]
\begin{document}


\begin{titlepage}
\setcounter{page}{1} \baselineskip=15.5pt 
\thispagestyle{empty}

\begin{center}
{\fontsize{19}{20}\selectfont Kosmic Field Theories: \\
Towards Holographic Duals for Unitary String Cosmologies}
\end{center}

\vskip 18pt
\begin{center}
\noindent
{\fontsize{11.5}{18}\selectfont Ayngaran Thavanesan\footnote{\tt \, at735@cantab.ac.uk}$^{,a,b,c}$ and Aron C. Wall\footnote{\tt \, aroncwall@gmail.com}$^{,a}$}
\end{center}

\begin{center}
\vskip 8pt
$a$\textit{ Department of Applied Mathematics and Theoretical Physics, University of Cambridge, \\ Wilberforce Road, Cambridge, CB3 0WA, UK.} \\
$b$\textit{ Kavli Institute for Theoretical Physics, Santa Barbara, CA 93106, USA.} \\
$c$\textit{ Laboratory for Theoretical Fundamental Physics, Institute of Physics, École Polytechnique Fédérale de Lausanne (EPFL), CH-1015 Lausanne, Switzerland.}
\end{center}


\vspace{1.4cm}

\begin{abstract}
\noindent
Recent work on cosmological amplitudes has established reality conditions (derived from unitarity) for general particle-creation processes in flat FLRW cosmologies, in the Bunch-Davies wavefunction.  In light of these results, we propose a new class of large-$N$ holographic gauge theories with $d$ spatial dimensions, which we call \emph{Kosmic Field Theories} (KFTs), of which only a subset are conformal (KCFTs).  By imposing an appropriate --- $d$ dependent --- complex phase for the number of colours $N$ and the t' Hooft coupling $\lambda$, we show that the complex phases of $n$-point functions in a KFT match the reality conditions required by $d + 1$ bulk unitarity, to all orders in bulk and boundary perturbation theory, including loop diagrams.  Since $N^2/\lambda$ is imaginary in all dimensions, the class of (parity-even) KFTs has no overlap with the class of Wick rotations of unitary QFTs.  We also make a preliminary investigation of positivity conditions, which might constrain the overall $\pm$ sign of $N^2$ and $\lambda$.

Although SU($N$) models with adjoint matter can be defined in generic real dimensions $d$, several features of the class of KFTs are nicer when $d$ is odd, allowing $N^2$ to be real.  This includes the possibility of non-oriented and/or open string duals, reality of the effective number of degrees of freedom, and closure under RG flow beyond the leading order in a $1/N$ expansion.  This could explain why we live in a cosmology with an even number of spacetime dimensions.
\end{abstract} 


\end{titlepage} 


\newpage
\setcounter{tocdepth}{3}
{
\hypersetup{linkcolor=black}
\tableofcontents
}

\newpage
\setcounter{footnote}{0}


\section{Introduction}\label{sec:Introduction}
\subsection{Holography and the Success of AdS/CFT}
Over the past twenty–five years, the gauge/gravity duality has provided the most compelling framework we possess for a non–perturbative definition of string theory. The AdS/CFT correspondence, conjectured by Maldacena in 1997~\cite{Maldacena:1997re}, posits a precise equivalence between a conformal field theory (CFT) living on the boundary of an asymptotically Anti–de Sitter (AdS) spacetime and a string/M-theory (or supergravity in the appropriate limit), propagating in the bulk of that spacetime. The duality relates strongly–coupled quantum field theory (QFT) dynamics to semiclassical gravity, providing calculational power as well as deep conceptual insights into black hole physics, thermalisation, and the emergence of spacetime itself.

The most famous concrete realisation of AdS/CFT is the correspondence between type IIB string theory on $\text{AdS}_5 \times \text{S}^5$ (supported by a self-dual 5-form flux) and $\mathcal N=4$ supersymmetric Yang–Mills theory in four dimensions. The gauge theory side is a well–defined local QFT with manifest unitarity (which implies reflection positivity in the Euclidean setting), and a rich set of global and supersymmetries. On the gravity side, the theory admits a controlled $1/N$ expansion at finite $\lambda$, where $N$ is the number of colours in the gauge group and $\lambda=g^2N$ the ’t Hooft coupling. The two sides of the correspondence match precisely: correlators of boundary operators reproduce scattering amplitudes of bulk fields, and partition functions agree when sources are appropriately identified.

The conceptual and computational utility of AdS/CFT has made it the dominant paradigm in holography~\cite{Maldacena:1997re,Gubser:1998bc,Witten:1998qj}. And yet, observational cosmology indicates that our universe is not asymptotically AdS, but approximately de Sitter (dS) at late times. Observational evidence from the cosmic microwave background~\cite{Planck:2018vyg}, large-scale structure surveys~\cite{SDSS:2006lmn,BOSS:2016wmc}, and the accelerated expansion of the universe inferred from Type Ia supernovae~\cite{SupernovaSearchTeam:1998fmf,SupernovaCosmologyProject:1998vns} all strongly indicate that de Sitter space is the appropriate maximally symmetric model for cosmology. The near scale-invariance of the CMB spectrum and the near-flat spatial curvature of the universe further support an early inflationary epoch consistent with quasi-de Sitter expansion~\cite{Guth:1980zm,Linde:1981mu,Starobinsky:1980te}. Thus a natural and pressing question arises: does there exist a de Sitter analogue of AdS/CFT — a holographic duality relating quantum gravity in de Sitter space to some boundary theory?

\subsection{History of de Sitter holography at Late Time}
Attempts to construct a dS/CFT duality date back to the late 90s / early 2000s, not long after Maldacena's original derivation of AdS/CFT. 

The first significant proposal we are aware of was by Hull~\cite{Hull:1998vg,Hull:1998ym}, who constructed nonunitary variants of string theory (the II$^*$ and $M^*$ theories), and noticed that they admit intrinsically Euclidean `E-branes' with $\text{dS}_p \times \text{H}_q$ compactifications (cf.~\cite{Hull:1999mt,Dijkgraaf:2016lym}). These examples were pathological since the bulk theories are nonunitary. There was also a related attempt by \cite{Balasubramanian:2001rb} to interpret the sphere factor in $\text{AdS}_p \times \text{S}_q$ as Wick rotated de Sitter universe.

After this, proposals were made by Witten~\cite{Witten:2001kn} and Strominger~\cite{Strominger:2001pn} for dS/CFT based on scattering in de Sitter from past to future infinity.  Strominger, emphasizing the importance of the conformal group Spin$(d+1,1)$, proposed that correlators of operators on $\mathcal{I}^+$ and (the antipodally identified) $\mathcal{I}^-$ could be interpreted as $n$-point functions of a Euclidean CFT.  Maldacena \cite{Maldacena:2002vr} modified this proposal by suggesting that the Hartle–Hawking wavefunction of the universe in de Sitter space could be interpreted as the generating functional of a Euclidean CFT living on the asymptotic future boundary $\mathcal{I}^+$, with no role for $\mathcal{I}^-$ (cf.~\cite{Arkani-Hamed:2015bza}).

Subsequent research groups have significantly advanced the study of holographic cosmology; here we can only give an inadequate summary.\footnote{Our historical review concerns itself mostly with ``late time'' holographic cosmology models where the dual field theory is related to $\mathcal{I}^+$, ignoring the long history of proposed duals to the de Sitter static patch or horizon.  Some specific proposals for de Sitter holography that do not involve $\mathcal{I}^+$, including constructions using timelike boundaries and horizon thermodynamics, hyperbolic and flux compactifications in string theory~\cite{Silverstein:2007ac,Silverstein:2008sg,Dong:2010pm,DeLuca:2021pej,Cotler:2024xzz}, dualities based on double-scaled SYK~\cite{Susskind:2021omt,Rahman:2022jsf,Susskind:2022bia,Narovlansky:2023lfz,Verlinde:2024znh,Yuan:2024utc,Verlinde:2024zrh}, $T\overline T+\Lambda$–type deformations of suitable seed CFTs~\cite{Gorbenko:2018oov,Lewkowycz:2019xse,Shyam:2021ciy,Coleman:2021nor,Silverstein:2022dfj,Batra:2024kjl,Silverstein:2024xnr,Aguilar-Gutierrez:2024nst,Philcox:2025faf,Shyam:2025ttb}, and other horizon-based constructions (see e.g.~\cite{Anninos:2012qw,Flauger:2022hie,Harlow:2022qsq,Galante:2023uyf} for reviews). These are conceptually distinct from the $\mathcal{I}^+$–based framework considered in this paper, although there might be ways of interpolating between them using a $T^2$ deformation~\cite{Chang:2025ays,Shyam:2025ttb}.} One line of research~\cite{Hertog:2011ky,Hartle:2012qb,Conti:2015ruo,Hertog:2015nia,Hawking:2017wrd,Hertog:2017ymy,Hertog:2024shf} is based on the complex gravitational path integral approach of Hartle, Hawking, and Hertog.  This approach considers a complex contour of the de Sitter gravitational path which looks like Euclidean hyperbolic space, and attempts to match it to known examples of supergravity with AdS/CFT duals (possibly with a minisuperspace truncation and/or complex field values).  A distinct approach~\cite{McFadden:2009fg,McFadden:2010na,McFadden:2010jw,McFadden:2010vh,McFadden:2011kk,Easther:2011wh,Bzowski:2011ab,Bzowski:2012ih,Bzowski:2013sza,McFadden:2013ria,Afshordi:2016dvb,Afshordi:2017ihr,Nastase:2019rsn,Bzowski:2019kwd,Penin:2021sry,Bzowski:2023nef}, pioneered McFadden and Skenderis\footnote{Some early comments are in Appendix A of~\cite{Skenderis:2002wp}.}, uses the ``domain wall/cosmology correspondence'' (developed in~\cite{Cvetic:1996vr,Skenderis:2006fb}) to investigate more general FLRW models.  In this version of holographic cosmology, one identifies a dual nonunitary ``pseudo-QFT'', either by analytically continuing AdS/CFT models, or else by postulating them directly.   To attempt to match experiment (e.g~cosmic microwave background fluctuations), these authors typically define the boundary in $d = 3$ by analytically continuing the number of colours as $N^2 \to -N^2$, and the momenta as $p_\mu \to ip_\mu$, which agrees with a special case of the manipulations that we will consider below.

In a separate line of research, a concrete realisation of dS/CFT was obtained via the vector Sp($N$) model\footnote{This model was initially studied by LeClair~\cite{LeClair:2006kb}, who later argued that the model has a pseudo-hermitian Hamiltonian~\cite{LeClair:2007iy,Robinson:2009xm}.  However, the proper signs in a theory with Grassmannian fields can be subtle, so it is not necessarily completely clear from this what is the most appropriate sign in the dS/CFT context.} in $d = 3$~\cite{Anninos:2011ui,Ng:2012xp,Das:2012dt,Anninos:2012ft,Anninos:2017eib}, but this involves an exotic Vasiliev ``higher spin'' gravity dual in $\mathrm{dS}_4$.  These may be regarded as an $N \to -N$ analytic continuation of the O($N)$ dual to higher spin gravity in AdS~\cite{Klebanov:2002ja,Sezgin:2002rt,Giombi:2009wh}, which have been speculated to arise from a tensionless limit of string theory~\cite{Sundborg:2000wp,Sezgin:2002rt,Giombi:2012ms}. More recent attempts to construct exact nonunitary boundary duals for Einstein gravity in $dS_3$ arise from taking a limit of a complex level $k$ in WZW models~\cite{Castro:2011xb,Castro:2011ke,Ouyang:2011fs,Castro:2012gc,Castro:2020smu,Hikida:2022ltr,Castro:2023dxp,Fliss:2023muk}; or from the complex Liouville model \cite{Collier:2024kmo,Collier:2024kwt,Collier:2024lys,Collier:2025pbm,Collier:2025lux} with central charge $c = 13 + i\nu$.

\subsection{Obstacles to de Sitter holography}\label{sec:obstacles}
The dS/CFT proposal is aesthetically appealing: it mirrors the AdS dictionary while respecting the fact that de Sitter has spacelike conformal boundaries.  However, significant obstacles have prevented the establishment of a concrete duality to Einstein gravity de Sitter backgrounds. In what follows, we restrict attention to holographic cosmology models formulated at future infinity $\mathcal{I}^+$, and we do not review approaches based on timelike or static–patch boundaries.

The first obstacle to a formulation of de Sitter holography at $\mathcal{I}^+$ is the absence of reflection positivity. Any candidate Euclidean boundary theory dual to a de Sitter bulk cannot satisfy all the Osterwalder–Schrader axioms: the would–be correlators derived from the bulk wavefunction fail to obey reflection positivity. This failure means that the boundary theory, if it exists, cannot be a conventional unitary Euclidean QFT. Unlike AdS/CFT, where unitarity and positivity on the boundary guarantee bulk consistency, in dS/CFT we are forced to relinquish these properties of the boundary theory at the outset.  Instead, we need to identify some novel property of the boundary dS/CFT field theory, which implies bulk unitarity.

The second obstacle is more conceptual. In AdS/CFT the radial direction corresponds to the energy scale of the CFT, and the Hamiltonian of the CFT generates time translations in the bulk. In de Sitter space, the situation is more subtle: the natural slicing relevant for cosmology is spatially flat FLRW coordinates, and the bulk time evolution is not generated by a boundary Hamiltonian in the same way. This complicates the interpretation of the bulk–boundary dictionary.

A third obstacle arises from the nature of the mass-dimension relation in dS/CFT.  Considering the case of scalars for specificity, in AdS/CFT we have the following relation:
\begin{equation}
\Delta(\Delta - d) = m^2 L_\text{AdS}^2
\end{equation}
and bulk stability requires that $m^2 L_\text{AdS}^2 \ge -d^2/4$, which is compatible with the conditions for unitary scalar representations ($\Delta \in \mathbb{R}$ and $\Delta \ge (d-2)/2)$) of the Lorentzian conformal group $\text{Spin}(d,2)$.  On the other hand, for dS/CFT we have the reversed rule
\begin{equation}\label{eqn:mass-relation}
    \Delta(\Delta - d) = -m^2 L_\text{dS}^2 \, ,
\end{equation}
and unitary scalar representations of the de Sitter conformal group $\text{Spin}(d+1,1)$ must either be in the range $0 \le \Delta \le d$ (complementary series) or $\Delta = d/2 + i\nu$ (principal series).  This means that any field theory with a ``usual'' spectrum (having scalar operators with increasing real values of $\Delta$) will necessarily have tachyons in the spectrum with $m^2 < 0$, signalling bulk instability.  This happens because the sign of the $m^2$ term gets reversed in the all-timelike ``Euclidean AdS'' contour of de Sitter spacetime \cite{Cadoni:2002xe}.  (Conversely, if the spectrum is entirely real, there will be no particles in the principal series, with mass above $m^2 L_\text{AdS}^2 = d^2/4$.)  This suggests that perhaps a theory with a normal spectrum will be dual to a negative tension string in dS.\footnote{Higher spin models of dS, which might correspond to tensionless strings, evade the problems listed above by having no single-trace scalar operators in their $\Delta > d$ spectrum --- only gauge fields of increasing spin, for which the unitary bounds are more forgiving.  However, unlike the AdS case, it does not seem to be possible to continuously take the tensionless limit of string theory in dS, because it requires passing over a gap in the allowed dimensions $\Delta$ of unitary operators \cite{Deser:2001us}.}  Such a string theory would have a spectrum of increasingly tachyonic particles, rather than a spectrum of increasingly massive ones.

On the bulk side, there are many restrictions concerning the existence of genuine de Sitter vacua in standard superstring theory. Several classical no--go theorems show that under certain fairly general assumptions (compact internal manifold, two-derivative supergravity, no orientifold or other negative-tension sources, standard form fields and fluxes) consistent compactifications to stable de Sitter are impossible; a canonical statement along these lines is the Maldacena–Nu\~nez no--go result for classical supergravity compactifications~\cite{Maldacena:2000mw}. See also the earlier constraints in the literature (e.g.\ Gibbons and related analyses) that highlight the difficulty of obtaining positive cosmological constant solutions in simple supergravity truncations. These theorems do not, however, rule out all constructions: by including ingredients outside their assumptions one may evade the conclusions.\footnote{Notable examples and strategies that circumvent the classical no--go restrictions include flux compactifications with orientifold planes and warping~\cite{Giddings:2001yu}, the KKLT construction employing non-perturbative effects plus anti-D3 brane uplifts~\cite{Kachru:2003aw}, and more recent proposals that exploit combinations of classical fluxes, metric fluxes, orientifolds and higher-derivative ($\alpha'$) corrections. For recent discussions and critical overviews of the status of de Sitter constructions in string theory (see e.g.~\cite{Danielsson:2018ztv,Obied:2018sgi}). Consequently, statements that rely on ``no de Sitter vacua in string theory'' must be qualified by the assumptions of the corresponding no--go theorem: in practice the space of consistent ingredients is large, and the existence (or otherwise) of fully controlled supersymmetric de Sitter vacua remains an active and subtle subject of research.}

Finally, explicit candidate boundary theories for dS/CFT have mostly remained elusive, with the exceptions mentioned above. Without supersymmetry, or top-down string derivations of the duality, it is difficult to constrain the spectrum and interactions of the hypothetical boundary theory. Consequently, while dS/CFT remains an alluring idea, concrete progress has been limited compared to the spectacular success of AdS/CFT.

\subsection{The Cosmological Optical Theorem}
In this article, we intend to study reality conditions for dS/CFT by placing the implications of \emph{bulk unitarity} in a central role.  We thus turn to reviewing these constraints as they have been studied in recent cosmology literature.

Recent work has made significant progress in identifying the constraints of unitarity for quantum field theory on cosmological spacetimes. First, constraints from unitarity on the perturbative wavefunction were derived, known as the \emph{Cosmological Optical Theorem}~\cite{COT,Cespedes:2020xqq,Melville:2021lst,Goodhew:2021oqg,Baumann:2021fxj}, as well as for the cosmological analogue of the flat space S-matrix~\cite{Marolf:2012kh,Melville:2023kgd,Melville:2024ove}.)  These constraints include:
\begin{itemize}
\item \textit{Hermitian Analyticity}: constraining the phase of an individual Feynman diagram, up to an overall minus sign (see~\cite{Stefanyszyn:2023qov,Stefanyszyn:2024msm,Stefanyszyn:2025yhq} for a complementary proof of Hermitian Analyticity for all tree-level diagrams involving massless and conformally-coupled scalar fields in the external legs of the Feynman diagram, as well as~\cite{Liu:2019fag,Cabass:2022rhr,Thavanesan:2025kyc} for the implications of these phases for cosmological observables).
\item \textit{Cutting Rules}: that relate the amplitude of a complex Feynman diagrams to simpler ones in which an edge is cut, similar to the flat-space cosmological optical theorem.
\end{itemize}
This first result, hermitian analyticity, was then extended to the case of non-perturbative bulk QFT by means of a \emph{Cosmological CPT Theorem} \cite{Goodhew:2024eup}.  This paper also derived perturbative constraints to all orders in loops.  

As discussed in \cite{Goodhew:2024eup}, unitarity has two aspects.  One aspect of bulk unitarity, which we call \emph{Reflection Reality} (\RR), is equivalent to the assertion that the bulk inner product is hermitian (though possibly with an indefinite norm).  \RR takes the form of a $\mathbb{Z}_2$ symmetry acting on bulk Euclidean correlators, and implies the hermitian analyticity conditions.  For a Lorentz-invariant local Lagrangian, this implies that the correlators also respect \CRT symmetry.\footnote{For Lorentz-violating Lagrangians, \cite{Goodhew:2024eup} identified distinct implications for \RR and \CRT.}

The second aspect of unitarity is that, conditional on the reality conditions encoded in \RR, we must also impose certain \emph{positivity conditions} to ensure that the bulk Euclidean correlators satisfy the \emph{Reflection Positivity} (\textbf{RP}) condition.  This ensures that the bulk theory has a positive norm, not just a real one.  Importantly, this second step (\textbf{RP} conditional on \RR) is a much milder constraint on the space of local bulk field theories, than the first step (\RR itself) is.  It was argued that the only local Lagrangian theories which violate \textbf{RP} but satisfy \RR, are theories in which certain bad \emph{discrete choices} are made (e.g.~the introduction of a fundamental field with a negative-norm propagator, or that violates spin-statistics).  Thus, among field theories satisfying \RR, an order unity fraction of these should also (``accidentally'') satisfy \textbf{RP} and be unitary.  

Therefore, imposing \RR symmetry on holographic cosmology, should already go 95\% of the way to identifying the class of theories with unitary duals.  Put another way, for any specific proposed dS/CFT model, for which the bulk dual is weakly coupled and satisfies \RR, it should be straightforward to manually check by hand whether any of the bulk fields have negative norm states, in their propagating degrees of freedom.  And in the case where the theory is unitary, this unitarity should be generically robust to any \RR-compliant changes in the coupling constants.

\subsection{Constaints on the Cosmological Wavefunction from Reflection Reality}
Under mild assumptions about the analytic structure of quantum field theory in curved spacetimes, the bulk \RR symmetry implies that the cosmological wavefunction must satisfy a non–perturbative conjugation symmetry.  Consider the wavefunctional coefficients $\psi_n$ that appear when we expand the late–time cosmological wavefunction in terms of boundary field configurations. For any flat FLRW background, \RR asserts that the Bunch-Davies wavefunction satisfies
\begin{equation}\label{eqn:HA}
    \big[\psi_n(\eta;\,\mathbf{x};\,\Omega)\big]^*
    \;=\;
    \psi_n\big(\eta;\,-e^{i\pi}\mathbf{x};\,e^{-i\pi}\Omega\big) \, ,
\end{equation}
where $\eta$ denotes conformal time, $\mathbf{x}$ the boundary coordinates, and $\Omega$ a Weyl factor specifying the bulk metric \cite{Goodhew:2024eup}. This condition holds non–perturbatively with respect to the bulk QFT,\footnote{We do not have a rigorous argument for gravity because of (i) the conformal mode problem, and (ii) the difficulty in defining quantum gravity non-perturbatively, but the perturbative relation below appears to hold for all graviton amplitudes that we have examined.} 
on any Cauchy slice of the bulk evolution, including at the asymptotic future boundary $\eta\to0^-$.  However, in this paper we will be interested in the expansion of \eqref{eqn:HA} in bulk perturbation theory:
\begin{equation}\label{eqn:HA_pert}
    \left[\psi^{(L)}_n(\eta; \textbf{k})\right]^* = e^{i\pi \left((d+1)L-1\right)} \psi^{(L)}_n(\eta; e^{-i\pi}\textbf{k}) \, , 
\end{equation}
where $L$ is the number of loops in the bulk and $d$ is the boundary dimension.  This formula was obtained by explicitly performing the bulk $\Omega$ rotation, which turns out to mainly just affect the factor of $\sqrt{g}$ in the front of the bulk action, producing the exponential prefactor in front of the RHS of \eqref{eqn:HA_pert}.

Furthermore, in the presence of bulk log divergences (or more generally, bulk quantum anomalies) at $L \ge 1$, there can be corrections to \eqref{eqn:HA_pert}, due to an anomalous dependence on the scale factor of the metric.  The simplest way to deal with them is by dimensional regularisation, in which case formula \eqref{eqn:HA_pert} holds exactly for the regulated theory in fractional $d$ dimensions.

Crucially, these conditions  do not rely on conformal invariance or reflection positivity of the \emph{boundary} theory.  Instead, they follow directly from bulk unitarity of the cosmological time evolution. Thus even though the boundary theory in holographic cosmology is necessarily non–reflection positive, the bulk \RR symmetry provides an exact constraint that any consistent holographic dual must satisfy.  
Because these results do not depend on conformal invariance, they may be applied to several different classes of holographic cosmology models.  This includes:
\begin{itemize}
    \item Traditional dS/CFT, in which the boundary theory is conformally invariant, and the bulk theory is exactly de Sitter up to small fluctuations.
    \item FLRW/QFT models, where the scale factor $a(\eta)$ takes some other form that differs from dS, discussed in e.g.~\cite{McFadden:2009fg} and its progeny.  This would include examples in which a dS/CFT is deformed by turning on various single-trace terms, resulting in a non-trivial $\beta$ function in the holographic RG flow.
    \item $T^2$/CSH models, in which a dS/CFT is deformed by a double-trace $T^2$ deformation~\cite{Zamolodchikov:2004ce,Smirnov:2016lqw,Aharony:2018vux}.  As the effect of the $T^2$ deformation is to move the holographic boundary theory inward to  a finite cutoff surface \cite{McGough:2016lol,Kraus:2018xrn,Taylor:2018xcy,Hartman:2018tkw,Shyam:2018sro}, it may be applied to a dS/CFT model to obtain a theory living on a Cauchy slice at finite $\eta$, which is known as Cauchy Slice Holography~\cite{Araujo-Regado:2022gvw,Araujo-Regado:2022jpj,Khan:2023ljg,Soni:2024aop,Araujo-Regado:2025elv,Shyam:2025ttb}.
\end{itemize}
In all of these cases, \eqref{eqn:HA} gives a powerful and rigid condition, for checking whether the boundary field theory is compatible with the reality conditions required by unitarity (and \textbf{CRT} symmetry). 

\subsection{Kosmic Field Theories}
In this paper we will identify a class of boundary gauge theories, which we call Kosmic Field Theories (KFT),\footnote{The term \emph{kosmic}, derived from the Greek $\upkappa \acute{{\rm o}} \upsigma\upmu {\rm o} \varsigma$ (kosmos), reflects both the cosmological nature of the dual spacetime, and the emergent structure of time in these theories.  We Romanise the kappa $\upkappa$ to the letter `K' because the abbreviation `CFT' is already taken.} which are constructed to satisfy \eqref{eqn:HA}, and thus have bulk duals that satisfy \RR.

A KFT is a large $N$ gauge theory which satisfies a different reality conditions on their parameters $N$ (number of colours) and $\lambda = g^2N$ (the 't Hooft coupling) than unitary field theories do.  In the ’t Hooft expansion of the boundary gauge theory, each Feynman diagram contributes with a factor
\begin{equation}
    \psi_n \sim N^{\,\chi}\,\lambda^{\,\bar{E}-\bar{V}},
\end{equation}
where $\chi=\bar{V}-\bar{E}+\bar{F}$ is the Euler characteristic of the ribbon graph. For a U($N$) or SU($N$) theory without quarks, $\chi$ is always even, so diagrams always come with even powers of $N$.  

This corresponds to the genus expansion of a hypothetical dual string theory (weakly coupled at large $N$), which we assume lives in a higher dimensional $d+1$ bulk for the usual reasons.  Although we do not know how to identify a bulk worldsheet model for this bulk string theory (or relate it to any other known string models e.g.~compactifications of $D = 10$ superstrings or $D = 26$ bosonic strings), we can treat it in a ``bottom-up'' fashion as an unknown bulk theory, which we will assume is described by a weakly coupled (and therefore nearly classical) bulk Lagrangian of some particle theory.  In other words, we assume that the string worldsheets have a well-defined limit where they look like a Feynman diagram expansion.

At this point, Eq. \eqref{eqn:HA} dictates a specific phase for each $\psi_n$, arising from the transformation of the bulk and boundary Weyl factors under conjugation. By matching these two structures, we find that the only way to satisfy Eq. \eqref{eqn:HA} consistently across all diagrams is to allow $N$ and $\lambda$ themselves to be complex, with the precise phases depending on the spacetime dimension $d$.  This matching leads us to the following phase choices:
\begin{eqnarray}
    N^2 &\;\propto\;& \pm e^{i\pi(d-1)/2} \, , \label{eqn:phaseN} \\
    \lambda &\;\propto\;& \pm e^{i\pi d/2} \, , \label{eqn:phaselambda}
\end{eqnarray}
where the $\propto$ symbol expresses agreement of the complex phase, and these formulae are derived in general (possibly fractional) boundary dimensions $d$.  One reason for considering fractional dimensions is that it enables easy use of dimensional regularisation (dim reg) to deal with bulk or boundary divergences.  Assuming we do that, \eqref{eqn:phaseN} and \eqref{eqn:phaselambda} give us the correct phases (up to a real $\pm$ sign) to match a unitary bulk theory, at least \emph{to all orders in bulk and boundary perturbation theory}, including diagrams with one or more bulk loops.\footnote{At least for processes not involving tachyons or principal series fields in the bulk.  While we hope that the work of \cite{Goodhew:2024eup} can be extended to principal series fields, we did not consider that case carefully.  Tachyons, on the other hand, might create problems with the definition of the Bunch-Davies vacuum, that again we did not address carefully.}  This is a significant extension of previous work on holographic cosmology which has mostly looked at tree-level amplitudes, see e.g.~the discussion in \cite{Bzowski:2023nef}.

Note that shifting $d \to d + \epsilon$ can introduce an extra complex phase into $N^2$ or $\lambda$, even though these quantities are purely real/imaginary at integer $d$.  These extra phases are more than just a formal nicety; as discussed in \cite{Goodhew:2024eup} they encode physics that is  \emph{required} for bulk unitarity, and thus in the absence of dim reg our rules would have to become considerably more complicated.  These shifts can only appear in the \emph{finite part} of a Feynman diagram (or subdiagram) which is log divergent, and they have an \emph{imaginary} sign relative to the log divergent term; thus we refer to them as ``imaginary shifts''.

Although analytically continuing $N$ may seem unusual, interpreting $N$ away from positive integers has already been studied fairly extensively in the mathematics literature\footnote{AT thanks Adrián López-Raven and Davide Gaiotto for making him aware of the existence of Deligne categories.} (see e.g.~\cite{Deligne1990TannakianCategories,Deligne2002TensorCategories,Knop2007TensorEnvelopes,Deligne2007RepsSt,ComesOstrik2011BlocksRepSt,Etingof2014RepComplexI,Etingof2014RepComplexII,EntovaHinichSerganova2015DeligneGLmnn,Harman2016DeligneLimits,BarterEntovaHeidersdorf2017InfiniteSymm,EntovaSerganova2018Periplectic,KhovanovOstrikKononov2020TQFT,Meir2021Interpolations,DeligneNotesOnline}) and has recently been explored in the context of large-$N$ matrix models~\cite{Cotler:2019nbi,Cotler:2024xzz}.  Analytically continuing $\lambda$ is more straightforward, as this is simply a modification of coefficients in the Lagrangian; the main concern here is that the the Euclidean path integral will not converge nonperturbatively, if it happens that the \emph{real part} of the Lagrangian is unbounded below.  However, arbitrary phases of $\lambda$ are always permitted at the level of naive perturbation theory, which is (by definition) analytic in the coupling constants.

Importantly, we do \emph{not} motivate these complex values of $N$ and $\lambda$ by analytic continuation from any known example of AdS/CFT, as there is no guarantee that a good example of dS/CFT will be continuously connected to an example of AdS/CFT, especially as a unitary dS bulk theory cannot be supersymmetric.  Instead, we obtain these phases by \emph{starting} on the cosmology side, and directly computing the constraints \eqref{eqn:HA} due to bulk unitarity.  Thus our methodology is quite different from some prior work, e.g.~\cite{Balasubramanian:2001rb,Hull:1998vg,Buchel:2002kj,Hoare:2014pna,Arutyunov:2014cra,Hertog:2024shf}, attempting to construct dS/CFT from AdS/CFT by analytically continuing $N$ and $\lambda$ so as to reverse the sign of the cosmological constant ($\Lambda \to -\Lambda$).  However, such continuations inevitably produce a non-unitary bulk theory, with e.g.~imaginary fluxes.  As our approach imposes (the \RR subset of) bulk unitarity from the start, we believe it is more promising than approaches which tamper with the unitarity of the bulk theory.

Regardless of $d$, \eqref{eqn:phaseN} and \eqref{eqn:phaselambda} imply that a KFT will always have
\begin{equation}\label{eqn:iR}
    \frac{N^2}{\lambda} \in i\mathbb{R} \, ,
\end{equation}
which implies that the KFTs that we consider will in fact never be unitary.\footnote{In this paper we consider only parity-even theories, in which the Euclidean coupling $\lambda$ should be real.  If we allow parity-odd theories, it might be possible to satisfy this condition if the Lagrangian is constructed exclusively from parity-odd terms, as in the case of e.g.~Chern-Simons theory.}  In even dimensions $\lambda$ is real but $N^2$ is imaginary, whereas in odd dimensions $N^2$ is real but $\lambda$ is imaginary. In no case are both real simultaneously. This is a striking departure from the familiar AdS/CFT case, where both parameters are positive integers or real couplings.  \eqref{eqn:iR} also implies that that the tree-level holographic counterterms are always imaginary, which is consistent with prior expectations for holographic cosmology~\cite{Skenderis:2002wp,Maldacena:2002vr,McFadden:2010na,McFadden:2010jw,McFadden:2010vh,McFadden:2011kk,Easther:2011wh,Bzowski:2011ab,Bzowski:2012ih,Bzowski:2013sza,Afshordi:2016dvb,Afshordi:2017ihr,Nastase:2019rsn,Bzowski:2019kwd,Penin:2021sry,Bzowski:2023nef,Araujo-Regado:2025elv}.\footnote{This may be most easily seen by considering the boundary conditions for a canonically normalised graviton field in the bulk.}

In a theory with multiple independent couplings $\lambda_1, \lambda_2 \ldots$, all of them must satisfy the condition \eqref{eqn:phaselambda}; or equivalently one can place a single complex factor of $1/\lambda$ outside the whole action, and introduce a set of real coefficients which vary from term to term.  Additionally, if we want the theory to have a weakly coupled semiclassical bulk limit, we should require $|N| \gg 1$, and also $|\lambda| \gg 1$ if we want to avoid the highly stringy regime.  (But, for reasons stated below, we also want to take $g \ll 1$ to avoid worrying about non-perturbative instanton effects which might spoil our analysis.)

\subsection{Regime of Validity}

As the argument for \eqref{eqn:phaseN} and \eqref{eqn:phaselambda} uses perturbation theory on \emph{both} the bulk and boundary side, one might worry about the validity of our expansion, e.g~for a weakly coupled bulk at large $|\lambda|$, where the boundary theory becomes strongly coupled.  However, it is not a requirement of our argument that the perturbation series in $\lambda$ actually be numerically tractable, or even within the (apparent or actual) radius of convergence.  All that is needed is that we are in the regime where the amplitude is an analytic continuation of the perturbation expansion, so that the phase expected in perturbation theory continues to hold.\footnote{The reason the analysis is so forgiving is that we are only trying to demonstrate a reality condition; and if a function $f(x)$ is e.g.~real in its Taylor expansion about $x = 0$, the analytic continuation of $f$ to arbitrary real values of $x$ will also be real, so long as one does not pass through a singularity on the real axis, and so long as the function really is analytic.  The same does not necessarily hold for positivity conditions, as the analytic continuation of a function which is positive on an interval, can easily become negative outside of it.}  

To be sure, this kind of analytic continuation might be spoiled by a truly non-perturbative effect, such as a boundary instanton, as these cannot be analysed using boundary Feynman diagrams.  However, a Yang-Mills instanton has an amplitude $\sim e^{-1/g^2} \sim e^{-N/\lambda}$, which means that suppressing them only requires $|g| \ll 1$ (assuming $\text{Re}(g^2) > 0$) which at large $N$ is compatible with $|\lambda | \gg 1$, suggesting however that we also need $\lambda > 0$ for a non-perturbatively well-defined theory.  For $d \ne \text{odd}$, this can be arranged by correctly choosing the $\pm$ sign for the coupling of the Yang-Mills action.  

On the other hand, in $d = \text{odd}$, $\lambda \in i\mathbb{R}$, and thus instantons are right on the edge of being problematic.  But this is just one manifestation of the fact that \emph{everything} about the path integral is now oscillatory.  This phenomenon is familiar in the case of Lorentzian path integrals, weighted by $e^{i{\cal S}}$, indicating that the path integral is ``on the edge'' of convergence, and thus needs to be defined by deforming it slightly into a convergent region, and then taking the limit back to the edge (an $i\epsilon$ expansion), similar to how the oscillatory Gaussian integral:
\begin{equation}
    \int^\infty_{-\infty}\!dx\,e^{ix^2/2a}
\end{equation}
may be defined by introducing a small damping factor as $|x|\to \infty$.  The exotic feature of the $d = \text{odd}$ KFT, is simply that it is oscillatory in Euclidean signature rather than Lorentzian signature.

Given the above, we believe that the KFT conditions above can be compatible with the holographic limit $|N| \to \infty$, fixed $|\lambda| \gg 1$.  This means than an inherently very strongly coupled dS/CFT (analogous to e.g. the ABJM model in $d = 3$, or the $d = 6$ (2,0) CFT in the case of AdS/CFT) might not be within our regime of validity.  It would be interesting to check whether our condition for bulk unitarity can be extended to situations with truly non-perturbative boundary or bulk effects, but such an analysis goes beyond the scope of this paper.

\subsection{Imposing Additional Conditions on the KFT}

In some wavefunction coefficients (e.g.~for 2-point functions and central charges), a particular choice of the real signs ($\pm$) will be required by imposing full bulk unitarity.  This is closely related to the criterion that the the cosmological wavefunction be normalisable.  But, such constraints require thinking about bulk Reflection Positivity (\textbf{RP}), so they do not follow from imposing \eqref{eqn:HA} alone.  These must be imposed separately.\footnote{We reserve the right (and give permission to others) to restrict the term KFT to those field theories satisfying these stronger positivity constraints, if that turns out to be convenient in the light of future developments.}  Assuming these $\pm$ choices are correct in every wavefunction coefficient, we expect that the bulk Local Lagrangian will be unitary, and  hence that all other implications of unitarity (including cutting rules) should also hold.  We will take a preliminary stab at identifying such positivity conditions in Section \ref{sec:positivity}.

As stated above, not every KFT is conformal (just like not every unitary QFT is conformal, and not every statistical field theory is conformal).  Those that are conformal, we refer to as KCFTs; these would be candidates for dS/CFT models.  This requires looking for KFTs that are also fixed points of the RG flow.  Conceptually, one expects such fixed points to appear generically at non-zero values of $\lambda$, only if the conditions for being a KFT are closed under RG flow.  We will show that this is the case at leading order in a 1/N expansion in Section \ref{sec:closure}.  But additional subtleties arise at subleading orders in when integrating out a handle of the worldsheet.  Thus, if we restrict attention to large $N$, then the identification of KCFTs (and thus dS/CFTs) at large values of $|\lambda|$ may be tricky, but becomes a well-defined question in principle.\footnote{Although lattice simulations are likely infeasible due to the presence of complex phases.}  

If the boundary field theory were unitary, the number of colours would be quantised ($N \in \mathbb{Z} \subseteq \mathbb{R}$), which is possible for a KFT only when $N^2 > 0$, implying $d = \text{odd}$.  Remarkably, it turns out that several other nice properties also only occur when $d = \text{odd}$.  As we discuss in Section \ref{sec:central}, these are the only dimensions where ``central charge''-like quantities, counting the number of degrees of freedom, are real.  Also, as we shall show in Section \ref{sec:closure}, when $d = \text{odd}$ the conditions for being a KFT are also closed under RG at subleading orders in $1/N$.  Perhaps it is only in these cases where it is nonperturbatively consistently take $N$ to be finite.  

Furthermore, it is also only possible to consistently include open or non-oriented strings when $d = \text{odd}$!  For other values of $d$, we cannot make the assignments work consistently for either non-oriented strings (with gauge group O($N$) or Sp($N$)), nor can we consistently add ``quarks'' in the fundamental representation.  The reason is that in string theory, open and non-orientable string diagrams are potentially ambiguous: the same value of $\chi$ can be interpreted as a Feynman diagram with different numbers of bulk loops $L$.  And the phase required by \eqref{eqn:HA} is only independent of $L$ when $d = \text{odd}$.

This suggests that perhaps it will be easier to find consistent examples of holographic cosmology with an even number of spacetime dimensions on the gravity side.  As the only experimentally known instance of quantum cosmology has $d = 3$, this conjecture is at least compatible with the experimental data, and might even help to explain it.

\subsection{The Sign of the String Tension}\label{sec:tension}

Since it is hard (although likely not impossible) to find string theory vacua in de Sitter, we expect a corresponding issue to arise when constructing ``good'' KFTs.  And, indeed, there does remain at least one serious unsolved problem from our list in Section \ref{sec:obstacles} --- although whether it is the ``same'' problem remains to be seen.

Specifically, the most important remaining issue that we will \emph{not} solve in this paper is the dual problems of how to: i) avoid tachyons in the bulk spectrum, and ii) obtain massive particles in the principal series.  Ideally, somebody else will write a paper solving these problems, and then hopefully their solution can be combined with a reality condition like the one in this paper, to provide a well-behaved class of holographic cosmology models.\nopagebreak[3]

While the existence of a tachyon field is actually compatible with bulk unitarity\footnote{This is true notwithstanding the absence of unitary representations of the dS group with $\Delta$ primaries corresponding to tachyonic values $m^2 < 0$.  As a tachyon field has some modes that exponentially grow or decay, there is no invariant notion of a 1-tachyon Hilbert space.  But this does not preclude an action of the de Sitter group on the complete Hilbert space of a free tachyon field, which can presumably be decomposed into the usual unitary representations.} --- in the sense of preservation of a positive norm, under finite amounts of Lorentzian time evolution --- it signals an instability of the vacuum, or rather an inability to invariantly define a vacuum state in the first place, as well as (relatedly) causing a new instability to appear in the Euclidean piece of the Hartle-Hawking contour.

As an 't Hooft string will typically have an infinite number of excitations with different $m^2$ values, the solutions to (i) and (ii) are likely to be the same as each other --- the point is that the bulk string needs to have positive tension instead of negative tension.  Regardless of the sign of $\lambda$, it does not appear to be possible to solve this problem for a weakly coupled gauge theory with adjoint fields, since in the free limit the value of single-trace operators will be given by their engineering dimensions, which are always real.  Worse still, one expects an infinite number of tachyon fields, whose $m^2$ is unbounded below.

It might be (in the sense that the authors are not aware of a theorem to the contrary) that a strongly coupled theory with large and complex $\lambda$ would solve this problem.  In ${\cal N} = 4$ Super Yang-Mills, operators whose dimension $\Delta$ is not protected all become heavy ($\Delta \gg 1$) in the holographic limit.  Perhaps, in the dS/CFT case, the holographic limit sends unprotected operators to large \emph{complex} dimensions, in a way that matches unstable massive particles.\footnote{It might also be that, in the limit $\lambda \gg 1$, the tachyonic masses get large $m^2 \to -\infty$, in which case it would still be possible to \emph{formally} integrate out the tachyon fields to obtain a (hopefully now stable) EFT limit.  But this is less satisfactory since at any finite value of $\lambda$, the tachyons would still be there.}

Or, it might instead be the case that some fundamentally new idea is required to obtain a field theory dual to a positive tension string in dS.  Whatever that new idea turns out to be, we expect that consistency with \eqref{eqn:HA} will remain an important check on what kinds of boundary duals are acceptable in holographic cosmology models.

That said, the existence of a tower of tachyons does not seem to lead to a direct mathematical inconsistency, in the context of this paper's results, at least not at the level of bulk perturbation theory.  If all operators in the CFT have positive dimension $\text{Re}(\Delta) > 0$, then on a technical level, the tachyon instability actually does not lead to divergent results as $\eta \to 0^-$ in dS-Poincar\'{e}, due to the postselection of the irrelevant ``source'' to be finite at late times.  (Assuming we interpret the Bunch-Davies initial boundary condition in such a way as to also disallow tachyon modes that exponentially grow to the past.)  And yet, the instability is still present at the level of unrestricted bulk unitary time evolution from given initial conditions.  Relatedly, if we turn on an irrelevant source with $\Delta > d$, then the amplitude of the dual bulk field grows with time.  Hence, at late enough times we expect that bulk perturbation theory should break down (at least when expanded around the original vacuum).  However the same is true for irrelevant sources in AdS/CFT, in the $z \to 0$ limit.

At a bare minimum, one can still use a dS/CFT dual to tachyons to calculate perturbative processes involving only non-tachyonic operators (e.g.~those in the complementary and/or discrete series, including e.g.~gravitons, bulk gauge bosons, and scalars in the range $0 \le \Delta \le d$).  We find it remarkable that, after imposing the KFT conditions, such amplitudes will perfectly satisfy the correct (often non-trivial) complex phases for a unitary bulk dual at tree level (modulo possible $\pm$ signs).\footnote{At higher loop orders, it will be more difficult to exclude the contributions of tachyons, since they might run around the loops.}  

\subsection{Plan of the Paper}
The plan of the rest of this article is as follows:  

In Section \ref{sec:constraints}, we will review recent work on reality and positivity conditions for the Bunch-Davies wavefunction. Our main results are in Section \ref{sec:reality}, where we apply these conditions to boundary gauge theories, and use them to determine the phases for $N^2$ and $\lambda$, up to an overall $\pm$ sign.  We will also discuss issues related to the choice of boundary gauge group (related to oriented vs.~non-oriented) and the matter sector (related to closed vs.~open), and the issue of when the KFT condition is closed under RG flow.  In Section \ref{sec:positivity}, we make a first stab at determining this $\pm$ sign from bulk unitarity, although (apart from the central charge) our analysis is limited to the case of weakly coupled boundary 2-point functions.  In Section \ref{sec:EvenvsOdd} we give five reasons for thinking that holographic cosmology makes more sense when $d = \text{odd}$ than when $d = \text{even}$. Finally, in Section \ref{sec:discussion} we discuss our results and possible future directions.

\section{Constraints for Holographic Cosmology}\label{sec:constraints}
\begin{figure}
    \centering
    \includegraphics[width=0.6\textwidth]{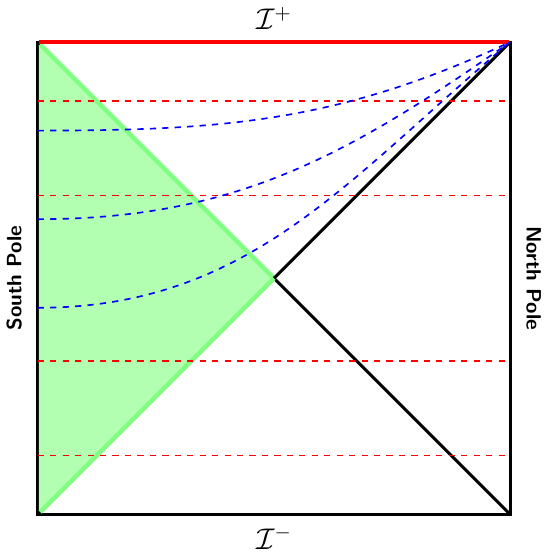}
    \caption{Penrose diagram of de Sitter spacetime illustrating multiple foliations and key geometric features. The blue dashed lines represent the Poincaré slicing, which is conformally flat, covers only half of the spacetime, and asymptotes to the future boundary (shared with the global slicing of de Sitter); this slicing is most relevant for inflationary cosmology, as it naturally describes an expanding universe with flat spatial sections. In this framework, we are metaobservers at $\mathcal{I}^+$, where late-time cosmological correlators are measured. The red dashed lines correspond to the global slicing, which foliates the entire spacetime into spatial $(D-1)$-spheres of constant global time. The thick red horizontal line at the top denotes the future de Sitter boundary $\mathcal{I}^+$, where the holographic dual in the dS/CFT correspondence is proposed to reside and encode information about cosmological correlators. The green shaded region indicates the static patch accessible to an observer at the South Pole. Each point in the interior of the diagram corresponds to a $(D-2)$-sphere whose radius varies across the diagram: it shrinks to zero at the left and right edges, representing the South Pole and North Pole respectively, which are the poles of the spatial slices. This shrinking radius captures the spherical geometry of spatial sections in global coordinates.}
    \label{fig:deSitter}
\end{figure}
Inspired by the AdS/CFT correspondence, holographic cosmology posits that quantum gravity in cosmological spacetimes --- particularly in de Sitter (dS) space --- can be described by a lower-dimensional conformal field theory (CFT), i.e.~a dS/CFT correspondence. An early version of the dS/CFT correspondence was proposed by Strominger in~\cite{Strominger:2001pn}. Strominger conjectured that quantum gravity in asymptotically Lorentzian de Sitter space is dual to a Euclidean CFT living on the spacelike future boundary $\mathcal{I}^+$ of dS (see Figure~\ref{fig:deSitter}). This conjecture was based on the matching of isometries between de Sitter space and the Euclidean conformal group, and supported by the asymptotic behaviour of bulk fields near $\mathcal{I}^+$, which Strominger interpreted as defining boundary operators in the dual CFT. However, he did not frame the correspondence in terms of the Wavefunction of the Universe (WFU) (sometimes referred to as the cosmological wavefunction) or a bulk gravitational path integral.

In contrast, Maldacena~\cite{Maldacena:2002vr} later proposed a refinement of dS/CFT, where the Hartle-Hawking (or Bunch-Davies) WFU is computed via an analytic continuation of a Euclidean AdS path integral. This approach provides a more concrete realisation of dS/CFT by leveraging the better-understood gravitational path integral in AdS, providing a more precise bulk gravitational definition of the duality. In this framework, the WFU is interpreted as the generating functional for the dual Euclidean CFT. In $d$ spatial dimensions (hence $D=d+1$ spacetime dimensions), the wavefunction $\Psi[\varphi]$ for a bulk field with boundary value $\varphi$ is given by
\begin{equation}\label{eqn:BDWFUGeneratingFunctional}
    \Psi[\varphi] \equiv Z_{\text{dS}}[\varphi]
    =\exp{-\sum_{n=2}^\infty \left[\prod_{a=1}^n \int \frac{\diff^d \bfk_a}{(2\pi)^d}\varphi (\bfk_a) \right] \psi_{n}(\bfk)} \, ,
\end{equation}
where the coefficients $\psi_n(\mathbf{k})$ are the boundary wavefunction coefficients at future infinity. The resulting wavefunctional $\Psi[\varphi]$ is typically taken to be the Bunch–Davies state~\cite{BunchDavies1978}, which corresponds to the inflationary vacuum in QFT on curved spacetime, but also arises naturally from the gravitational path integral in quantum cosmology. In particular, it arises as the late-time semiclassical limit of the Hartle–Hawking ``no-boundary’’ wavefunction~\cite{Hartle:1983ai,Halliwell:1984eu}, which is a proposed solution to the WDW equation, 
defined via a Euclidean path integral over compact 4-geometries with a specified boundary 3-metric $\gamma$: 
\begin{equation}
    \Psi_{\text{HH}}[\gamma, \phi] = \int \mathrm{D}g \, \mathrm{D}\varphi \, e^{-S_E[g, \varphi]} \, .
\end{equation}
This defines a wavefunctional over spatial geometries and field profiles on a final spacelike slice. In the semiclassical (WKB) regime where the path integral is dominated by classical saddles, one can write
\begin{equation}
    \Psi[h_{ij}, \phi] \approx \exp\left( \frac{i}{\hbar} W[h_{ij}, \phi] \right) \, ,
\end{equation}
where $W$ is the Hamilton–Jacobi functional evaluated on the classical solution~\cite{Pimentel:2013gza,Khan:2023ljg}. This connects the no-boundary proposal to the Bunch–Davies wavefunctional, providing a conceptual bridge between quantum cosmology and quantum field theory in curved spacetime (see also~\cite{deAlwis:2018sec} for a recent attempt to make this equivalence more precise). The Bunch–Davies wavefunctional is evaluated in terms of a field basis $\varphi$,\footnote{Here $\varphi$ can be any spinning field, hence for the case of pure gravity $\varphi$ would correspond to boundary gravitons/metric fluctuations, with their indices suppressed.} usually living at the late-time inflationary boundary at $\eta_0=0^-$ (but in principle can be defined at any finite time slice $\eta_0$), i.e. it is given by the following QFT in curved spacetime path integral
\begin{equation}\label{eqn:IntroWFU}
    \Psi_{\text{BD}}[\eta_0;\varphi] = \int_{\phi_{\text{BD}}(-\infty)=0}^{\phi(\eta_0)=\varphi} \mathrm{D}\phi \, e^{iS[\phi]} =\exp{-\sum_{n=2}^\infty \left[\prod_{a=1}^n \int \frac{\diff^d \bfk_a}{(2\pi)^d}\varphi (\bfk_a) \right] \psi_{n}(\eta_0;\bfk)} \, ,
\end{equation}
and is conjectured to have the form of a generating functional of a Euclidean CFT, as expressed in the right hand side of \eqref{eqn:BDWFUGeneratingFunctional}. Notice that this parameterisation does not require any saddle-point approximation of the bulk path integral that defines $\Psi$. In fact, the wavefunction coefficients $\psi_n$ can be found non-perturbatively from
\begin{equation}
        \psi_n(\eta_0;\bfk) \equiv \psi'_n(\eta_0;\bfk) (2 \pi)^d \delta^d \bigg(\sum_{a=1}^n \bfk_a \bigg)=- \frac{\delta^n \log\Psi [\eta_0;\varphi]}{ \delta \varphi_{\bfk_1}\cdots \delta \varphi_{\bfk_n}  } \bigg|_{\varphi  = 0} \, ,
\end{equation}
where the prime in $\psi'_n$ is used to explicitly highlight that these wavefunction coefficients contain the momentum-conserving $\delta$-function, which comes from the Fourier transform of translation invariance.

\subsection{Relation to the Cosmological CPT Theorem}
The Cosmological CPT Theorem~\cite{Goodhew:2024eup,Thavanesan:2025kyc} provides a foundational framework for understanding the interplay of the discrete symmetry of \CRT with Scale Invariance and Unitarity in realistic cosmological settings. (This work builds on the foundation of the perturbative unitarity constraints derived in the Cosmological Optical Theorem~\cite{COT,Cespedes:2020xqq,Goodhew:2021oqg,Melville:2021lst}.) The key insight underlying the theorem is that the Poincaré patch of de Sitter allows for a natural analytic continuation between expanding and contracting patches. A global $\text{SO}^+(1,1) \subset \text{SO}^+(d+1,1)$ Lorentz boost in de Sitter becomes, when restricted to a single Poincaré patch, a scaling transformation $\mathbf{D}_{\lambda}$ acting as 
\begin{equation}
    (\eta, \mathbf{x}) \mapsto (\lambda \eta, \lambda \mathbf{x}) \, .
\end{equation}
The discrete $180^\circ$ rotation corresponds to an analytic continuation $\mathbf{D}_{-1} : (\eta, \mathbf{x}) \mapsto (e^{\pm i\pi}\eta, e^{\pm i\pi}\mathbf{x})$, which we term Discrete Scale Invariance. In a realistic inflationary context where full de Sitter invariance (specifically SCTs), is broken, this discrete remnant survives and becomes central to identifying symmetry constraints on observables.\footnote{In~\cite{Boyle:2018rgh,Boyle:2018tzc,Boyle:2021jej,Turok:2022fgq,Boyle:2022lyw,Turok:2023amx,Deng:2024uuz} the authors assume the existence of a CPT-reflected universe to derive observational constraints. The Cosmological CPT theorem not only proves that \CRT \emph{is} a symmetry of the expanding universe, but also enables us to determine non-trivial constraints in the expanding Poincaré patch without any need for analytic continuation to a contracting one.} At the level of the cosmological correlators these symmetries act as a $\mathbb{Z}_2 \times \mathbb{Z}_2$ structure generated by \CRT and $\mathbf{D}_{-1}$, but at the level of the wavefunction this extends to a non-abelian cover group $\text{Aut}(\mathbb{Z})$.

The Cosmological CPT Theorem relates these discrete symmetries via the following logic:
\begin{center}
    \text{Scale Invariance} + \text{Unitarity} $\Longrightarrow$ \text{\CRT invariance} \, ,
\end{center}
\begin{center}
    \text{Discrete Scale Invariance} + \text{Reflection Reality} $\Longrightarrow$ \text{\CRT invariance} \, ,
\end{center}
\begin{center}
    \text{\CRT invariance} + \text{Discrete Scale Invariance} $\Longrightarrow$ \text{Reflection Reality} \, ,
\end{center}
\begin{center}
    \text{\CRT invariance} + \text{Reflection Reality} $\Longrightarrow$ \text{Discrete Scale Invariance} \, .
\end{center}
Remarkably, the \CRT symmetry results in a constraint on the phase of the wavefunction coefficients $\psi_n$ in dS/CFT at $\eta = 0$.\footnote{In some dimensions the phase at higher loop order is different from the phase at tree-level.  There are also additional subtleties associated with bulk UV and IR divergences.} This constrains the wavefunction coefficients to be some specific complex phase (up to a real sign) which is only purely real or purely imaginary in some special cases.

On the other hand, Reflection Reality (\RR) is a fundamental requirement of bulk unitarity that all physically consistent theories must satisfy.  This too fixes the complex phase of the wavefunction coefficients, but even at $\eta = 0$ it is necessary to do some analytic continuations. (In~\cite{Goodhew:2024eup} this was achieved by finding the form \CRT at the late-time boundary where $\eta = 0$, which can be derived independently by combining \RR with the boundary form of scale invariance, i.e.~what CFT people call dilatations.) However, the advantage of \RR is that it can be applied to \emph{any} flat Friedmann-Lemaître-Robertson-Walker (FLRW) cosmology, even without scale invariance.\footnote{We believe that the constraints in our paper probably also apply to non-flat cosmologies, but our current derivation uses flatness.}  

For ease of notation we will refer to this constraint on $\psi_n$ as \RR below.  However in a holographic cosmology context (e.g.~dS/CFT), it should always be remembered that when we say \RR, this refers to the implications of the \emph{bulk} \RR symmetry on the boundary theory.  The boundary field theory itself is neither reflection-positive, nor typically even reflection-real!  

\subsection{Non-perturbative Constraints on the Cosmological Wavefunction}
We will now examine how these symmetries act on the WFU $\Psi$ and particularly on its Taylor-expanded coefficients $\psi_n$ in \eqref{eqn:IntroWFU}, which encode bulk correlation functions (see e.g.~\cite{Maldacena:2002vr,Pimentel:2013gza,Bzowski:2011ab,Bzowski:2012ih,Bzowski:2013sza,Bzowski:2019kwd,Bzowski:2023nef,WFCtoCorrelators1,WFCtoCorrelators2,Abolhasani:2022twf,BaumannJoyce:2023Lecs,Cespedes:2023aal}). A key result of the Cosmological CPT framework is that these constraints of $\psi_n$ hold non-perturbatively, allowing us to go beyond previous results reliant on perturbative expansions~\cite{COT,Cespedes:2020xqq,Goodhew:2021oqg,Melville:2021lst}. 

Throughout this section, we use ``non-perturbative'' to refer to results that hold non-perturbatively within a bulk QFT on a \emph{fixed} curved spacetime background, as we don't know how to treat gravitons non-perturbatively.\footnote{Although our derivation of the discrete symmetry is not completely rigorous for the case of perturbative gravitons because of the conformal mode problem in Euclidean signature, an examination of their wavefunction coefficients shows that they do in fact obey exactly the same conditions as perturbative matter fields.}  Of course these results can be expressed at each order in perturbation theory, but they have additional content if the amplitude is non-analytic in the coupling constant. That is, the constraints are not derived by truncating at any particular loop order, but rather follow from symmetry principles or the analytic structure of QFT in curved spacetime. However, we do not claim to include genuinely non-perturbative \emph{quantum gravitational} effects --- such as gravitational instantons, vacuum tunnelling, or large-$N$ resummation --- unless explicitly stated.

The action of the discrete symmetries on the wavefunction coefficients is:
\begin{align}
    \mathbf{CRT}:& \quad \left[\psi_n(\eta; \textbf{x}; \Omega)\right]^* = \psi_n(e^{-i\pi}\eta; -\textbf{x}; e^{-i\pi}\Omega) \, , \\
    \mathbf{D}_{-1}^{\pm}:& \quad \psi_n(\eta; \textbf{x}; \Omega) = \psi_n(e^{\pm i\pi} \eta; e^{\pm i\pi} \textbf{x}; \Omega) \, , \\
    \mathbf{RR}:& \quad \left[\psi_n(\eta; \textbf{x}; \Omega)\right]^* = \psi_n(\eta; -e^{i\pi} \textbf{x}; e^{-i\pi}\Omega) \, , \label{eqn:NonPertRR}
\end{align}
where the complex analytic continuation of the Weyl factor $\Omega$ ensures these transformations stay within the original dS space (or, for the case of \RR, any general flat FLRW spacetime). 

Importantly, because $\psi_n$ arises from a single-sheeted path integral (unlike in-in correlators $B_n$),\footnote{Some similar, yet distict, analytic continuations in the in-in formalism were performed in~\cite{Sleight:2019hfp,Sleight:2019mgd,Sleight:2020obc,Sleight:2021iix,Sleight:2021plv,Sleight:2023ojm,Chopping:2024oiu,Pacifico:2025emk,MdAbhishek:2025dhx,Sleight:2025dmt}.} it can carry non-trivial phases under these continuations due to monodromies, especially around branch points in the complex $\eta$- and \textbf{x}-planes. Given that \RR represents a non-perturbative constraint that \emph{any wavefunction arising from bulk unitary time evolution is required to satisfy}, this is the cosmological or dS/CFT analogue of reflection positivity.  This answers one of the outstanding fundamental questions of how bulk unitary time evolution manifests itself in holographic cosmology, where the boundary theory is seemingly ``non-unitary'', but still dual to a unitary bulk.

\subsubsection*{Perturbative Constraints}
We can also expand the non-perturbative identities above to arbitrary loop order in perturbation theory. In this context, we write the loop-expanded wavefunction coefficient as
\begin{equation}
    \psi_n = \sum_{L=0}^\infty \psi^{(L)}_n \, ,
\end{equation}
where $\psi^{(L)}_n$ denotes the contribution from $L$-loop diagrams in the bulk. Taking the Fourier transform, we find:
\begin{align}
    \text{\CRT} :& \quad \left[\psi^{(L)}_n(\eta_0; \textbf{k})\right]^* = e^{i\pi (d+1)(L-1)} \psi^{(L)}_n(e^{-i\pi}\eta_0; \textbf{k}) \, , \label{eqn:PertCRT} \\ 
    \mathbf{D}_{-1}^{\pm} :& \quad \psi^{(L)}_n(\eta_0; \textbf{k}) = e^{\pm i\pi d} \psi^{(L)}_n(e^{\pm i\pi}\eta_0; e^{\mp i\pi}\textbf{k}) \, , \\
    \text{\RR} :& \quad \left[\psi^{(L)}_n(\eta_0; \textbf{k})\right]^* = e^{i\pi \left((d+1)L-1\right)} \psi^{(L)}_n(\eta_0; e^{-i\pi}\textbf{k}) \, . \label{eqn:PertRR}
\end{align}
These relations hold for UV-finite amplitudes, or for power-law divergent diagrams if you throw out such divergences (this can always be done, without introducing a new scale, by choosing an appropriate RG scheme).  Log divergent diagrams are more subtle, but these formulae can be applied directly if the divergence is regulated using dimensional regularisation (dim-reg) to non-integer $d$.  However, just as in flat space QFT, one must be careful and consistent with the choice of renormalisation scheme.  In loop integrals, there can be non-trivial finite terms arising from the Taylor expansion to non-integer $d$, if these terms multiply with a pole (log divergence) in the $d \to \text{integer}$ limit; hence even if one is ultimately interested in integer $d$, it is necessary to know the correct phases at non-integer values of $d$.  On the other hand, tree-level diagrams ($L = 0$) are UV divergence-free, and therefore respect the above identities exactly regardless of the choice of renormalisation scheme.  

Crucially, for odd $d$, the $\Omega$-dependence in the non-perturbative symmetries becomes trivial, and so (apart from the non-trivial shifts described above, arising from UV log divergences) in this case the phase is independent of the loop order $L$.

These results provide a precise and universal classification of the phases up by the cosmological wavefunction coefficients under discrete symmetries. They can be used to predict interference patterns, identify allowed EFT interactions, and exclude certain non-local or non-unitary theories at the level of the wavefunction~\cite{Cabass:2022rhr}.  This lends hope that preserving these discrete symmetries may also help resolve some of the deeper questions concerning in quantum gravity in de Sitter space.

\paragraph{Phase of the de Sitter Wavefunction.}
The \CRT formula \eqref{eqn:PertCRT} has significant implications for holographic cosmology. By pushing the wavefunction coefficients to the future boundary where, for light fields, the time dependence factorises straightforwardly, and the resulting boundary wavefunction coefficient, $\overline{\psi}^{(L)}_n$, no longer depends on $\eta$. Thus, we find it is possible to deduce the phase of an arbitrary wavefunction coefficient in de Sitter spacetime:
\begin{equation}\label{eqn:CPTPhase} 
    \arg(\overline{\psi}^{(L)}_n)
    = -\frac{\pi}{2}\!\left((d+1)(L-1)+dn-\sum_{\alpha} \Delta_{\alpha}\right) + \pi \mathbb{N} \, ,
\end{equation}
where the last term indicates that \CRT can only fix the phase up to an overall $\pm$ sign. $\Delta$ is the conformal dimension of the $n$ operators $\overline{\phi}_{+}$ used to differentiate the boundary wavefunction $\overline \Psi[\overline{\phi}_{-}]$, which is written in the basis of the conjugate operators $\overline{\phi}_{-}$ with dimension $d-\Delta$.

From a holographic cosmology or dS/CFT perspective, Eq.~\eqref{eqn:CPTPhase} allows us to determine the phases of arbitrary $n$-point functions in the dual CFT. In dS/CFT, the objects with dimension $d-\Delta$ are the ``sources'' of the CFT partition function $Z_\text{CFT} = \overline{\Psi}$, while the ones with dimension $\Delta$ are the ``operators''. But from a cosmology perspective, both are operators in the bulk Hilbert space, and we have to choose a basis. In cosmology, one normally chooses this basis by writing $\overline \Psi$ as a function of whichever field component falls off more slowly, which would imply that $\Delta \ge d/2$, but the phase formula above would still be valid if we make the opposite choice.\footnote{While an extension of these results to heavy fields in the principal series, where $\Delta = d/2 + i\nu$, we leave this for future work. Doing so would require making the unpleasant decision to either sacrifice self-adjointness, or conformal invariance of the boundary conditions. See e.g.~\cite{Anous:2020nxu,Anninos:2023lin,Joung:2006gj,Sengor:2019mbz,Sengor:2021zlc,Sengor:2022hfx,Sengor:2022lyv,Sengor:2022kji,Sengor:2023buj,Dey:2024zjx} for discussions regarding the principal series.} This result holds for any boundary wavefunction coefficients that are UV and IR-finite (i.e. free from logarithmic divergences in $\eta$ or the UV cutoff $\epsilon$), as well as the coefficient in front of the leading $\log(\eta)$ divergence for any IR-divergent wavefunction coefficients, or the leading $\log(\epsilon)$ divergence for any UV-divergent diagram.\footnote{There is also an extra correction for diagrams with spinor propagators, see~\cite{Goodhew:2024eup} for more details.}  

If we restrict attention to the case where all particles have $\Delta=d$, as for the stress-tensor and massless scalar fields, we find the phase \eqref{eqn:CPTPhase} is independent of $n$; hence massless scalar fields and gravitons have real boundary wavefunction coefficients to all loop order in even spacetime dimension, provided they don't diverge in the UV.  For tree-level, IR-finite and scale-invariant Feynman-Witten diagrams in cosmology, this was recently found independently~\cite{Stefanyszyn:2023qov,Stefanyszyn:2024msm}, with additional implications explored in \cite{Thavanesan:2025kyc}.\footnote{For UV divergences more care is needed, see e.g.~\cite{Thavanesan:2025kyc,Melville:2021lst,Lee:2023jby,Senatore:2009cf}.}

In summary, the Cosmological CPT Theorem provides a symmetry-based foundation for understanding the analytic structure of cosmological wavefunctions. Its implications range from the consistency of the dS/CFT correspondence, to constraints on the phase structure of cosmological correlators, to no-go theorems for certain parity-violating interactions. This framework therefore offers a robust and theoretically checkable approach to quantum gravity in spacetimes with a positive cosmological constant.

\section{Reality Conditions for Kosmic Field Theories}\label{sec:reality}
Although the boundary theory is non–reflection positive in dS/CFT and thus unusual compared to the familiar AdS/CFT examples, the Cosmological CPT Theorem~\cite{Goodhew:2024eup} implies non-perturbative constraints that we can impose on the putative dual. In particular, we can ask: \emph{what classes of holographic field theories are consistent with} \textbf{bulk unitarity}?

\subsection{Bulk Unitarity in Holographic Cosmology}
A non-perturbative constraint corresponding to unitary bulk time evolution in any flat FLRW background was established in~\cite{Goodhew:2024eup}. Importantly, this \RR condition does not presuppose conformal invariance of the bulk spacetime and thus the dual boundary theory: it applies more generally to large-$N$ holographic field theories, with conformal field theories (CFTs) arising only as the special subclass realised at conformal fixed points. In terms of wavefunctional coefficients,
\begin{equation}\label{eqn:BulkRR}
    \textbf{RR:}\qquad \left[\psi_n(\eta;\mathbf{x};\Omega)\right]^* =\psi_n\left(\eta;-e^{i\pi}\mathbf{x};e^{-i\pi}\Omega\right) \, ,
\end{equation}
which holds on any Cauchy slice, i.e.~for any value of conformal time $\eta$ (including the asymptotic future boundary $\eta \to 0^-$, where the dual theory is proposed to live in dS/CFT).  Regardless of whether $\eta = 0^-$ or $\eta = \text{finite}$, we will hold $\eta$ fixed in what follows.  Accordingly, we will not explicitly write the dependence on $\eta$ in the functional form of $\psi_n$ (although it is implicitly there).

Instead, we will consider the effects of performing a uniform complex Weyl transformation $\bar{\Omega}$ of the \emph{boundary} theory.  Thus we write: 
\begin{equation}\label{eqn:BoundaryRR}
    \textbf{RR:}\qquad \left[\psi_n(\mathbf{x};\bar{\Omega};\Omega)\right]^* = \psi_n\left(-e^{i\pi}\mathbf{x};\bar{\Omega};e^{-i\pi}\Omega\right) \, ,
\end{equation}
with $\bar{\Omega}$ and $\Omega$ denoting the Weyl factors of the boundary and bulk metrics, respectively. A boundary Weyl rotation by $e^{i\pi}$, accompanied by a complex diffeomorphism\footnote{The boundary Weyl rotation will have the effect of also rotating any boundary operator insertions. Since we are assuming our boundary theory depends covariantly on the background metric, all diffeomorphism transformations are automatically preserved in our covariant theory. Hence, we can use diffeomorphism transformations to rotate the operator insertions back!}, can be used to remove the $e^{i\pi}\mathbf{x}$ rotation, yielding the equivalent relation:
\begin{equation}\label{eqn:BoundaryRR2}
    \textbf{RR:}\qquad \left[\psi_n(\mathbf{x};\bar{\Omega};\Omega)\right]^* = \psi_n\left(-\mathbf{x};e^{i\pi}\bar{\Omega};e^{-i\pi}\Omega\right) \, .
\end{equation}
For parity-even theories, this further simplifies to
\begin{equation}\label{eqn:BoundaryRR3}
    \textbf{RR:}\qquad \left[\psi_n(\mathbf{x};\bar{\Omega};\Omega)\right]^* = \psi_n\left(\mathbf{x};e^{i\pi}\bar{\Omega};e^{-i\pi}\Omega\right) \, ,
\end{equation}
which is the form we will use in the analysis which follows.  

This relation may also be directly justified by noting that the combination of the bulk $\Omega$ and the boundary $\bar{\Omega}$ is simply to deform the $\sqrt{g_{tt}}$ component of the metric, from the value $+i$ to $-i$, which is equivalent to replacing an expanding universe with a contracting universe\footnote{The bulk Weyl factor $\Omega$ and boundary Weyl factor $\bar{\Omega}$ must rotate in opposite directions under \RR to ensure that only the time direction is Wick-rotated, while spatial directions remain spacelike. Concretely, the bulk metric takes the conformally flat form $\diff s^2 = \Omega^2(-\diff\eta^2 + \diff\mathbf{x}^2)$, whereas the boundary metric is induced as $\bar{g}_{ij} \sim \bar{\Omega}^2 \delta_{ij}$. If both $\Omega$ and $\bar{\Omega}$ were rotated by the same phase, all coordinates — including the spatial ones — would be Wick-rotated, rendering the boundary metric fully timelike or spacelike. By instead taking
\begin{equation}
    \Omega \to e^{-i\pi}\Omega \, , \qquad \bar{\Omega} \to e^{+i\pi}\bar{\Omega} \, ,
\end{equation}
the analytic continuation affects only the time coordinate, keeping spatial slices spacelike and preserving the physical interpretation of \RR as a reflection through the light cone.}. By \CRT, the resulting Bunch-Davies amplitudes must then be the complex conjugate of the original amplitude.\footnote{This result is for a Lorentz-invariant, parity-even theory.  In a theory with terms in the Lagrangian that are odd under parity or time reversal, one must use \eqref{eqn:BoundaryRR2}, and the analysis of \cite{Goodhew:2024eup} shows that the result really requires the local Lagrangian to satisfy \RR symmetry.}  Although in \cite{Goodhew:2024eup} we derived the \RR relation for flat FLRW cosmologies in the Bunch-Davies state, this justification appears to work for the Bunch-Davies state in any perpetually expanding cosmology.

From the point of view of complex analysis, it is important to note that this rotation of $\sqrt{g_{tt}}$ stays in the region where ordinary QFT path integrals should converge.  This argument is valid at the level of bulk QFT.  Although as usual the graviton is subject to the conformal mode problem, inspection of simple amplitudes indicates that perturbative graviton amplitudes appear to obey the same relationship.\footnote{In the previous argument, we were not taking into account gravitational back-reaction, but just doing bulk QFT in curved spacetime.  Oddly, if we \emph{do} take into account gravitational back reaction on the geometry (without worrying too much about the conformal mode problem) then the final result \eqref{eqn:BoundaryRR3} \emph{also} appears to also apply to global de Sitter in the Hartle-Hawking state.  Although one might worry at first that the bulk $\Omega$ rotation will do bad things in the Euclidean part of the Hartle-Hawking contour, this is compensated for by the fact that an $e^{i\pi/2}\bar{\Omega}$ half Weyl rotation, also changes the (gravitationally back-reacted) scale factor $a(t)^2$ from $\cosh^2(t)$ to $\sinh^2(t)$, resulting in a hyperbolic geometry of fixed signature.  Then, performing the remaining half of the $\bar{\Omega}$ rotation, one comes back to $\cosh^2(t)$ again, but now with the opposite sign of the Euclidean piece of the Hartle-Hawking contour.  Upon doing the $\Omega$ rotation, one ends up with the correct signs for the usual contracting Hartle-Hawking contour.}

\begin{figure}[ht]
    \centering
    \includegraphics[width=.9\textwidth]{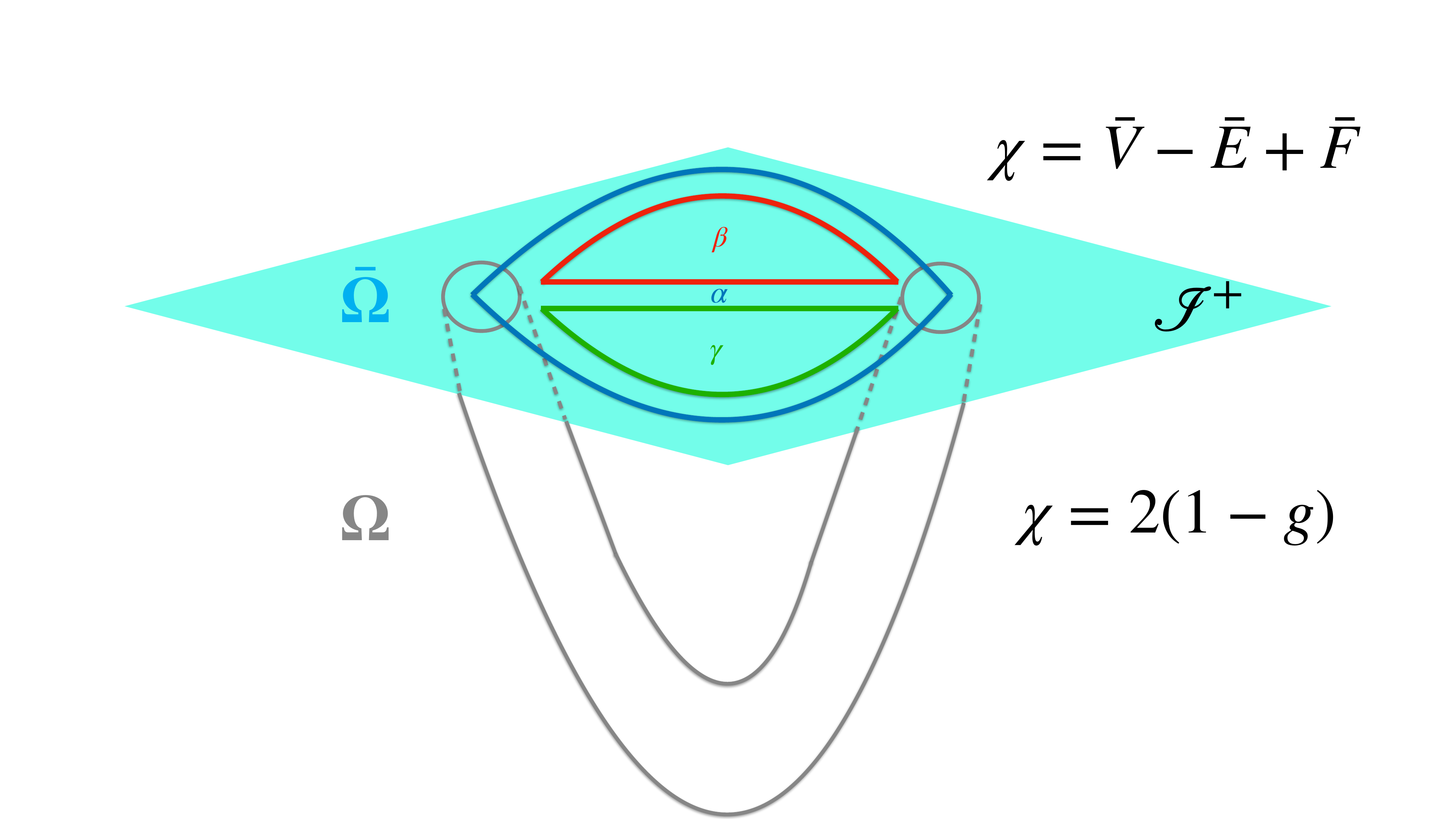}
    \caption{A connected Feynman diagram with Euler characteristic $\chi=\bar{V}-\bar{E}+\bar{F}$ on the boundary metric $\bar{g}_{ab}$ and Weyl factor of $\bar{\Omega}$ is dual to a string worldsheet of genus $g$ with $\chi=2(1-g)$ with bulk metric $g_{ab}$ and Weyl factor of $\Omega$. This motivates the interpretation of large-$N$ gauge theories in the ’t Hooft limit as dual to closed strings, even in cosmological settings. Schematic correspondence between a 2-loop boundary ribbon graph at $\mathcal{I}^+$ and bulk closed-string worldsheet sphere (genus $g=0$) diagram. The simplest ribbon Yang-Mills diagram (with at least 1 vertex) that contributes at bulk tree-level is shown.}
    \label{fig:BulkStringBoundaryGenusZero}
\end{figure}

\begin{figure}[ht]
    \centering
    \includegraphics[width=\textwidth]{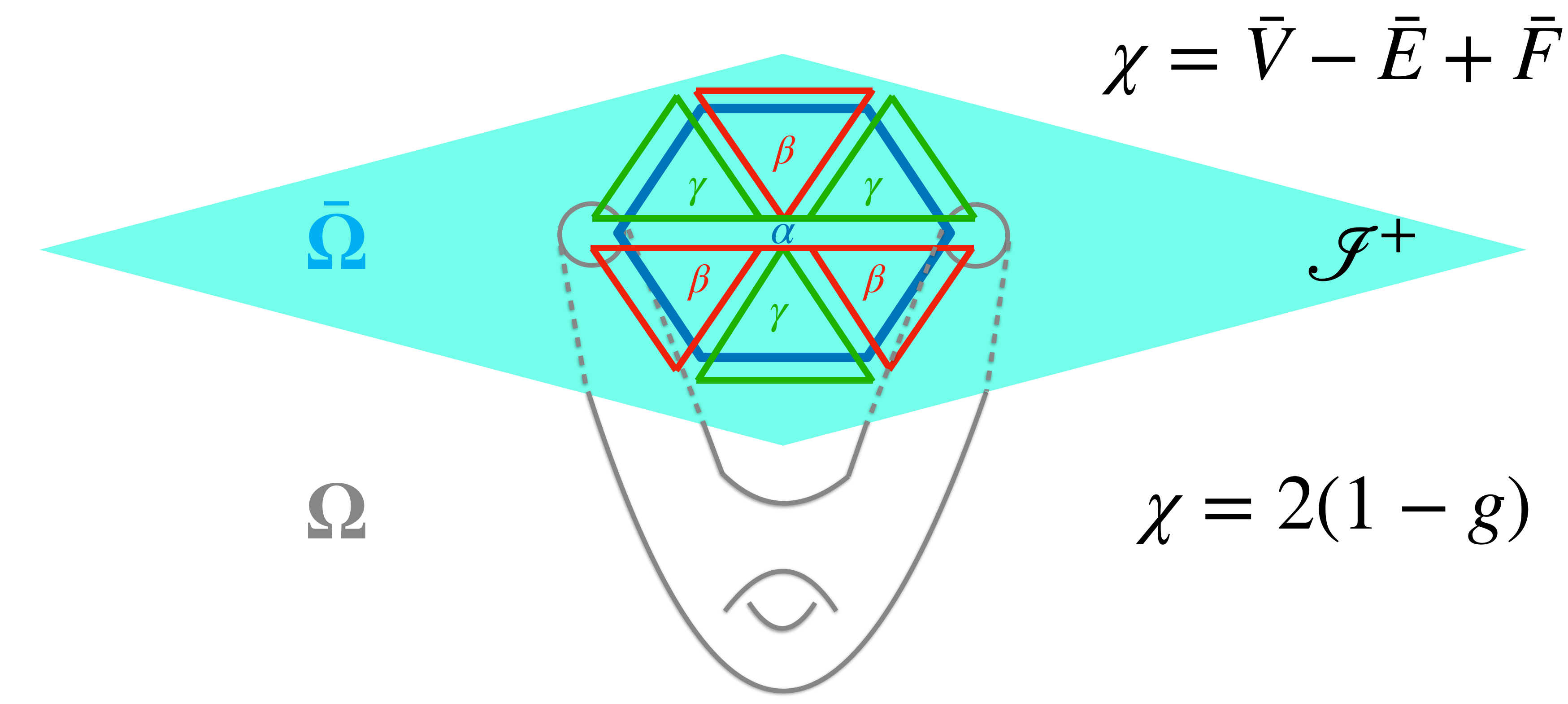}
    \caption{The genus 1 analogue of Figure \ref{fig:BulkStringBoundaryGenusZero}, showing the schematic correspondence between a 4-loop boundary ribbon graph at $\mathcal{I}^+$ and bulk closed-string worldsheet torus (genus $g=1$) diagram. This is the simplest non-planar Feynman diagram with $\bar{F} \ge 3$, allowing purely SU($N$) fields to interact (it vanishes in pure Yang-Mills due to Jacobi identities, but could contribute in other theories).  Here we have chosen to interpret 2 of the 6 trivalent vertices as single-trace operator insertions (shown as circles) while the other 4 arise from expanding the boundary Lagrangian.}
    \label{fig:BulkStringBoundaryGenusOne}
\end{figure}

\subsection{Closed Oriented Gauge Theories}\label{sec:consider}
Large-$N$ gauge theories in the ’t Hooft limit \cite{tHooft:1973alw,tHooft:2002ufq,Polyakov:2001af} provide a mechanism for a \emph{stringy} reorganisation of perturbation theory in terms of ribbon graphs (see Figures~\ref{fig:BulkStringBoundaryGenusZero} and~\ref{fig:BulkStringBoundaryGenusOne}).\footnote{Not every ’t Hooftian gauge theory is conformal: the large-$N$ limit with fixed $\lambda=g^2N$ merely ensures a string-like genus expansion of Feynman diagrams. Conformal invariance requires the theory to sit at a fixed point of the renormalisation group (as in $\mathcal{N}=4$ SYM), while generic large-$N$ gauge theories such as SU($N$) QCD are not conformal.}

We will consider a boundary theory defined with respect to a fixed background metric $\bar{g}$, whose dynamical matter fields can include gauge fields $A_\mu$, Dirac spinor fields $\psi$, and/or scalars $\phi$, but possibly other types of matter as well.  As the boundary theory need not be unitary, it is possible that some fields might have negative norm or even violate spin-statistics (as long as all gauge-invariant combinations satisfy spin-statistics\footnote{This is so that the bulk theory satisfies spin-statistics, and thus has a chance at being unitary.}).  Hence, in what follows we distinguish ``spinor'' from ``fermion'', and ``integer-spin'' from ``boson''.\footnote{As spin-statistics violating fields effectively count as a negative number of degrees of freedom, this possibility may be useful in obtaining theories with a net negative central charge, as required for dS/CFT in $d = 4n - 1$ dimensions, whereas a net positive number of degrees of freedom is required in $d = 4n+1$ dimensions.}

For the time being, we assume that the gauge group of the gauge bosons $A_\mu$ is either U($N$) or SU($N$), and that all fields transform in the adjoint representation thereof.  This means that the dual worldsheet theory will be oriented.  (It is also permissible to consider a product gauge group $G_1 \times G_2 \ldots$ as long as each of their $N_1, N_2, \ldots$ scales with the appropriate phase, and the matter fields each carry a colour and anticolour, which may however be charged in distinct gauge groups.  It is also possible to introduce uncharged particles by a trick that will be described below.)

We will also assume that the boundary and bulk field theories are defined over a continuous range of dimensions $d$ (so that the loop expansion of the \RR phase formula can be safely applied via dim reg).  For simplicity, we also assume that the boundary theory is parity-even (as parity-violating theories are more complicated to dimensionally regulate), in the sense that we can flip any or all of the coordinates without changing the unperturbed Lagrangian, and we assume that this property is inherited by the Lagrangian of the dual bulk theory; including (by bulk Lorentz invariance) the bulk imaginary time coordinate $\tau = it$.  Because the boundary theory is non-chiral, there is an action of spatial reflections on the spinor fields, which means that they are really \emph{pinor} fields.  This in turn implies that all traces of an odd number of their $\gamma^\mu$ matrices vanish, as even in odd dimensions the  permutation symbol $\epsilon_{\mu\nu\sigma\ldots}$ with $d$ indices is not available.  This will eliminate an annoying sign issue later on.  The same applies to any half-integer spin fields appearing in the bulk dual.\footnote{Except that in that case the forbidden permutation symbol would have $d+1$ indices.}

In the simplest type of 't Hooftian gauge theory, the Lagrangian consists exclusively of single-trace terms, of the schematic form
\begin{equation}\label{eqn:tHooftAction}
    I= \frac{N}{\lambda}\int d^d x \, \sqrt{\bar{g}} \; \Tr\!\left(M[A,\phi,\psi;\bar{g}]\right) \, ,
\end{equation}
where $\lambda=g^2 N$ is the ’t Hooft coupling, $M$ is an $N \times N$ matrix obtained by summing over some matrix products of the $N \times N$ fields ($A$, $\phi$ and $\psi$), and $\Tr$ is the trace over the $N$ colour indices.  We assume that this action is covariant and gauge-invariant.  It is acceptable to have multiple independent couplings, as long as each term scales like $1/\lambda$.  \emph{The coefficients in our Lagrangian shall be defined so that all amplitudes are real, apart from the dependence on $\lambda$ and $N$, which are allowed to be complex.}\footnote{I.e. we choose them so that if $\lambda$ and $N$ were real, the amplitudes would be real.  This would be a necessary although not sufficient condition to define a Statistical Field Theory; in an SFT the amplitude for each history (which is now interpreted as a probability weight) must additionally be non-negative.}  This will allow us to derive the equation for $N$ and $\lambda$ from bulk unitarity considerations.

The cosmological wavefunction $\Psi$ (equivalently, the generating functional) is then given by the holographic dictionary as follows:
\begin{equation}\label{eqn:tHooftPartitionFunction}
   \Psi[\bar{g}_{\mu\nu},J_{\mathcal{O}},\ldots] \equiv Z[\bar{g}_{\mu\nu},J_{\mathcal{O}},\ldots] = \int \mathcal{D}A\,\mathcal{D}\phi\,\mathcal{D}\psi\; \exp\!\left(-I[A,\phi,\psi;\bar{g}] + \int \frac{N}{\lambda} d^dx\sqrt{\bar{g}} \, \sum_{i} J_i \cdot \mathcal O_i\right) \, ,
\end{equation}
where each $J_i$ represents a source for the single-trace matter field operators $\mathcal O_i$, which we assume are normalised with the same $N/\lambda$ scaling, and hence have the same reality conditions, as the parameters in the unperturbed Lagrangian.  Thus, the reality condition for our boundary sources $J$ is the same as the reality condition for the boundary Lagrangian, and this does not need to be chosen independently.  This in turn implies a reality condition for the correlators obtained by differentiating with respect to the $J$'s.  It is also permitted to differentiate with respect to $g_{ab}$ to obtain the stress-tensor $T^{ab}$.

In order to make the effects of the $\bar{\Omega}$ rotation completely clear, we adopt the convention that each source $J$ takes the form of an \emph{unweighted tensor}, dotted into an $\mathcal{O}$ with the same number of indices.  Hence, the density of each term comes entirely from the explicit factor of $\sqrt{\bar{g}}$ out front.

For a half-integer spin source, the sources must additionally depend on a spinor index in an internal spin space.  This requires introducing a spin-connection.  Even on a curved background, all correlators with an odd number of half-integer spins vanish.  Correlators with an even number of spins, are implicitly defined by parallel transporting the spins into some point $p$, so that all spin-gauge-invariant information may be encoded by integer-spin tensors at $p$.  Thought of in this way, there is no such thing as a correlator with uncontracted spins --- they always contract in a loop, allowing some $\gamma^\mu$ trace identity (tracing over the spinor index, not $N$) to be used.

Terms involving bosonic fields can be contracted by an integer number of powers of the metric $g^{ab}$ in $\cal O$, but these will drop out when doing the full $\Omega$ or $\bar{\Omega}$ rotation, at least in the absence of log divergences or quantum anomalies.  Terms involving spinor propagators (like $\overline{\psi}(\gamma^\mu \partial + iA_\mu)\psi$) are a bit more subtle since they require a (bulk or boundary) vielbein $e_\mu^a$ to be defined.  This counts as ``half'' of a metric, and would result in an extra minus sign, any time there are an odd number of $\gamma^\mu$ symbols, in \emph{either} the boundary Feynman diagram (because of the $e^{i\pi} \bar{\Omega}$ rotation) \emph{or} the bulk Witten diagram (because of the $e^{-i\pi}\Omega$ rotation).  This will in turn lead to an extra $i$ sign in the reality condition for $\psi_n$ (see Eq. \eqref{phase} below).  Fortunately, this annoyance is absent in a parity-even theory, since traces of an odd number of $\gamma^\mu$ symbols always vanish for the reason stated previously.\footnote{Chiral theories are likely to be more subtle, and so we leave a careful analysis of these for future work.  The boundary spinor issue can presumably be dealt with by making a judicious choice for the sign of the spinor propagator terms in the boundary Lagrangian.  But the bulk spinor issue might be more subtle, since it is not immediately apparent how to count the total number of spinor propagator terms in a \emph{particle limit} of the \emph{bulk worldsheet}, simply by staring at the dual boundary Feynman diagram.}


\begin{figure}[ht]
    \centering
    \includegraphics[width=.7\textwidth]{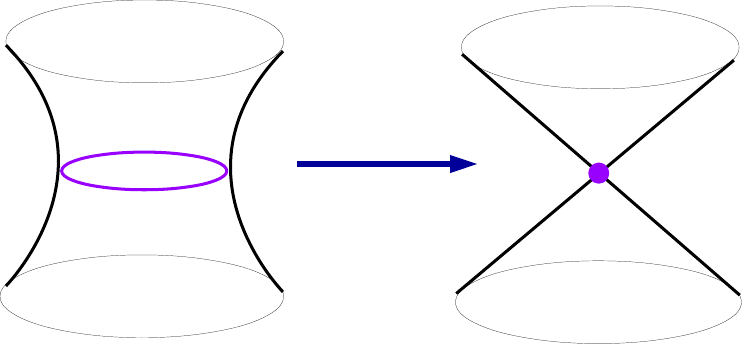}
    \caption{A worldsheet is pinched on a circle to form a double-trace vertex that creates/annihilates 2 strings.  We normalise such vertices in such a way that the $N$ scaling of both diagrams is the same.}
    \label{fig:pinch}
\end{figure}

It is also permissible to add terms to the Lagrangian with an arbitrary number of traces, so long as each term comes with appropriate powers of $N$ and $\lambda$:
\begin{equation}\label{eqn:multitrace}
    I= \int d^d x \, \sqrt{\bar{g}} \;
\left[
\frac{N^2}{\lambda} CT 
+ \frac{N}{\lambda} \Tr\!\left(\ldots\right)
+ \frac{1}{\lambda} \Tr\!\left(\ldots\right)\Tr\!\left(\ldots\right)
+ \frac{1}{N \lambda} \Tr\!\left(\ldots\right)\Tr\!\left(\ldots\right)\Tr\!\left(\ldots\right) + \ldots
\right] \, ,
\end{equation}
where here $CT$ is a pure counterterm (e.g. an Einstein-Hilbert term $R$) with no dependence on the fields, and a term with $t$ traces has a power of $N^{2-t}$ out front.  A vertex with $t$ traces represents a process that creates or annihilates $t$ different strings.  If such multi-trace vertices are interpreted by smoothing them out a bit into connected tree-level worldsheets (Figure~\ref{fig:pinch}), then the genus expansion for single-trace terms (described in more detail below) will continue to hold.  This is also the appropriate scaling w.r.t.~$N$ for multi-trace terms that are generated by a $T^2$ deformation of the boundary field theory, when moving the boundary a finite distance into the bulk as in \cite{McGough:2016lol,Kraus:2018xrn,Taylor:2018xcy,Hartman:2018tkw,Shyam:2018sro,Araujo-Regado:2022gvw,Araujo-Regado:2022jpj,Khan:2023ljg,Soni:2024aop,Araujo-Regado:2025elv,Shyam:2025ttb}.  Similarly, one can also define correlators of multi-trace operators by inserting multi-trace sources $J_i$; these must be inserted with the same factors of $N$ and $\lambda$ as in \eqref{eqn:multitrace}; otherwise the same conventions hold as in the single-trace case.

In order to write down an SU($N$) theory, it is necessary to restrict attention to the traceless part of the $N\times N$ matrix.  For a given field $\phi$, this can be done by taking the traceless combination:
\begin{equation}
\tilde{\phi} = \phi - \frac{1}{N} \Tr(\phi).
\end{equation}
This projection onto SU($N$) will introduce multi-trace terms into the Lagrangian, and it can be checked that such terms scale with $N$ in a way compatible with \eqref{eqn:multitrace}.  Similarly, it is also possible to incorporate fields transforming in the trivial representation of SU($N$), by thinking of them as a matrix for which only the trace (the U(1) factor of some U($N$) adjoint) appears in the action.  But be warned, such neutral particles will count as an \emph{entire} string for purposes of the genus expansion defined below; so for example a circular 1-loop Feynman diagram would have genus ${\rm g} = 1$, like a torus.

For Yang-Mills, or more generally any SU($N$) theory where interactions only involve commutators, the amplitudes are actually exactly the same as those of U($N$) Yang-Mills, apart from the U(1) photon's 1-loop circular diagram with no vertices.  Hence, the U($N$) theory automatically factorises into the interacting SU($N$) and free U(1) component.  But the same may not hold for theories with other types of Lagrangians, e.g.~if matter has a $\Tr(\Phi^4)$ interaction term.

In either the U($N$) case or the SU($N$) case, there is always a factor of $N^2-1$ for any pure Yang-Mills vacuum bubble (with at least one vertex), ensuring that the amplitude vanishes for the trivial $N=1$ case.  This factor arises because the total number of colour loops (and hence the $N$ scaling) is not always preserved by the commutator.  The subleading -1 term can be thought of either as a non-planar contribution, or else as coming from projecting out the U(1) contribution along a single edge in the vacuum bubble.  On either interpretation, in what follows we must regard the subleading -1 term as having a genus g one higher than the leading $N^2$ term.\footnote{Apart from this overall $N^2 - 1$ factor, Jacobi identities require a minimum of 5 loops to have a non-planar contribution to a Yang-Mills vacuum bubble!  This is analogous to the fact that there is no non-planar contribution to the beta function until four loops \cite{vanRitbergen:1997va,Czakon:2004bu,Herzog:2017ohr}.} 


\subsection{Reality Conditions for $N$ and $\lambda$}
With the normalisation in \eqref{eqn:tHooftAction}, the connected $n$-point functions of single-trace operators admit the standard ribbon-graph expansion
\begin{equation}\label{eqn:tHooftFeynmanDiagramScaling}
    \psi_n \equiv \langle \mathcal{O}_1\cdots\mathcal{O}_n\rangle_{\rm conn} \sim N^{\bar{V}-\bar{E}+\bar{F}}\lambda^{\,\bar{E}-\bar{V}}
    = N^{\chi}\lambda^{\,\bar{E}-\bar{V}} \, ,
\end{equation}
where $\bar{V}, \bar{E}, \bar{F}$ are the numbers of vertices, internal edges/lines, and faces (index loops), and
\begin{equation}\label{chi}
\chi=\bar{V}-\bar{E}+\bar{F} = 2(1-\mathrm{g})
\end{equation}
is the Euler characteristic of the ribbon graph, and g is the genus of the associated closed and orientable string worldsheet.  We place a bar\footnote{As a mnemonic, you can think of this bar as being the horizontal line drawn on the top of the Penrose diagram of de Sitter, to represent ${\cal I}^+$.} over boundary quantities to distinguish them from bulk quantities with the same names, e.g.~the edges $E$ and vertices $V$ of the \emph{bulk} Feynman diagram (treating strings as particles), for which we instead have:
\begin{equation}
\chi = 2(V-E) = 2(1-L),
\end{equation}
where $L$ is the number of bulk loops,\footnote{This should not be confused with the number of bulk (momentum) loops, which is given by $1 - \bar{L} = \bar{V} - \bar{E}$; hence $\bar{L} = \bar{F} + 1 - \chi$.} $V$ is the number of vertices in the bulk Witten diagram (including the boundary operator insertions) and $E$ is the number of bulk edges (including bulk-boundary propagators).\footnote{On the condition that we insert only single-trace operators on the boundary, we can exclude the boundary insertions and bulk-boundary propagators from the count, and get the same net answer.}  

\paragraph{Boundary Weyl rotation.}
The boundary Weyl rotation $e^{i\pi}\bar{\Omega}$ now has the effect of rotating the boundary metric components and the volume factor
\begin{equation}
    \bar{g}_{ab} \to e^{2\pi i}\bar{g}_{ab} \implies \sqrt{\bar{g}} \to e^{i\pi d} \sqrt{\bar{g}} \, ,
\end{equation}
which in turn rotates a (covariant, parity-even) integer spin boundary action by
\begin{equation}
   I \to e^{i\pi d} I \, .
\end{equation}
In any Feynman diagram, this provides an $e^{i\pi d}$ factor for every vertex $\bar{V}$, and an extra $e^{-i\pi d}$ factor for every internal edge $\bar{E}$, since we invert the action for the propagator.  (Additionally, for any spinor field propagator term with an odd number of derivatives, we would obtain an extra minus sign due to the contraction of the derivative with the bulk vielbein, but again for a parity-even theory we may neglect this phase for the reasons  stated above.)  This yields
\begin{equation}\label{eqn:BoundaryRR4}
    \textbf{RR:}\qquad \left[\psi_n(\mathbf{x};\bar{\Omega};\Omega)\right]^* = e^{\,i\pi d\,(\bar{V}-\bar{E})} \psi_n(\mathbf{x};\bar{\Omega};e^{-i\pi}\Omega) \,.
\end{equation}

\paragraph{Bulk Weyl rotation.}

For the same reasons, the bulk Weyl rotation $\Omega \to e^{-i\pi}\Omega$ gives us a factor of $e^{-i\pi D}$ per bulk vertex, and $e^{i\pi D}$ per bulk propagator.  Taking into account that $D = d+1$, this gives us a factor $e^{-i\pi(d+1)(V-E)}$, where 
\begin{equation} 
V - E = \chi/2 = (\bar{V} - \bar{E} + \bar{F})/2.
\end{equation}
Hence, the reflection-reality condition becomes, in terms of boundary variables:
\begin{equation}\label{eqn:BoundaryRR5}
    \textbf{RR:}\;\; \big[\psi_n(\mathbf{x};\bar{\Omega};\Omega)\big]^* =
    \exp\!\left\{\frac{i\pi}{2}\Big[2d(\bar{V}-\bar{E})-(d+1)(\bar{V}-\bar{E}+\bar{F})\Big]\right\}\psi_n(\mathbf{x};\bar{\Omega};\Omega) \, ,
\end{equation}
where again we neglect the extra minus sign from spinor propagators for the same reason as stated above.
 
From \eqref{eqn:BoundaryRR5} we can solve directly for the phase of $\psi_n$:
\begin{align}
    e^{i\arg(\psi_n)} 
    &= \pm \left(\frac{\psi_n}{\psi_n^*}\right)^{\!1/2} \label{phase}
    \\
    &= \pm \exp\!\left\{\frac{i\pi}{4}\Big[(d+1)(\bar{V}-\bar{E}+\bar{F})-2d(\bar{V}-\bar{E})\Big]\right\} \\
    &= \pm \exp\!\left\{\frac{i\pi}{4}\Big[(d-1)(\bar{V}-\bar{E}+\bar{F})+2(\bar{V}-\bar{E}+\bar{F})+2d(\bar{E}-\bar{V})\Big]\right\} \\
    &= \pm \exp\!\left\{\frac{i\pi}{4}\Big[(d-1+2c_1)(\bar{V}-\bar{E}+\bar{F})+2(d+2c_2)(\bar{E}-\bar{V})\Big]\right\} \, , 
    \label{eqn:WFUcoefficientPhase}
\end{align}
where in the last line we used that for closed string diagrams the Euler characteristic
\begin{equation}
    \chi = 2(1-{\rm g}) = \bar{V}-\bar{E}+\bar{F} \in 2\mathbb{Z}
\end{equation}
is always even, introducing integers $c_1,c_2\in\mathbb Z$ to parametrise the degeneracy of the phase.

\paragraph{Phases and the effective large-$N$ parameters.}
Comparing \eqref{eqn:tHooftFeynmanDiagramScaling} with \eqref{eqn:WFUcoefficientPhase} one may \emph{package} the phase as complex redefinitions of $N$ and $\lambda$:
\begin{equation}\label{midpaper_phases}
    N \propto e^{(i\pi/4)(d-1+2c_1)} \, , \qquad \lambda \propto e^{(i\pi/2)(d+2c_2)}\,,
\end{equation}
in agreement with the results \eqref{eqn:phaseN} and \eqref{eqn:phaselambda} quoted in the introduction.  Due to the integer shifts of $c_1$ and $c_2$, $N$ has 4 possible phases and $\lambda$ has 2 possible phases.  Note, however, that for a U($N$) theory, all perturbative amplitudes are invariant under $N \to -N$, so it might be better to think in terms of $N^2$ and $\lambda$, as shown in \eqref{eqn:phaseN}.

This makes manifest that \RR generically forces at least one of $N,\lambda$ to be complex, since $\lambda \in \mathbb{R} \Longleftrightarrow d \in 2\mathbb{Z}$, but $N \in \mathbb{R} \Longleftrightarrow d \in 2\mathbb{Z}+1$, i.e. the boundary theory is non-unitary in the usual sense. Hence, in holographic cosmology, \emph{no unitary} ’t Hooftian gauge theory satisfies bulk \RR!  (At least in the parity-even case.)\footnote{Of course all unitary boundary theories satisfy \RR \emph{on the boundary}, because reflection positivity is implied by reflection positivity.  But throughout this paper, \RR means reflection reality of the \emph{bulk} Euclidean theory.}

\subsection{Closure Under RG flow}\label{sec:closure}

The next thing to check is whether our proposed phases for $N$ and $\lambda$ result in a class of theories which is closed under RG flow.  If not, it would raise the question of whether our assignments are physically meaningful, as imposing them at one value of the UV cutoff $\epsilon_1$ would lead to a different result than imposing them at another value $\epsilon_2$.

More precisely, the question is whether there \emph{exists a natural choice of RG scheme} for which this is the case.  Power law divergences can always be removed by a suitable choice of scheme (in fact, this happens automatically in RG schemes based on analytic continuation, such as dim reg or zeta function).  Therefore, at least perturbatively, it is only necessary to check whether the right phases are obtained for log divergent diagrams.  

We will find that this is in fact the case for all diagrams that do not change the topology of the 't Hooft worldsheet (and thus are independent of $N$).  For diagrams that \emph{do} change the topology, there can sometimes be an unexpected phase, which we attempt to interpret at the end of this subsection.

In general, (non-exact) renormalisation proceeds by identifying a divergent subdiagram of the Feynman graph, containing one or more loops, and replacing it with a counterterm vertex.\footnote{If this subdiagram itself contains a divergent subdiagram, there can be nested divergences (giving rise to e.g.~products of logs).  Here we assume such cases can be dealt with iteratively, and thus consider a simple diagram which is either log divergent or not.}  Now in an 't Hooftian gauge theory, a loop can be contracted to a vertex in multiple ways, depending on the relation to the colour structure (see Figure~\ref{fig:BulkStringBoundaryGenusOne}).  The following examples illustrate the range of possible processes:
\begin{enumerate}
    \item One or more colour cycles can be contracted.  If all the vertices involved are single-trace, this generates a $\beta$-function for another single-trace vertex, which is independent of $N$ because the topology does not change.\footnote{If any of the operators are multi-trace, the extra strings produced simply ``go along for the ride'' without participating in a meaningful way.  Hence, the resulting counterterm has as many ``extra traces'' (beyond 1) as the sum of the extra traces that went into it.}
    \item A loop not containing a colour cycle can be ``pinched'' down to a point, as in Figure \ref{fig:pinch} shown previously.  This results in a $\beta$ function for a multi-trace operator.  Given the $N$ dependence of the multi-trace terms in \eqref{eqn:multitrace}, such pinching processes are \emph{not} regarded as changing the worldsheet topology, and as a result we do not need to worry about the $N$ dependence of such $\beta$ functions.
    \item An entire sphere is integrated out, to form a counterterm at a single point.  In the case where all vertices are single-trace, this results in a ``pure counterterm''.\footnote{If there are vertices with ``extra traces'', the resulting counterterm has that many total traces.}  This is a zero-trace operator, and yet because there is already the correct factor of $N^2$ in the CT factor of \eqref{eqn:multitrace}, again, this already has the right factors of $N$.
    \item An entire handle is integrated out.  This increases the value of $\chi \to \chi + 2$ and thus results in a subleading $1/N^2$ correction to the $\beta$ function.    
    \item An entire Riemann surface $\Sigma$ with genus g $\ge 1$ is integrated out.  This is a combination of the previous 2 cases, and thus also results in a subleading $1/N^2$ correction.
\end{enumerate}
Let us refer to processes \#1-3 as ``topologically trivial'' (since they don't change the $\pi_1$ homotopy group) and processes \#4-5 as ``topologically nontrivial'' (because they do).

\paragraph{Topologically Trivial RG.}
For topologically trivial RG, we can ignore $N$ as a parameter and simply think about $\lambda$, which as stated above scales as 
\begin{equation}\label{phaseoflambda}
\lambda \,\propto\, \pm e^{i\pi(d/2)}.
\end{equation}
In this case, the RG flow is closed in arbitrary dimension $d$.  In dim reg of a weakly coupled perturbation expansion, log divergences manifest as poles in the dependence of the amplitude on $d$.  Such poles can appear only when $d = \text{rational}$.  At the leading order where log divergences appear, the statement of closure is that when all vertices and edges in the diagram receive the appropriate phases given \eqref{phaseoflambda}, the resulting counterterm vertex also does.  This requires that
\begin{equation}\label{closure}
\lambda^{\bar{E} - (\bar{V} - 1)} = \lambda^{\bar{L}} \propto \pm 1,
\end{equation}
whenever $\bar{L}$ is such as to permit a log divergence.  To build intuition let us look at a scalar field theory with an arbitrary potential:
\begin{equation}
I= \frac{1}{\lambda}\int d^d x \, \sqrt{\bar{g}}
\big[\partial_\mu \phi \;\! \partial^\mu \phi + V(\phi)  \big].
\end{equation}
In this case, the superficial degree of divergence of a subdiagram $\Gamma$ is  
\begin{equation}
D[\Gamma] = \bar{L}d - 2\bar{E} - 2n_{\partial},
\end{equation}
where the first term comes from the $\bar{L}$ integrals over $d$ components of momentum $p$, $\bar{E}$ counts the number of massless $1/p^2$ propagators, and $n_\partial$ is the number of extra derivatives appearing in the resulting counterterm (which is always even by covariance). 
In fact the $\bar{E}$ term doesn't matter, because the $n_\partial$ term already means we can only constrain $D$ mod 2.\footnote{For the same reasons, adding covariant higher derivative terms to the Lagrangian won't change the story.}  The existence of a log divergence requires that
\begin{equation}\label{eq:logzero}
D = 0 \,\implies\, \bar{L}d \text{ mod } 2 = 0.
\end{equation}
Hence, $\bar{L} = 2n/d$ for some integer $n$, and plugging into \eqref{phaseoflambda} we obtain
\begin{equation}
\lambda^{\bar{L}} \,\propto\, \pm e^{i\pi n} = \pm1,
\end{equation}
which verifies the closure relation \eqref{closure}.\footnote{In fact, this closure argument works for \emph{any} linear dependence of $\text{arg}(\lambda)$ on $d$, as long as $\lambda \in \mathbb{R}$ whenever $d = \text{even}$.}  A special case of this result is that when $d = \text{odd}$, log divergences only occur with an even number of loops, so that a pure imaginary action is closed under RG.

At first sight, the addition of particles with spin complicates the narrative, but in fact the same result holds, at least given our simplifying assumption of a parity-even theory.  For example, vector bosons can allow couplings with an odd number of derivatives,  e.g.~$A_\mu \phi\:\!\partial^\mu \phi$ (scalar QED) or $[A_\mu, A_\nu] \partial^\mu A^\nu$ (Yang-Mills).  But, at least in a parity-even theory, the same results follow.  A covariant vertex will always satisfy $n_\partial + n_A = \text{even}$, due to the fact that all indices must be contracted with the boundary metric $\bar{g}_{\mu\nu}$.\footnote{If we were not imposing the parity-even assumption, then there would also be the possibility of contracting with the permutation symbol $\epsilon^{\mu\nu\ldots}$, which has an odd number of indices when $d = \text{odd}$.  This might affect the appropriate reality conditions for parity-odd terms in a KFT.}  Since $n_A$, the number of external $A_\mu$ lines,\footnote{More generally, the number of fields whose polarisation tensor has an odd number of indices.} is always conserved mod 2 by RG flow (our only options are to Wick-contract them or else have them stick out of the final counterterm vertex), the condition \eqref{eq:logzero} is the same as the scalar case.  

The same conclusion holds for half-integer spin fields. In this case, spinor propagator terms like $\overline{\psi}\gamma^\mu (\partial_\mu + iA_\mu)\psi$ have $n_\partial + n_A = \text{odd}$, but these cancel in a parity-even theory due to the vanishing of traces of odd number of $\gamma^\mu$ matrices.  Hence, for a general parity-even theory, the complex phase for $\lambda$ is closed under general topologically-trivial RG flows, i.e.~at leading order in a $1/N$ expansion.

There is another, more abstract, argument which establishes the same conclusion.  This argument notes that the phase in \eqref{phaseoflambda} originally arises from a \emph{complex} rescaling of the ${\bf x}$ coordinate in \eqref{eqn:BulkRR}.  Now RG flow is just a \emph{real} rescaling of ${\bf x}$.  Any two such rescalings commute with each other.  Hence --- at least to all orders in boundary perturbation theory where the continuations are analytic --- the flow must be closed under RG.  Of course, this  argument requires that the \emph{original} theory with real $\lambda$ be closed under RG, but this is obvious since if all amplitudes are real then there is no opportunity for an imaginary component to arise.

An advantage of this more abstract argument is that it makes it clear that the closure continues to hold even in the presence of imaginary shifts arising if we do dim reg, in a manner compatible with our $d$-dependent phases (see the discussion under \eqref{eqn:phaselambda}).  Although the resulting imaginary shifts are finite in the presence of a \emph{single} log divergence, they might potentially contribute to a divergence in the case of a doubly divergent Feynman diagram with a $\log^2$ divergence (i.e.~at second order in the RG flow).

\paragraph{Topologically Nontrivial RG.}
Let us now consider the topologically nontrivial case.  Of course $N$ is not itself renormalised, but it can enter into the $\beta$-functions of vertices.  In general (except perhaps in a superrenormalisable theory) there is no way to preclude $1/N^2$ corrections from entering into the $\beta$-functions at sufficiently high order in perturbation theory, and their log divergences do not appear to be constrained by any similar parity argument.  Thus, if we also demand that renormalisation also respects our reality condition, we need
\begin{equation}
N^2 \,\propto\, \pm 1  \;\implies\; d = \text{odd}.
\end{equation}
It would appear that our condition for being a KFT is not closed for e.g.~$d = \text{even}$ dimensional boundary field theories.

We are not certain how to interpret this fact, given that the complex $N^2$ rotation \emph{is} apparently required for bulk unitarity.  But it could be that the genus expansion must be treated in a fundamentally different way than the $\lambda$ expansion, and that the differing phases in the higher genus expansion keep track of some important physical effect.  In the context of dS/CFT, a boundary UV divergence should correspond to a bulk IR divergence; thus we should look for a physical bulk phenomenon that can only occur for \emph{IR divergences involving loops}.\footnote{IR divergences in dS/CFT were recently studied in \cite{Bzowski:2023nef}, although they did not explicitly discuss issues related to postselection or decay processes.}

An example of such a phenomenon in de Sitter is \emph{particle decay}.  If e.g.~a $\phi$ particle is unstable, it will decay into $n > 1$ other particles before reaching ${\cal I}^+$, hence its probability to \emph{not} decay should diminish exponentially with proper time.  This reduction of the amplitude for the 1-particle state, is given by a diagram involving $\bar{L} = n$ loops (which destructively interferes with the process where no decay occurs).  From a mathematical perspective, nothing prevents us from post-selecting onto a scenario in which the particle reaches ${\cal I}^+$ without decaying, even though this process occurs with 0 probability in the limit $\eta_0 \to 0^-$.  However, an anomalous dimension is introduced into the dual associated boundary operator $\cal O$, as a result of the need to introduce a normalisation factor which diverges as $\eta_0 \to 0^-$.\footnote{We can apply the same postselection procedure to cases in which the de Sitter vacuum is itself metastable, and has a small nonzero probability $p$ to decay in any unit de Sitter volume.  This leads to an exponential decay for survival in all space regions, which is formally $\exp(- p\text{Vol}[dS])$.  Such a term may be absorbed into a $d$-dimensional cosmological constant in the effective action $\log Z$ of the boundary theory --- so there is still a well-defined boundary dual --- but the sign of this counterterm is \emph{real}, not imaginary, and therefore does not correspond our condition \eqref{eqn:multitrace} for zero-trace counterterms.  This merely \emph{apparent} deviation from the requirements imposed by bulk unitarity, is physically explained by the fact that some probability ``went elsewhere'' (to whatever the final state of vacuum decay is) and was not captured by our choice of boundary field theory at ${\cal I}^+$.  The general lesson is that contexts involving postselection should lead to corrections to the reality conditions defining a KFT.  A similar vacuum decay phenomenon should arise in unstable vacua, from loop amplitudes involving tachyons (whose existence is hard to avoid in weakly coupled boundary duals, as discussed in Section \ref{sec:tension}) but we do not consider such amplitudes carefully in this article.}

In general, decay leads to an imaginary shift in the mass of the unstable particle: $m^2 \to m^2 - i\varepsilon$, which is an IR divergent phenomenon because (in a given scattering process) it can occur anywhere on the external particle leg.  By the mass relation \eqref{eqn:mass-relation}, $m^2 L_{dS}^2 = \Delta(d-\Delta)$, this leads to a shift in $\Delta$ \emph{away} from a unitary representation of the conformal group Spin($d+1,1$).  Let us work to first order in $\varepsilon$.  In the principal series, this corresponds to an increase of the real part: $\text{Re}[\Delta] = d/2 + O(\varepsilon)$, which has a nice semiclassical interpretation as representing the additional exponential factor due to decay.  In the complementary series, it should lead to an imaginary $O(\varepsilon)$ shift.  Either way, it leads to a result not permitted by the naive unitary single-particle representations (not because the dS bulk theory is truly non-unitary, but because the probability leaked into multi-particle states).  Hence, there \emph{should} exist a bulk 1-loop process which has a complex sign relative to the tree-level expectations for unitarity, and this sign might be provided by integrating out handles in the 't Hooft worldsheets.\footnote{Even when $d = \text{odd}$, such a sign could appear if the IR divergent diagram also has a UV log divergence, necessitating the use of dim reg, and hence (according to \eqref{eqn:phaseN}) a slightly non-real value for $N^2$.}

It might also be relevant that in a \emph{renormalisable} theory,\footnote{For the usual reasons, a nonrenormalisable theory is potentially ill-defined in the UV, unless it can be given a UV completion by a better behaved theory.} these topologically nontrivial divergences disappear \emph{if we fix the boundary momenta} $p_\mu$ going around all noncontractible $\pi_1$-cycles of the 't Hooft worldsheet.  (Perhaps, with the motivation that the integration over such momenta is more easily dealt with on the bulk side of the duality, rather than on the boundary side.)  This stipulation means that we can only integrate momentum around the colour faces.  As an example, consider $\phi^n$ theory in $d$ dimensions.  In any vacuum bubble (with fixed $\pi_1$ momenta), the number of $d$-momentum integrals $\int d^dp$ is given by $\bar{F}-1$ (the last one doesn't count because it is redundant with the others).  Of the edges $\bar{E}$, each one has a factor of $1/p^2$. 

Since each edge has two faces next to it, it is clever to associate one factor of $1/p$ with each face, so that a face with exactly $d$ edges contributes nothing to the degree of divergence $D$.  Then, in order for a subdiagram $\Gamma$ to be divergent, it needs to satisfy the criterion
\begin{equation}\label{sides}
s(\Gamma) := \text{Average(Edges per Face in $\Gamma$)} < d,
\end{equation}
where the inequality on the sides $s$ is strict because either (i) the subdiagram is incomplete, in which case the edges on its boundary (which are connected to only one face in the divergent subdiagram) have an extra $1/p$ associated with them, or (ii) the diagram is the entire bubble,\footnote{If some of the vertices are multi-trace, the ``extra traces'' may simply be ignored in this argument.} in which case the final face is redundant as stated above.  Now the marginally renormalisable theories ($\phi^6$ in $d = 3$, $\phi^4$ in $d = 4$, and $\phi^3$ in $d = 6$) lie at precisely the threshold for a tiling with $n$ different $d$-gons lying flat in a plane.  That is, the conditions of renormalisability for general $d$ is
\begin{equation}\label{tiling}
\frac{1}{n} + \frac{1}{d} \,\ge\, \frac{1}{2},
\end{equation}
which is the same as the conditions for a non-negatively curved tiling if we substitute $s$ for $d$.  Saturation of this inequality corresponds to a graph $\Gamma$ with $\chi = 0$ (e.g.~a torus), but as \eqref{sides} is strict, this cannot actually be divergent.   Hence, a divergent diagram always has $\chi > 0$, and as a result it must be simply connected.\footnote{Here we are assuming orientability, but even in the non-orientable case $\chi = 1$, the resulting homotopy group is $\pi_1 = \mathbb{Z}_2$, which has finite order and thus does not permit a net flow of momentum around a noncontractible cycle.}  For a superrenormalisable theory, we have the stronger result that the average curvature of the diagram is bounded below by some positive coefficient, and this is precisely why such theories are superrenormalisable, i.e.~only a finite number of divergent diagrams exist.  This explains why the same formula \eqref{tiling} applies to both tilings and renormalisation.\footnote{In fact, the five interacting superrenormalisable $\phi^n$ theories (with integer $n>2$ and $d>2$) are in 1-to-1 correspondence with the five Platonic solids.  In each case, the maximal sized divergent Feynman graphs $\Gamma$ have exactly half the element count of the corresponding Platonic solid.  The reason for the ``half'' is that actually the non-orientable case $\chi = 1$ works best because $\bar{F} = \bar{L} + \chi - 1$.  E.g.~a maximally divergent diagram for $\phi^3$ theory in $d = 5$ would have $\bar{V} = 10$, $\bar{E} = 15$, and $\bar{L} = 6$; and one example of such a diagram would be an antipodally identified dodecahedron.  (There are others, because the faces are only required to be pentagons on average.)}  

This analysis may be extended to spinor and vector couplings.  Although the relationship to tilings becomes more complicated, fixing the $\pi_1$-momenta still suffices to eliminate topologically nontrivial RG, in any renormalisable theory.  Although massless spinor propagators contribute only $1/p$ to momentum integrals, the valence of renormalisable vertices becomes correspondingly more limited, in a way that continues to enforce $\chi > 0$ for all divergent subdiagrams.

\subsection{Open and Non-Oriented Cases}\label{sec:ONO}
It seems natural to ask if there is a similarly nice choice of phases for open and/or non-oriented strings, ensuring the satisfaction of bulk \RR to all orders.  The answer appears to be no, at least for general dimension $d$.  For generic real $d$, there appears to be no choice of phase that works for all bulk diagrams, regardless of the order of loops $L$.  However, again for $d = \text{odd}$, the requisite phase is independent of $L$, and we find that the same phases work as before.

In the 't Hooft duality, open strings are obtained by adding quarks in the fundamental representation, while non-oriented strings are obtained by replacing the gauge-group with an orthogonal group O($N$) or a symplectic group Sp($N$).  In all cases, we can continue to define the Euler number from the boundary as:
\begin{equation}
    \chi = \bar{V} - \bar{E} + \bar{F},
\end{equation}
and we discover that $\chi$ is odd in some cases, leading to terms in the expansion with odd powers of $N$.  But the relationship of $\chi$ to the bulk Witten diagrams becomes more fraught.  (Clearly, $L = \chi/2 - 1$ must fail, since $L$ must be a natural number.)

In general, a Riemann surface may have g $\ge$ 0 handles, b $\ge$ 0 open boundary circles, and c $\ge$ 0 cross-caps.  The allowed topologies form a monoid under the connected sum operation, subject to a single relation: 3 crosscaps $=$ 1 handle $+$ 1 crosscap. The Euler characteristic is:
\begin{equation}
    \chi = 2 - 2{\rm g} - {\rm b} - {\rm c},
\end{equation}

These may be interpreted as string diagrams, which degenerate to Feynman diagrams along various limits~\cite{Polchinski:1998rq}, or Witten diagrams if we allow particles to reach an asymptotically dS boundary.  However, if b $+$ c $\ge$ 2, then there exist multiple interpretations of the \emph{same} string diagram as Witten diagrams with \emph{different} numbers of loops $L$.  For example, the Riemann surfaces with b $+$ c $=$ 2, $\chi = 0$, are composed as follows:
\begin{itemize}
\item 2 boundaries $=$ Cylinder, 
\item $2$ cross-caps $=$ Klein Bottle, \&
\item 1 boundary $+$ 1 cross-cap $=$ M\"{o}bius Strip. 
\end{itemize}
In each of these cases, there exist particle interpretations which have both $L = 0$ (tree-level) and $L = 1$ (loop-level).  For example, in the case of the Cylinder (perhaps dressed with vertex operators going to the boundary), we may interpret it either as an open string ($L = 1$) or as a closed string that is created, propagates along an edge, and then is annihilated ($L = 0$).  This is the famous open-closed string duality; see e.g. Figure 3.6 of Polchinski \cite{Polchinski:1998rq}.  This is quite unlike the torus where we could only interpret it as being $L = 1$.\footnote{Assuming of course that we do not allow an entire handle to be compressed into a single vertex of the bulk theory.}  Similarly, although the Klein Bottle and M\"{o}bius Strip may be interpreted as loops of a closed and open string respectively (glued with a twist), they may also be viewed as tree-level processes by thinking of the cross-cap(s) as the creation or annihilation of a closed string.

Now, let us compare to the effects of the bulk $\Omega$ rotation, and its relationship to $N$.\footnote{There is no corresponding issue with $\bar{\Omega}$ rotation, as this can be calculated directly from $\bar{E} - \bar{V}$.}  As shown above, \RR requires each bulk Witten diagram to have a contribution to its phase of
\begin{equation}
e^{(i\pi/2)(d+1)(V-E)} = e^{(i\pi/2)(d+1)(1-L)},
\end{equation}
which, if it is to be explained by a rotation of $N$, requires that
\begin{equation}
N^{\chi} = \pm e^{(i\pi/2)(d+1)(1-L)}
\end{equation}
for \emph{all} possible choices of $L$ associated with a given string diagram.  As $L$ differs by units of 1, consistency therefore requires
\begin{equation}
d = \text{odd}, \quad N \in \mathbb{R},
\end{equation}
where the second condition is on $N$ rather than $N^2$ because $\chi$ can be odd.  Hence, we find the very surprising result that (assuming holographic duality to an 't Hooftian gauge theory), open and/or non-oriented strings are allowed only in an \emph{even} dimensional de Sitter spacetime!\footnote{Since the moduli space of the full worldsheet string theory is ambiguous with respect to $L$, one might wonder what the phase of the string diagram itself is.  However, this is itself an ambiguous question, because the contour of integration over the worldsheet moduli parameters can be deformed in various ways.  All that really matters is that the contour has the right asymptotics at the boundary of the moduli space.  In order to obtain a Lorentzian string S-matrix in e.g.~flat spacetime, the $i\varepsilon$ prescription for string theory~\cite{Witten:2013pra} states that as the worldsheet degenerates along any direction, this degeneration should (asymptotically) be integrated along a \emph{Lorentzian} proper time contour.  At the boundary of the moduli space, the $L$ value of the worldsheet becomes unambiguous, and in this limit string theory gives the same factors of $i$ that are expected from ordinary particle field theory.}  

Even then, we are not completely out of the woods, because the restriction on $d$ means we cannot sidestep issues related to bulk quantum log divergences, by doing dimensional regularisation of both sides (as we proposed for the U($N$) adjoints case).  The proper treatment of such divergences might then lead to additional consistency conditions, which we leave for future work.\footnote{The example of AdS/CFT suggests that the UV finiteness of the \emph{bulk} string theory could be an automatic consequence of finding a fully valid dS/CFT correspondence, in which case there might well be no need to dimensionally regulate the bulk theory.  (This should not be confused with the story of \emph{boundary} UV divergences, which are dual to bulk IR divergences.)}

\paragraph{Open-Closed Euler Number.}
A few more words are helpful concerning the open-closed case.  The simplest types of bulk vertex each have a calculable contribution to the Euler characteristic.  This can be calculated to be:
\begin{equation}
\chi = 2(V_{\rm cl} - E_{\rm cl}) + (V_{\rm op} - E_{\rm op}),
\end{equation}
where ``cl'' (closed) represents a vertex or edge involving closed strings only, while ``op'' (open) represents a vertex or edge involving some open strings traced together in a circle (plus any number of closed strings).  From this perspective, we can immediately see the problem.  For a diagram involving open strings only, we could make things work by selecting
\begin{equation}\label{eqn:N4quarks}
    N \propto \pm e^{i\pi d/2},
\end{equation}
while for closed strings we want $N^2$ on the LHS as stated above.  These conditions are only compatible when $d = \text{odd}$, as stated above.  (For bulk tree-level correlators, we can make everything work by also taking $N_f$, the number of quark flavours, to scale as $N_f \propto \pm N$, thus ensuring that all tree-level correlators scale like $N^2$.  But this does not work in the general case, as can be seen from the fact that b is a topological invariant of the Riemann surface, while $L$ is not.)

\paragraph{Vector Models.}
However, \eqref{eqn:N4quarks} does give the right scaling for vector models, such as the analytically continued O($N$) model, in which all of the boundary fields are in the fundamental representation.  The only ``single-trace'' combinations are doublets (schematically of the form $\phi \cdot \phi$, but possibly with derivatives).  These seem to be dual to higher-spin gravity models, with the best studied being the duality \cite{Anninos:2011ui} between an Sp($N$) model of scalar fermions in $d = 3$, and Vasiliev gravity in $D = 4$.  Here, the boundary theory can be free (which means that $\lambda$ plays no role in perturbation theory).  In the bulk, all correlators are rational, and thus seem to come from the tree-level contribution alone, presumably implying that loop contributions cancel \cite{Anninos:2019nib}.
 
\section{A Glance at Positivity Conditions}\label{sec:positivity}
Although the \RR condition is necessary for bulk unitarity, it is not sufficient.  It needs to be supplemented with the statement that the norm of every bulk state is positive.  

We do not yet have a good understanding of what this positivity condition should look like as applied to the boundary theory, but we can check some specific features. (Some constraints from only having states with positive norm have been explored in~\cite{Hogervorst:2021uvp,DiPietro:2021sjt,Penedones:2023uqc,Loparco:2023rug,Green:2023ids,Loparco:2023akg,Loparco:2024ibp,Loparco:2025azm,Chakraborty:2025mhh}, and explained in a particularly pedagogical way in~\cite{SalehiVaziri:2022sdr}.)  We will do this under the dS/CFT assumption that the boundary is conformally invariant.  

In this section we will look at the sign imposed by unitarity for the central charge $c$ (measuring the net number of degrees of freedom of the model) and the 2-point function of scalar operators in the complementary series $0 \le \Delta \le d$.

\subsection{Sign of the Central Charge}\label{sec:central}

Although in general \cite{Goodhew:2024eup} only derived a reality condition, the sign of the tree-level central charge $c$ in a dS/CFT can be determined from the positivity of the Newton constant $G > 0$.  

By the central charge $c$ of a CFT, we mean any overall measure of the degrees of freedom, whether via a trace anomaly (when $d = \text{even})$, the sphere partition function (when $d \ne \text{even}$), the 2-point entropy of the stress-tensor (in any $d$) or the universal piece of the boundary entanglement entropy, in each case defined so that unitary theories have $c > 0$.  But even for dS/CFT, in cases where two different measures of $c$ exist, their phase agrees in every example that we have checked.

For dS with $G > 0$, the central charge has the sign:
\begin{equation}\label{c}
c \propto e^{i\pi (d-1)/2}.
\end{equation}
(Note the absence of the $\pm$ symbol; the whole point of this section is to try to pin down the sign of some things completely.)  This is in agreement with bulk calculations in various dimensions \cite{Cotler:2019nbi,Collier:2024kmo,Freidel:2008sh,Anninos:2011ui,Maldacena:2002vr,Araujo-Regado:2025elv,Thavanesan:2025tha}, and implies a fascinating $d$ mod 4 pattern in the characteristic properties of the theory.  When $d = \text{even}$ we need $c$ to be imaginary at tree-level (but in $d = 2$, \cite{Cotler:2019nbi,Collier:2024kmo} find a real correction at $L=1$, in agreement with expectations).   In $d = 4n + 3$ we need the number of degrees of freedom to be negative (which explains the spin-statistic violation in the $d = 3$ dual to Vasiliev).  On the other hand, a dS/CFT in $d = 4n+1$ dimensions would require a net positive number of degrees of freedom.

Since in an adjoint Yang-Mills model, the U($N$) gauge field $A_\mu$ has $N^2$ degrees of freedom, this suggests that we should use the + sign in \eqref{eqn:phaseN}, that is:
\begin{equation}\label{Nplus}
N^2 = i^{d-1},
\end{equation}
where here we restrict to integer dimensions, since full unitarity (unlike \RR) is only possible in that case.
That would mean that the number of colours can be an integer $N \in \mathbb{Z} \subseteq \mathbb{R}$ only when $d = 4n + 3$, corresponding to a positive number of degrees of freedom.\footnote{This is a bit too hasty, since it could be that there are also an effectively a \emph{negative} number of flavours of matter fields $\psi$ and $\phi$ in the action!  These could overcome the positive gauge degrees of freedom.  However, a possible argument against this scenario, from looking at the free limit, will be presented in Section \ref{sec:2ptsign}.}  

For a free theory, $c$ is independent of $\lambda$.  Thus, if we \emph{start} with a weakly coupled unitary theory prior to doing the $N$ and $\lambda$ rotations, then \eqref{Nplus} gives the correct sign for $N^2$ (or for $N$ in a vector model).  At strongly coupled $\lambda$ one might guess we use the same sign, but it could be that there is some critical value of $\lambda$ at which $c$ changes sign, in which case the weakly coupled behaviour may be misleading.

\subsection{Sign of the Two-Point Function}\label{sec:2ptsign}
It is also instructive to check what conditions are required for the phase of the two-point function and the normalisability condition (that the wavefunction $\Psi\equiv Z$ is normalisable) for bulk-unitary theories.  Although it is dubious that a weakly–coupled boundary field theory could provide a consistent realisation of holographic cosmology, in this section we will nevertheless check positivity conditions from bulk unitarity in the free limit.  In this section we will we work with our bulk \CRT constraint \eqref{eqn:PertCRT} rather than the \RR constraint, as the two are equivalent for scale-invariant cosmologies.\footnote{While a Yang-Mills gauge field in $d \ne 4$ is not conformally-invariant even in the free limit, it \emph{is} scale-invariant, modulo issues involving quantisation of charge which do not affect the flat space $n$-point functions.} 

We will first consider a scalar operator $\cal O$ in the complementary series, whose boundary operator satisfies $0 \le \Delta \le d$.  In position space, the (tree-level contribution to the) 2-point function at the boundary CFT has the form:\footnote{When $\Delta \ge d/2$, this is modulo contact terms at ${\bf y}_1 = {\bf y}_2$, whose purpose is to ensure finiteness when integrating with a smooth test function.}
\begin{equation}\label{psi2_y}
-\psi_2^{(0)}(\eta = 0^-; {\bf y}_1, {\bf y}_2) 
\quad\propto\quad
f(\Delta)\,|{\bf y}_1 - {\bf y}_2|^{-2\Delta},
\end{equation}
where $f(\Delta)$ is a phase to be determined, and the minus sign on the LHS is to cancel out the minus sign in \eqref{eqn:BDWFUGeneratingFunctional} that is conventionally introduced in the cosmology literature, but is not present in the usual field theory definition of the 2-pt function.  Up to a positive factor, the Fourier transform to momentum space is (with the overall $\delta^d({\bf k}_1 - {\bf k}_2)$ delta function stripped off):
\begin{equation}\label{psi2_k}
-\psi_2^{(0)}(\eta = 0^-; {\bf k})
\quad\propto\quad
\Gamma(\tfrac{d}{2}-\Delta)
f(\Delta)\,|{\bf k}|^{2\Delta - d},
\end{equation}
as can be seen from the fact that there are poles in the Fourier transform whenever $\Delta = \tfrac{d}{2} + n$, for $n \in \mathbb{N}$.  Whenever $\Delta$ is sitting at one of these poles, a nontrivial renormalisation is required that introduces an arbitrary scale $k_0$, leading to a factor of $(-1)^{n+1} \log(|{\bf k}|/k_0)$ in the RHS of \eqref{psi2_k}, replacing the divergent $\Gamma$ factor, where the sign as $|{\bf k}| \to \infty$ matches the case of slightly higher $\Delta$, and the sign as $|{\bf k}| \to 0$ matches the case of slightly lower $\Delta$. The reason this happens is that $|{\bf k}|^{2n}$ is a polynomial in the ${\bf k}$ vector, and without the log term it would Fourier transform to a pure contract term, failing to match \eqref{psi2_y}.

For a unitary bulk theory (and for a non-tachyon field), the Bunch-Davies wavefunction $\Psi[J]$  should be normalisable.  This is easiest to check in momentum space,\footnote{The reason for this, is that a ${\bf k}$ momentum state is an eigenvalue mode of the 2-pt function, and so the normalisation criterion is simple.  The position space correlator with ${\bf y_1} \ne {\bf y_2}$ is an off-diagonal element, and so its relation to normalisability is more obscure.} where it requires (for a free theory) that the phase of the $-\psi_2$ coefficient is negative:
\begin{equation}\label{nml}
\text{Re}[-\psi_2] = 
\text{Re}[\Gamma(\tfrac{d}{2}-\Delta)
f(\Delta)] < 0.
\end{equation}
Even in an interacting theory, \eqref{nml} should hold (at least weakly as $\le 0$) if we demand that we are Taylor expanding the cosmological wavefunction around a peak of its absolute value (rather than a saddle point or minimum).\footnote{From the bulk perspective where $\Psi[J]$ is a wavefunction (i.e.~the square-root of a probability distribution), this ``peak'' is somewhat dependent on the measure used to evaluate $\Psi[J]$, but in the large $N$ limit one gets tightly peaked Gaussians, and so reasonably smooth choices of measure should agree.  While from a boundary perspective, $Z[J]$ is defined as a functional over $J$ from the beginning.  Evidently, some sort of measure on the sources $J$ enters into the dS/CFT dictionary, or it would not be possible to compare these two objects.}

Meanwhile our \CRT reality condition \eqref{eqn:CPTPhase} implies that
\begin{equation}\label{reality}
f(\Delta) 
\;\propto\; 
\pm e^{(i\pi/2)((d+1) - 2d + 2\Delta)}
= \pm e^{i\pi(\Delta - (d-1)/2)}.
\end{equation}
Combining \eqref{nml} and \eqref{reality}, we deduce that in momentum space, we have
\begin{equation}
\Gamma(\tfrac{d}{2}-\Delta) f(\Delta) 
\; \propto \; 
(-1)^{\big\lfloor \Delta - \tfrac{d}{2}\big\rfloor}
e^{i\pi(\Delta - (d-1)/2)},
\end{equation}
where the sign prefactor ensures that the real part is non-positive.\footnote{That is, the real part has the opposite sign as would be expected in a Statistical CFT, for a 2-pt function without UV divergences ($\Delta < d/2$).}  The jumps in the sign occur whenever $f(\Delta)$ is purely imaginary, and are always associated with log divergences in holographic renormalisation.

Hence, the correct position space phase is:
\begin{equation}\label{fphase}
f(\Delta) \;\propto\;
ie^{i\pi(\Delta - (d-2)/2)}
\end{equation}
whenever $\Delta > (d-2)/2$, which is the case of interest when rotating from a unitary theory.\footnote{Whenever $\Delta < (d-2)/2 - n$, $n \in \mathbb{N}$, there is a sign jump in position space.  Although bulk unitarity in dS does not prohibit operators with $\Delta < (d-2)/2$, in general these are more annoying because they require introducing boundary counterterms proportional to products of dynamical boundary operators, rather than just fixed boundary sources.  So at least from a bulk perspective, you would generally want to consider the Fourier transformed basis in which the roles of the source and operator are swapped.  This would, however, correspond to a modified dual boundary CFT.}  As a sanity check, for $\Delta < d/2$, the signs are the same in position and momentum space, and hence in the window 
$(d-2)/2 < \Delta < d/2$, the real part of the phase \eqref{fphase} is negative.  And the signs are real in the conformally coupled cases $\Delta = (d \pm 1)/2$, as this is equivalent to a flat spacetime, where the vacuum state should have time-reversal symmetry.

Next, we check what values of $N$ and $\lambda$ are required when the boundary theory is weakly coupled, and arises from rotating a unitary theory.  (Neither of these assumptions needs to be true in general; this is merely a first pass at identifying the correct positivity constraint.)  Consider first operators of the form
\begin{equation}\label{O_p}
    \mathcal{O}_p \;=\; \frac{N}{\lambda}\,
    :\!\Tr(\phi^p)\!:\,,
\end{equation}
where $\phi$ is a boundary adjoint scalar, $p \ge 2$ in an SU($N$) theory, and for the moment we assume a conventional structure of the particle spectrum (no negative flavours, no negative norm states, etc.), so that all phases (including minus signs relative to unitary models) originate purely from the complex values of the parameters $N$ and $\lambda$.

We now examine the boundary position space 2 point function:
\begin{equation}
\langle {\cal O}_p({\bf y}_1) {\cal O}_p({\bf y}_2)\rangle \;\propto\; N^2\lambda^{p-2},
\end{equation}
because in the free limit the phase comes from a simple Wick contraction of all $\phi$'s in one operator with all $\phi$'s in the other.  Since $\phi$ is a free field, the weight of ${\cal O}_p$ is just $p$ times the unitary bound: $\Delta_p = p(d-2)/2$, and thus plugging into \eqref{fphase} we obtain
\begin{equation}
-\psi_2 
\;\propto\; 
\frac{N^2}{\lambda} \lambda^{p-1}
\;\propto\;
ie^{i\pi(p-1)(d-2)/2}.
\end{equation}
Now taking the ($p-1$) root of both sides, we have the solution:
\begin{equation}
    N^2 \;=\; i^{d-1} \, , \qquad \lambda \;=\; i^{d-2} \, ,
\end{equation}
where this is the only solution that works for both $p=2$ and $p=3$ (the latter of which is in the complementary series range for $d \le 6$, where $\phi^3$ is renormalisable).\footnote{If the the theory is such that the operator spectrum can be restricted to include only traces with $p = \text{even}$, there is a second solution sending $\lambda \to -\lambda$; this reverses the signs of Feynman diagrams with an odd number of vertices.}  Hence the phase is fixed uniquely by the boundary dimension $d$ mod 4, with no residual $\pm$ ambiguity. This is the key point of the exercise: in this simplified free-field limit, consistency with bulk unitarity and normalisability of the wavefunction enforces definite complex phases for both $N^2$ and $\lambda$.\footnote{It is worth noting that a gauge theory is most likely to be stable nonperturbatively if $\text{Re}[\lambda] \ge 0$.  This suggests that dS/CFT is most likely to be defined in the ranges $1 \le d \le 3$ and $5 \le d \le 7$, with $d = 4$ (the critical dimension for perturbative Yang-Mills) excluded.  Nevertheless, at the level of a formal perturbation expansion, a $d = 4 + \epsilon$ expansion to $d = 5$ might be a possible way to access strongly coupled gauge theories in $d = 5$ dimensions, where $N^2 > 0$ is permitted.  See \cite{Morris:2004mg,DeCesare:2021pfb} for discussion of the $4+\epsilon$ expansion for Yang-Mills (in the unitary case).}

If we replace two of the $\phi$'s in the trace \eqref{O_p} with two spinors $\psi^2$, this adds +1 to the dimension of $\Delta_p$, because each free spinor has a different dimension $\Delta_\psi = (d-1)/2$.  This is presumably dealt with by the 2 extra factors of $i$ appearing upon doing the $\bar{\Omega}$ rotation of the boundary vielbein $\overline{e_\mu^a}$ in the spinor propagator (this did not matter in Section \ref{sec:consider} because there we were not concerned with an overall minus sign).

\paragraph{Beyond Scalar Operators.}
The above analysis can also be generalised beyond the case of scalar operators.  In the case of a stress-tensor 2-point function $\langle T T \rangle$, the condition is the same as for scalar operators.  Indeed, the overall stress-tensor coefficient was already fixed by $c$, in full agreement with \eqref{Nplus}.  Actually, in a free theory there are \emph{multiple} stress-tensors, one for each gauge-distinguishable species of boundary field.  Enforcing positivity for \emph{each} of them would additionally require the number of each type of species to be positive.  This would naively rule out a scenario with $d = 4n + 3$ and yet $N^2 > 0$, in which the required $c < 0$ arises from a negative number of $\phi$ or $\psi$ matter fields.  However, when interactions are introduced with $|\lambda| > 0$, only one of these stress-tensors will remain conserved, and so it is not completely clear that this scenario can be ruled out in the case of a strongly interacting theory.

If the boundary theory has a continuous global symmetry, we can also look at the correlators of the associated conserved Noether current $J^\mu$, which (if the matter sector it acts on is a CFT) has dimension $\Delta_J = d-1$, and is a conformal primary (distinct from the derivative $\partial {\cal O}$ of a scalar operator, which is not).  In this case, similar arguments show that for a unitary bulk dual (containing a Maxwell field) the 2-point function of $J^\mu$ also has the phase
\begin{equation}
\langle J^\mu(({\bf y}_1) J^\nu({\bf y}_2) \rangle 
\;\propto\; 
ie^{i\pi(\Delta_J - (d-2)/2)} = i^{d-4}
\end{equation}
relative to the unitary case (with the index structure being the only one allowed by conservation and scaling).  Since a conserved current is bilinear in the free fields (e.g.~$J^\mu \sim \phi^* \partial^\mu \phi$ or $\overline{\psi}\gamma^\mu \psi$), it scales as $N^2$.  Thus, it turns out that (starting with the unitary case) this requires the opposite sign of $N^2$ as the spin 0 and spin 2 operators!\footnote{This is not true of $\partial {\cal O}$, the derivative of a scalar, as in this case an appropriate phase is inherited from that of the scalar, whose $\Delta$ is one less than that of $\partial {\cal O}$.}  This gives rise to an apparent inconsistency.  This suggests the need for either (i) some additional nonunitary feature in the boundary theory, such as rotating $J^\mu$ by an additional factor of $i$, or else (ii) odd index currents must be gauged in order to have a unitary bulk dual, or else (iii) perhaps the sign can change at strong coupling.  This is another puzzle which might help to illuminate the constraints required for a consistent example of dS/CFT.  

\section{Even vs Odd Dimensions}\label{sec:EvenvsOdd}
A striking outcome of our analysis is the sharp distinction between even- and odd-dimensional cosmological spacetimes in the formulation of Kosmic Field Theories (KFTs). Several structural features make the case of $d = \text{odd}$ boundary dimensions  significantly more natural, and more likely to be consistent.  If any of these features play an important role in our own universe, this could potentially explain why the cosmology we observe has an even number ($d+1 = 4$) of spacetime dimensions.

\paragraph{1.~Central charge and large-$N$ scaling.}
As discussed in Section \ref{sec:central} and \cite{Goodhew:2024eup}, the number of degrees of freedom of a KFT has to scale as $i^{d-1}$, where $d$ is the boundary dimension.  This means that for even-dimensional de Sitter spacetimes ($d = \text{odd}$) the analogue of the central charge, scaling as $N^2$ at large-$N$, is real.  This also makes it possible that $N \in \mathbb{Z}$, thus satisfying the usual quantisation condition for the number of colours.  By contrast, in odd-dimensional de Sitter ($d = \text{even}$), the central charge is imaginary at tree-level and at $L = \text{even}$ loop order. The physical meaning of an imaginary central charge, which nominally counts degrees of freedom, is obscure, and undermines any straightforward interpretation in terms of the number of degrees of freedom, for KFTs in even $d$.  

On the other hand, it is easy to construct field theories with a positive number of degrees of freedom in $d = 4n + 1$, or a even negative number of degrees of freedom, in $d = 4n - 1$.  The latter can be done by adding spin-statistics violating ghosts, which effectively count as negative degrees of freedom. (This is most familiar from their use as Faddeev-Popov ghosts to cancel unwanted degrees of freedom, but in this case we be introducing them for other reasons.)  This would be the case corresponding to our own physical universe $d = 3$, and also holds for the Sp($N$) duals to Vasiliev gravity, although our analysis of the 2 point function in Section \ref{sec:2ptsign} suggests that it may not be possible to make this work for weakly coupled adjoint models.

\paragraph{2.~Closure under RG flow.}
As shown in Section \ref{sec:closure}, when $d = \text{odd}$, our conditions for a KFT are closed under RG flow.  For other dimensions $d$, this only seems to be true for topologically trivial RG flows, i.e.~at leading order in a $1/N$ expansion.  This seems to imply that, while one could potentially identify strongly coupled KFT fixed points in the large $N$ limit, worldsheet loop corrections would need to be handled separately without absorbing them into the parameters of the boundary Lagrangian.

\paragraph{3.~Open and Non-orientable String Duality.}
Consistency of string perturbation theory requires open--closed duality: surfaces with boundaries in the open-string channel must be equivalent to closed-string diagrams of corresponding genus. For example, the annulus is simultaneously a one-loop open-string amplitude and a tree-level closed-string exchange between branes.  In even bulk spacetime dimensions, Reflection Reality (\RR) ensures this duality holds consistently, as shown in \ref{sec:ONO}.  In other dimensions, \RR requires relative phase factors that differ by loop order, spoiling the necessary identification of open- and closed-string channels   Again, this seems to favour the case of even dimensional bulk cosmology, i.e.~$d = \text{odd}$.

An important caveat to the above is that we do not have access to a conformal worldsheet theory in the bulk.  The usual form of open–closed worldsheet duality arises from the conformal invariance of the worldsheet, which implies that it is invariant under modular transformations.  In our duality statement, we are simply \emph{assuming} that there exists a dual string theory description, with the usual notion of modular invariance, and associated Feynman diagrams.  The problem is simply that the bulk loop order $L$ of an open (or non-orientable) worldsheet is fundamentally ambiguous, because it could potentially be associated with worldsheets with different numbers of $L$.  Unless that ambiguity can be resolved in a way that is \emph{local} with respect to the \emph{boundary} Feynman diagrams, there is simply no way to consistently assign amplitudes so as to always result in the correct phase.

\paragraph{4.~A Unitary Euclidean Boundary Dual.}
While the boundary field theory is not reflection-positive, at some level this is not a surprise since it does not arise by Wick rotating a boundary theory.  Instead, it arises as a temporal boundary to a Lorentzian spacetime.  This naturally leads to a different choice of adjoint ($\dagger$) operation than for the Wick rotation of a unitary theory.\footnote{We are grateful to Santiago Agüí Salcedo for making this point.}  This is because in a usual Euclidean field theory, the $\dagger$ obtained from Wick rotation reverses the sign of one of the space dimensions $\tau$ (whichever one you are thinking of as imaginary time), while in holographic cosmology, the adjoint $\dagger$ inherited from the Lorentzian bulk dual should not do so.

For example, in the case of dS/CFT, the conformal generators acting on the Lorentzian de Sitter group should be self-adjoint.  But these are related by holographic duality to the generators of the conformal group on the boundary Spin($d+1,1$).  It is thus natural to propose that the boundary theory is also unitary under those same generators.  In other words, the Euclidean dual theory should satisfy ``Euclidean Unitarity'' ({\bf EU}) --- {\emph{not}} in the sense of being the Wick rotation of a QFT that is unitary in Lorentzian signature --- but rather in the sense native to Euclidean signature, in which the Euclidean generators act in a unitary way.  Even in the case of a holographic dual which is not conformal, we could still demand unitarity with respect to the generators of the Euclidean group ISO($d$).

As the boundary theory's Hamiltonian $\bar{H}$ is one of these generators, {\bf EU} would require it to act in a unitary way.  If we now make the simplifying assumption that the boundary theory was reflection-positive \emph{before} rotating $N$ and $\lambda$,\footnote{This assumption could be violated by e.g.~a boundary scalar Lagrangian with more than 2 derivatives, because then the configuration space would have variables that are odd under reflection, which would imply the existence of states with negative norm.} then there already exists a positive-definite Hilbert space defined on ($d - 1$)-dimensional slices of the boundary theory, and the condition for unitarity is simply that the Hamiltonian must be anti-self adjoint rather than self-adjoint:
\begin{equation}\label{eq:antiself}
    \bar{H} = -\bar{H}^{\dagger_E} = -\bar{H}^*,
\end{equation}
where $\dagger_E$ means the \emph{usual Euclidean dagger}, and in the second line we are using the fact that in a parity-even theory (the only kind we consider in this paper), the Hamiltonian is always transpose symmetric: $\bar{H} = \bar{H}^{T}$.  Also, we are defining time evolution by the usual Euclidean convention:
\begin{equation}
\bar{\Psi}(\tau) = e^{- \tau \bar{H}} \bar{\Psi}(0).
\end{equation}
Then $\bar{H}\Psi = E\Psi$ with $E \in i\mathbb{R}$, whenever $\Psi$ is a $\delta$-function normalisable energy eigenstate $\bar{\Psi}$ of $\bar{H}$.\footnote{For radial evolution, the energy eigenvalues that evolve unitarily correspond to $\Delta \in i\mathbb{R}$.  While this might seem to conflict at first sight with $\Delta$ associated with the principal ($\Delta = d/2 + i\mathbb{R}$) or complementary ($0 \le \Delta \le d$) series, in fact there is no contradiction here.  The usual values of $\Delta$ quoted are for a \emph{primary} state in the representation, and (unlike AdS/CFT) these states are badly non-normalisable, and thus exist off the real axis for unitary evolution.  By analogy, in one particle quantum mechanics, it is possible to find wavefunctions with arbitrary complex values of the momentum operator $\hat{p}$ if you don't care about whether $\Psi(x) = e^{i\hat{p}x}$ is normalisable, but only the states with real $\hat{p}$ form a good basis in general, in the absence of an analyticity condition imposed on the states.}  It follows from \eqref{eq:antiself} that $\bar{H}$ is purely imaginary, and hence the boundary Lagrangian is also imaginary.  As we shall see, this should be interpreted as requiring $\lambda$ to be imaginary.

We now claim that a KFT satisfies {\bf EU} only when $d = \text{odd}$, or equivalently that {\bf EU} $\Leftrightarrow$ \RR only in odd dimensions.  Since in some cases $N \notin \mathbb{R}$, we need to be careful, since the interpretation of the KFT in terms of gauge fields is obscure.  To be sure we are imposing {\bf EU} correctly, let us instead consider specifically how the Hamiltonian $\bar{H}$ acts on a 1-string (single-trace) state.  As long as no topology change occurs on the worldsheet, i.e.~in the planar limit of 1 string $\to$ 1 string processes, the string undergoes pure propagation rather than splitting or joining, so $N$ plays no role.  Consequently, the dynamics of a single-string state is governed entirely by the ’t~Hooft coupling $\lambda$, independent of $N$.  Now a KFT must satisfy
\begin{equation}
    \lambda \;\propto\; \pm (i^{\,d}) \,,
\end{equation}
and thus the expectation of an imaginary sign for bulk evolution {\bf EU} appears to be met only in the case where $d = \text{odd}$. Only then, can we simultaneously have unitarity \emph{both} on the boundary ({\bf EU}), and have the bulk wavefunction satisfy the conditions required by \RR.

It is surprising that these two principles come apart in general.  That is, it seems like a bit of a miracle that a KCFT which is not Euclidean unitary with respect to the boundary conformal group, can nevertheless give rise to amplitudes which satisfy the \RR condition coming from unitarity with respect to bulk de Sitter generators.

\paragraph{5.~Volume of the Conformal Group.}
We will mention one more way in which $d = \text{odd}$ dimensions is simpler, concerning the absence of log divergences in the conformal group at infinity.\footnote{If we consider fractional dimensions, this one actually only requires $d \ne \text{even}$, so its conditions are actually less restrictive than the others.  Perhaps it is conceptually unrelated to the others.}

At the end of the day, we would like to use the cosmological wavefunction to evaluate the expectation values for some cosmological observable $X$.  This requires doing a path integral over 2 copies of the boundary field theory:
\begin{equation}
    \int \frac{D\bar{g}\,DJ}{\text{Vol}[G]}  \Psi[\bar{g}, J] \Psi^*[\bar{g}, J] X,
\end{equation}
where $\bar{g}$ is the boundary metric, $J$ represents all other boundary sources, and $G$ is the gauge group of bulk canonical symmetries.

Let us suppose we are considering a dS/CFT dual, so the boundary theory at ${\cal I}^+$ is conformally invariant; then even after gauge fixing $G$ must include a factor of 1/Vol[CKG], the group of conformal Killing symmetries.  If we ignore the finite Spin$(d)$ rotational contribution, this is the same as the volume of a hyperbolic space $H_{d+1}$ with the same number of dimensions as the bulk de Sitter.  This is divergent, necessitating the introduction of a small cutoff $\epsilon$ (which looks like an IR cutoff in the bulk, and a UV cutoff in the boundary).

After throwing out power law divergences, we obtain a divergent answer as $\epsilon \to 0$ for when $d = \text{even}$, and a finite one when $d = \text{odd}$:
\begin{equation}\label{eq:volumes}
\text{Vol[Spin($d+1,1$)]}
\propto
\begin{cases}
+\log(\epsilon^{-1}),&\quad d \text{ mod } 4 = 0, \\
\quad -1, &\quad d \text{ mod } 4 = 1,\\
-\log(\epsilon^{-1}),&\quad d \text{ mod } 4 = 2, \\
\quad +1, &\quad d \text{ mod } 4 = 3.
\end{cases}
\end{equation}
Here, we see a $d$ mod 4 pattern, so which case arises is in synch with the value of the dS/CFT central charge $c$.

The proper treatment of the cases $d=1$ and $d = 2$ is important in string theory, as the conformal killing symmetries of a disk and sphere worldsheet respectively.  See \cite{Eberhardt:2021ynh} and \cite{Ahmadain:2022tew} for recent discussion of the role of the respective $H_2$ and $H_3$ hyperbolic geometries.  In the case of the sphere, the resulting log divergence causes the string action to vanish on-shell, and requires the use of a special $\partial / \partial (\log \epsilon)$ prescription to evaluate amplitudes \cite{Tseytlin:1988tv}, the effect of which is that the amplitude for an exact CFT actually \emph{vanishes} in the absence of vertex operator insertions, and only the off-shell effective string action (meaning that there is a nonzero $\beta$ function in the worldsheet theory) is nonzero.  On the other hand, for the disk a finite result is obtained \cite{Liu:1987nz,Witten:1992qy,Witten:1992cr,Tseytlin:2000mt} even for an exact CFT, due to the finiteness of the volume after removing the power law divergence, with a finite \emph{negative} amplitude as predicted by \eqref{eq:volumes}.

By analogy with the string theory case, when $d = \text{even}$ one would expect that the probability density must vanish for boundary sources $J$ with exact conformal invariance.  Additionally, the $d = 4n + 1$ case gives us a minus sign which is also hard to interpret from a probabilistic perspective.  However, both of these pathologies might be avoided by considering only relational observables defined away from exact conformal invariance, as proposed recently by \cite{Chakraborty:2025izq} based on previous work in \cite{Chakraborty:2023yed,Chakraborty:2023los}.

\medskip
Taken together, these observations strongly suggest that even-dimensional cosmological spacetimes provide the natural setting for KFT holography, while odd-dimensional cases face intrinsic obstacles ranging from the meaning of imaginary central charges to the breakdown of open--closed string duality.

\section{Discussion}\label{sec:discussion}
In this work, we have proposed and analysed a new class of large-$N$ field theories, which we have called \emph{Kosmic Field Theories} (KFTs), as candidate holographic duals to strings propagating on cosmological (time-dependent) spacetimes such as de Sitter and, more generally, flat FLRW geometries. The key technical input was the non-perturbative constraint of bulk Reflection Reality (\RR), derived from the Cosmological CPT Theorem~\cite{Goodhew:2024eup}, which encodes bulk unitarity at the level of the cosmological wavefunction. We have shown how this constraint shapes the complex phases appearing in the dual boundary theory and, in particular, forces the large-$N$ expansion parameters $N$ and $\lambda$ to take complex values depending on the spacetime dimension. This leads to a new and unusual class of large-$N$ theories which extend the familiar ’t Hooftian gauge theories of AdS/CFT. In this discussion we summarise our results, highlight the implications, and outline several pressing open questions.

\subsection{Summary of Results}
Our starting point was to formulate the action for a KFT in the ’t Hooftian normalisation
\begin{equation}
   I = \frac{N}{\lambda}\int d^d x\,\sqrt{\bar g}\,\Tr\!\big(M[A,\phi,\psi;\bar g]\big)\,,
\end{equation}
and to analyse the large-$N$ expansion of connected correlation functions,
\begin{equation}
   \psi_n \equiv \langle \mathcal O_1\cdots \mathcal O_n\rangle_{\rm conn}
   \;\sim\; N^{\chi}\,\lambda^{\,\bar{E}-\bar{V}}\,,
\end{equation}
where $\chi=\bar{V}-\bar{E}+\bar{F}$ is the Euler characteristic of the ribbon graph. By imposing \RR, which enforces consistency with bulk unitarity, we determined the phase factors multiplying amplitudes of a given loop order or genus, and we solved for the allowed phases of $N$ and $\lambda$ in terms of the spacetime dimension $d$. 

One striking outcome is that there is a fundamental obstruction to having both $N$ and $\lambda$ be real simultaneously. In other words (at least in the parity-even case, which we are considering for simplicity) no unitary reflection-positive large-$N$ theory can ever satisfy the \RR condition while being dual to a unitary Lorentzian bulk spacetime.  Hence, in accordance with previous expectations, holography in cosmological spacetimes cannot be formulated in terms of ordinary unitary large-$N$ CFTs of the familiar AdS/CFT type. Instead, one is naturally led to either complex rank gauge theories or complex ’t Hooft coupling.  (Or, upon dimensional regularisation to fractional $d$, both at the same time!)

Our proposal identifies a mathematical framework in which to understand dS holography and holographic cosmology more generally: large-$N$ theories analytically continued to complex $N$ and $\lambda$, with correlation functions defined by the bulk \RR constraint. These are the Kosmic Field Theories we have studied.

\subsection{What is the String Dual?}
A natural and pressing open question is: what precisely is the string theory dual to a KFT?  For KFTs, at least one of $N$ and $\lambda$ are complex, and hence so too are the effective couplings of the dual string theory.  This raises several questions about the target-space description.

In the well-studied AdS/CFT duality of $\mathcal{N}=4$ SYM and type IIB string theory on $\text{AdS}_5 \times \text{S}_5$ we have a clear dictionary relating $N$ and $\lambda$ to the Planck length $\ell_p$ and string length $\ell_s$ in AdS units:
\begin{equation}\label{eq:AdS5}
     N^{-\frac{2}{3}} \sim \left(\frac{\ell_p}{L_{\text{AdS}}}\right),
\qquad
    \lambda^{-\frac{1}{4}} \sim \left(\frac{\ell_s}{L_{\text{AdS}}}\right)
\end{equation}
while analogous expressions exist for dualities in other dimensions.  

In the case of \eqref{eq:AdS5}, reversing the sign of the cosmological constant in planck units $\Lambda/(\ell_p)^2 \to - \Lambda/(\ell_p)^2$ requires solving $N^2 = (-1)^{3/4}$, of which one of the 4 possible interpretations is compatible with our proposed rotation $N^2 \to -i N^2$.\footnote{If this idea were correct, presumably the correct choice of fourth root would be determined by a contour deformation argument.}  However, the justification of expressions like \eqref{eq:AdS5} depend in a detailed way on e.g.~the number of KK fibre directions, so formulas like these cannot be used as a safe basis for extrapolating to dS/CFT dualities.

At present we do not have a useful worldsheet description.  By construction we expect the bulk theory to at least satisfy \RR, and hopefully be fully unitary.  However, we expect that the worldsheet description of strings in time-dependent (cosmological) backgrounds will involve studying nonunitary worldsheet theories, given that the $X^0$ temporal coordinate of the string has negative norm.\footnote{Of course this is always true even in Minkowski, but in a time-symmetric background it is easy to Wick rotate $X^0$ to obtain a positive norm scalar field.}

\subsection{Why Even Spacetime Dimensions are Preferred}
Our analysis in Section \ref{sec:EvenvsOdd} also revealed a striking difference between even-dimensional spacetimes, and other dimensions. Several features conspire to make $d + 1 =$ even-dimensional cosmology the more natural setting for holography:

\begin{enumerate}
   \item The analogue of the central charge (or number of degrees of freedom) is real, though possibly negative, at all loop orders. In odd dimensional dS, by contrast, the central charge is imaginary at tree level and at even loop orders, with no clear physical interpretation.
   \item Even-dimensional bulk spacetimes are closed under renormalisation group flow in the sense that perturbations remain within the class of KFTs. In odd dimensions, complex phases appear in the large-$N$ expansion that obstruct such closure.
   \item Open–closed string duality, a cornerstone of perturbative string consistency, is maintained in even spacetime dimensions but obstructed in odd dimensions by relative phase factors induced by \RR.
   \item  Naively, unitary representations of the boundary conformal group would require an \emph{anti-self-adjoint} boundary Hamiltonian, ruling out reflection-positive or unitary field theories as duals.  This suggests that the boundary theory should be unitary with respect to \emph{Euclidean} evolution, but we find this result only for even spacetime dimensions.
   \item For odd $d + 1$, there is a log divergence in the renormalised volume of the conformal group Spin($d+1$,1), which is absent in other dimensions.
\end{enumerate}

These arguments suggest that the holographic duality between KFTs and strings on cosmological spacetimes can be consistently formulated most easily in even-dimensional cosmologies.  Perhaps, this might even explain why our own universe is even-dimensional.

\subsection{Entropy and Microstate Counting}
A natural question is whether KFTs can provide a microscopic counting of the entropy associated with de Sitter horizons and with black holes in asymptotically de Sitter spacetimes. 

In AdS/CFT, the central charge and Cardy-like formulas underpin successful entropy counts. In dS holography, by contrast, the horizon entropy is finite but lacks a universally accepted microstate interpretation. Since KFTs are large-$N$ theories, one might hope that the density of states they describe could account for the de Sitter entropy. However, the fact that $N$ and $\lambda$ are complex introduces new subtleties: the state-counting formulae may produce phases or cancellations that must be carefully interpreted. 

A central open direction is to construct partition functions of KFTs on compact Euclidean manifolds and investigate whether their asymptotics reproduce the Bekenstein–Hawking entropy of de Sitter space. If successful, this would provide a microscopic underpinning of de Sitter entropy in a holographic framework, much as AdS/CFT has done for black hole entropy.

\subsection{RG Flows and General FLRW Backgrounds}
Another avenue is to interpret KFTs in the context of renormalisation group flows. In AdS/CFT, deformations of the boundary CFT correspond to changes in the bulk geometry away from pure AdS. By analogy, we expect KCFTs to be dual to de Sitter, while more general flat FLRW cosmologies correspond to RG flows away from such fixed points.

This picture is attractive: it would mean that the rich structure of cosmological spacetimes could be understood in terms of flows in the space of complexified large-$N$ gauge theories. It would be useful, however, to clarify this dictionary more precisely.  It would also be interesting to try to derive an analogue of the $c$-theorem in this setting, perhaps as boundary duals to General Relativity theorems (assuming a reasonable energy condition for the bulk cosmology).

\subsection{Outlook}
These results open a novel window on the problem of de Sitter holography. Rather than seeking a conventional reflection–positive CFT, our analysis suggests that the appropriate boundary duals are complexified large–$N$ gauge theories constrained by Reflection Reality. This perspective reconciles bulk unitarity with the non–unitarity of the boundary, and provides a precise dictionary between phases in the cosmological wavefunction and the parameters of the boundary theory.  Much remains to be done.  Some of the most urgent tasks are as follows:
\begin{itemize}
   \item Constructing explicit examples of KFTs, either by analytic continuation of known gauge theories or by novel constructions, and computing their correlation functions.  
   \item Exploring RG flows of KFTs and their bulk interpretation in terms of FLRW cosmologies; with special attention to fixed points, which would correspond to scale-invariant cosmologies.  In the case of a conformal fixed point (a KCFT), this would give rise to a dS/CFT model.  
   \item Determine whether it is possible (at strong coupling) to eliminate tachyons and/or introduce principal series particles.
   \item As a $d = 5$ KCFT (dual to $dS_6$) would be compatible with a positive number of degrees of freedom ($c > 0$), a $d = 4+\epsilon$ expansion of Yang-Mills \cite{Morris:2004mg,DeCesare:2021pfb} (but with KFT reality conditions) seems like a possible avenue, although for pure (unitary) Yang-Mills it appears that a fixed point exists in $d = 5$ (and possibly $d = 6$) only when $N$ is small.  For the O($N$) model there is a fixed point in $d = 5$ which is nonunitary at the nonperturbative level \cite{Giombi:2019upv}; but this would not necessarily be an obstacle to interpreting it in terms of dS/CFT.
   \item Developing the dictionary between KFT parameters and bulk string theory couplings, particularly the worldsheet description of 't Hooft strings with complex $N$ or $\lambda$.
   \item Investigating the entropy and state-counting problem in de Sitter holography from the perspective of KFTs.

\end{itemize}
Answering these questions will help to determine whether KFTs can provide a complete holographic description of cosmological spacetimes. Regardless of the eventual outcome, our results confirm that time-dependent spacetimes must go beyond ordinary unitary, reflection-positive field theories.  Accommodating ourselves to this fact will be essential for developing a consistent theory of quantum gravity in general spacetimes.

\section*{Acknowledgements}
AT is particularly grateful to Evan Craft for his engaging discussions on gauge theories and analytic continuations, Barak Gabai for his helpful insights on gauge theories and string field theory, as well as Maciej Kolanowski, Shota Komatsu and Victor Rodriguez for in-depth conversations regarding the possible string duals to KFTs. We are also grateful for useful conversations with Amr Ahmadain, Nima Arkani-Hamed, Santiago Agüí Salcedo, Dionysios Anninos, Tarek Anous, Gonçalo Araújo Regado, Alejandra Castro, Luca Ciambelli, Joydeep Chakravarty, Carlos Duaso Pueyo, Lorenz Eberhardt, Davide Gaiotto, Victor Gorbenko, Sean Hartnoll, Thomas Hertog, Michael Imseis, Rohit Kalloor, Diego Liska, Arthur Lipstein, Adrián López-Raven, Alex Maloney, Juan Maldacena, Donald Marolf, Robert Myers, Enrico Pajer, Sabrina Pasterski, João Penedones, Harvey Reall, Jiaxin Qiao, Bharathkumar Radhakrishnan, Veronica Sacchi, Kamran Salehi Vaziri, Jorge Santos, Atul Sharma, Steve Shenker, Vasudev Shyam, Ronak Soni, Eva Silverstein, Xi Tong and Mattia Varrone.

AT was supported in part by the Heising-Simons Foundation, the Simons Foundation, the Bell Burnell Graduate Scholarship Fund, the Cavendish (University of Cambridge) and a KITP Graduate fellowship. AT also gratefully acknowledges hospitality from the Perimeter Institute while working on this article.  AT also acknowledges support from the SNF starting grant “The Fundamental Description of the Expanding Universe. AT and AW were supported by the AFOSR grant FA9550-19-1-0260 “Tensor Networks and Holographic Spacetime”, the STFC grant ST/X000664/1 “Quantum Fields, Quantum Gravity and Quantum Particles”, and grant no. NSF PHY-2309135 to the Kavli Institute for Theoretical Physics (KITP).

 For the purpose of open access, we have applied a CC BY public copyright licence to any Author Accepted Manuscript version arising.


\bibliographystyle{JHEP}
\bibliography{refs}

@article{Ouyang:2011fs,
    author = "Ouyang, Peter",
    title = "{Toward Higher Spin dS3/CFT2}",
    eprint = "1111.0276",
    archivePrefix = "arXiv",
    primaryClass = "hep-th",
    month = "11",
    year = "2011"
}

@article{Sleight:2019mgd,
    author = "Sleight, Charlotte",
    title = "{A Mellin Space Approach to Cosmological Correlators}",
    eprint = "1906.12302",
    archivePrefix = "arXiv",
    primaryClass = "hep-th",
    doi = "10.1007/JHEP01(2020)090",
    journal = "JHEP",
    volume = "01",
    pages = "090",
    year = "2020"
}

@article{Sleight:2019hfp,
    author = "Sleight, Charlotte and Taronna, Massimo",
    title = "{Bootstrapping Inflationary Correlators in Mellin Space}",
    eprint = "1907.01143",
    archivePrefix = "arXiv",
    primaryClass = "hep-th",
    reportNumber = "PUPT-2590",
    doi = "10.1007/JHEP02(2020)098",
    journal = "JHEP",
    volume = "02",
    pages = "098",
    year = "2020"
}

@article{Sleight:2020obc,
    author = "Sleight, Charlotte and Taronna, Massimo",
    title = "{From AdS to dS Exchanges: Spectral Representation, Mellin Amplitudes and Crossing}",
    eprint = "2007.09993",
    archivePrefix = "arXiv",
    primaryClass = "hep-th",
    month = "7",
    year = "2020"
}

@article{Sleight:2021iix,
    author = "Sleight, Charlotte and Taronna, Massimo",
    title = "{On the consistency of (partially-)massless matter couplings in de Sitter space}",
    eprint = "2106.00366",
    archivePrefix = "arXiv",
    primaryClass = "hep-th",
    month = "6",
    year = "2021"
}

@article{Sleight:2021plv,
    author = "Sleight, Charlotte and Taronna, Massimo",
    title = "{From dS to AdS and back}",
    eprint = "2109.02725",
    archivePrefix = "arXiv",
    primaryClass = "hep-th",
    doi = "10.1007/JHEP12(2021)074",
    journal = "JHEP",
    volume = "12",
    pages = "074",
    year = "2021"
}

@article{Sleight:2023ojm,
    author = "Sleight, Charlotte and Taronna, Massimo",
    title = "{Celestial Holography Revisited}",
    eprint = "2301.01810",
    archivePrefix = "arXiv",
    primaryClass = "hep-th",
    doi = "10.1103/PhysRevLett.133.241601",
    journal = "Phys. Rev. Lett.",
    volume = "133",
    number = "24",
    pages = "241601",
    year = "2024"
}

@article{Chopping:2024oiu,
    author = "Chopping, Alistair J. and Sleight, Charlotte and Taronna, Massimo",
    title = "{Cosmological correlators for Bogoliubov initial states}",
    eprint = "2407.16652",
    archivePrefix = "arXiv",
    primaryClass = "hep-th",
    doi = "10.1007/JHEP09(2024)152",
    journal = "JHEP",
    volume = "09",
    pages = "152",
    year = "2024"
}

@article{Pacifico:2025emk,
    author = "Pacifico, Francesca and Sleight, Charlotte and Taronna, Massimo",
    title = "{Conformal Partial Wave Expansion of Celestial Correlators}",
    eprint = "2502.03087",
    archivePrefix = "arXiv",
    primaryClass = "hep-th",
    month = "2",
    year = "2025"
}

@article{MdAbhishek:2025dhx,
    author = "Abhishek, Md. and Sleight, Charlotte and Taronna, Massimo",
    title = "{Cosmological Correlators in Gauge Theory and Gravity from EAdS}",
    eprint = "2509.09536",
    archivePrefix = "arXiv",
    primaryClass = "hep-th",
    month = "9",
    year = "2025"
}

@article{Sleight:2025dmt,
    author = "Sleight, Charlotte and Taronna, Massimo",
    title = "{(Non-)Conserved Currents and Cosmological Correlators}",
    eprint = "2509.18888",
    archivePrefix = "arXiv",
    primaryClass = "hep-th",
    month = "9",
    year = "2025"
}

@article{Morris:2004mg,
    author = "Morris, Tim R.",
    title = "{Renormalizable extra-dimensional models}",
    eprint = "hep-ph/0410142",
    archivePrefix = "arXiv",
    reportNumber = "CERN-PH-TH-2004-120, SHEP-0420",
    doi = "10.1088/1126-6708/2005/01/002",
    journal = "JHEP",
    volume = "01",
    pages = "002",
    year = "2005"
}

@article{DeCesare:2021pfb,
    author = "De Cesare, Fabiana and Di Pietro, Lorenzo and Serone, Marco",
    title = "{Five-dimensional CFTs from the {\ensuremath{\varepsilon}}-expansion}",
    eprint = "2107.00342",
    archivePrefix = "arXiv",
    primaryClass = "hep-th",
    doi = "10.1103/PhysRevD.104.105015",
    journal = "Phys. Rev. D",
    volume = "104",
    number = "10",
    pages = "105015",
    year = "2021"
}

@article{Giombi:2019upv,
    author = "Giombi, Simone and Huang, Richard and Klebanov, Igor R. and Pufu, Silviu S. and Tarnopolsky, Grigory",
    title = "{The $O(N)$ Model in $ {4 < d < 6} $ : Instantons and complex CFTs}",
    eprint = "1910.02462",
    archivePrefix = "arXiv",
    primaryClass = "hep-th",
    reportNumber = "PUPT-2599",
    doi = "10.1103/PhysRevD.101.045013",
    journal = "Phys. Rev. D",
    volume = "101",
    number = "4",
    pages = "045013",
    year = "2020"
}

@article{vanRitbergen:1997va,
    author = "van Ritbergen, T. and Vermaseren, J. A. M. and Larin, S. A.",
    title = "{The Four loop beta function in quantum chromodynamics}",
    eprint = "hep-ph/9701390",
    archivePrefix = "arXiv",
    reportNumber = "UM-TH-97-01, NIKHEF-97-001",
    doi = "10.1016/S0370-2693(97)00370-5",
    journal = "Phys. Lett. B",
    volume = "400",
    pages = "379--384",
    year = "1997"
}

@article{Czakon:2004bu,
    author = "Czakon, M.",
    title = "{The Four-loop QCD beta-function and anomalous dimensions}",
    eprint = "hep-ph/0411261",
    archivePrefix = "arXiv",
    reportNumber = "DESY-04-223, SFB-CPP-04-62",
    doi = "10.1016/j.nuclphysb.2005.01.012",
    journal = "Nucl. Phys. B",
    volume = "710",
    pages = "485--498",
    year = "2005"
}

@article{Herzog:2017ohr,
    author = "Herzog, F. and Ruijl, B. and Ueda, T. and Vermaseren, J. A. M. and Vogt, A.",
    title = "{The five-loop beta function of Yang-Mills theory with fermions}",
    eprint = "1701.01404",
    archivePrefix = "arXiv",
    primaryClass = "hep-ph",
    reportNumber = "NIKHEF-2017-001, LTH-1117",
    doi = "10.1007/JHEP02(2017)090",
    journal = "JHEP",
    volume = "02",
    pages = "090",
    year = "2017"
}

@article{tHooft:1973alw,
    author = "'t Hooft, Gerard",
    editor = "Taylor, J. C.",
    title = "{A Planar Diagram Theory for Strong Interactions}",
    reportNumber = "CERN-TH-1786",
    doi = "10.1016/0550-3213(74)90154-0",
    journal = "Nucl. Phys. B",
    volume = "72",
    pages = "461",
    year = "1974"
}

@inproceedings{tHooft:2002ufq,
    author = "'t Hooft, G.",
    title = "{Large N}",
    booktitle = "{The Phenomenology of Large N(c) QCD}",
    eprint = "hep-th/0204069",
    archivePrefix = "arXiv",
    reportNumber = "SPIN-2002-08, ITF-2002-14",
    doi = "10.1142/9789812776914_0001",
    pages = "3--18",
    month = "4",
    year = "2002"
}

@article{Polyakov:2001af,
    author = "Polyakov, Alexander M.",
    editor = "Duff, M. J. and Liu, J. T.",
    title = "{Gauge fields and space-time}",
    eprint = "hep-th/0110196",
    archivePrefix = "arXiv",
    reportNumber = "PUPT-2010",
    doi = "10.1142/S0217751X02013071",
    journal = "Int. J. Mod. Phys. A",
    volume = "17S1",
    pages = "119--136",
    year = "2002"
}

@article{Cadoni:2002xe,
    author = "Cadoni, Mariano and Carta, Paolo",
    title = "{Tachyons in de Sitter space and analytical continuation from dS / CFT to AdS / CFT}",
    eprint = "hep-th/0211018",
    archivePrefix = "arXiv",
    reportNumber = "INFNCA-TH0209",
    doi = "10.1142/S0217751X0401821X",
    journal = "Int. J. Mod. Phys. A",
    volume = "19",
    pages = "4985--5002",
    year = "2004"
}

@article{Deser:2001us,
    author = "Deser, Stanley and Waldron, A.",
    title = "{Partial masslessness of higher spins in (A)dS}",
    eprint = "hep-th/0103198",
    archivePrefix = "arXiv",
    reportNumber = "BRX-TH-486",
    doi = "10.1016/S0550-3213(01)00212-7",
    journal = "Nucl. Phys. B",
    volume = "607",
    pages = "577--604",
    year = "2001"
}

@article{Hull:1998vg,
    author = "Hull, C. M.",
    title = "{Timelike T duality, de Sitter space, large N gauge theories and topological field theory}",
    eprint = "hep-th/9806146",
    archivePrefix = "arXiv",
    reportNumber = "QMW-PH-98-28",
    doi = "10.1088/1126-6708/1998/07/021",
    journal = "JHEP",
    volume = "07",
    pages = "021",
    year = "1998"
}

@article{Hull:1998ym,
    author = "Hull, C. M.",
    title = "{Duality and the signature of space-time}",
    eprint = "hep-th/9807127",
    archivePrefix = "arXiv",
    reportNumber = "QMW-PH-98-30",
    doi = "10.1088/1126-6708/1998/11/017",
    journal = "JHEP",
    volume = "11",
    pages = "017",
    year = "1998"
}

@article{Hull:1999mt,
    author = "Hull, Christopher M. and Khuri, Ramzi R.",
    title = "{World volume theories, holography, duality and time}",
    eprint = "hep-th/9911082",
    archivePrefix = "arXiv",
    reportNumber = "QMW-PH-99-17, BCUNY-HEP-99-02",
    doi = "10.1016/S0550-3213(00)00057-2",
    journal = "Nucl. Phys. B",
    volume = "575",
    pages = "231--254",
    year = "2000"
}

@article{Dijkgraaf:2016lym,
    author = "Dijkgraaf, Robbert and Heidenreich, Ben and Jefferson, Patrick and Vafa, Cumrun",
    title = "{Negative Branes, Supergroups and the Signature of Spacetime}",
    eprint = "1603.05665",
    archivePrefix = "arXiv",
    primaryClass = "hep-th",
    doi = "10.1007/JHEP02(2018)050",
    journal = "JHEP",
    volume = "02",
    pages = "050",
    year = "2018"
}

@article{Balasubramanian:2001rb,
    author = "Balasubramanian, Vijay and Horava, Petr and Minic, Djordje",
    title = "{Deconstructing de Sitter}",
    eprint = "hep-th/0103171",
    archivePrefix = "arXiv",
    reportNumber = "CITUSC-01-005, RUNHETC-2001-09, UPR-929-T",
    doi = "10.1088/1126-6708/2001/05/043",
    journal = "JHEP",
    volume = "05",
    pages = "043",
    year = "2001"
}

@article{Arkani-Hamed:2015bza,
    author = "Arkani-Hamed, Nima and Maldacena, Juan",
    title = "{Cosmological Collider Physics}",
    eprint = "1503.08043",
    archivePrefix = "arXiv",
    primaryClass = "hep-th",
    month = "3",
    year = "2015"
}

@inproceedings{Witten:2001kn,
    author = "Witten, Edward",
    title = "{Quantum gravity in de Sitter space}",
    booktitle = "{Strings 2001: International Conference}",
    eprint = "hep-th/0106109",
    archivePrefix = "arXiv",
    month = "6",
    year = "2001"
}

@article{Castro:2011xb,
    author = "Castro, Alejandra and Lashkari, Nima and Maloney, Alexander",
    title = "{A de Sitter Farey Tail}",
    eprint = "1103.4620",
    archivePrefix = "arXiv",
    primaryClass = "hep-th",
    doi = "10.1103/PhysRevD.83.124027",
    journal = "Phys. Rev. D",
    volume = "83",
    pages = "124027",
    year = "2011"
}

@article{Castro:2011ke,
    author = "Castro, Alejandra and Lashkari, Nima and Maloney, Alexander",
    title = "{Quantum Topologically Massive Gravity in de Sitter Space}",
    eprint = "1105.4733",
    archivePrefix = "arXiv",
    primaryClass = "hep-th",
    doi = "10.1007/JHEP08(2011)040",
    journal = "JHEP",
    volume = "08",
    pages = "040",
    year = "2011"
}

@article{Castro:2012gc,
    author = "Castro, Alejandra and Maloney, Alexander",
    title = "{The Wave Function of Quantum de Sitter}",
    eprint = "1209.5757",
    archivePrefix = "arXiv",
    primaryClass = "hep-th",
    reportNumber = "NSF-KITP-12-165",
    doi = "10.1007/JHEP11(2012)096",
    journal = "JHEP",
    volume = "11",
    pages = "096",
    year = "2012"
}

@article{Castro:2020smu,
    author = "Castro, Alejandra and Sabella-Garnier, Philippe and Zukowski, Claire",
    title = "{Gravitational Wilson Lines in 3D de Sitter}",
    eprint = "2001.09998",
    archivePrefix = "arXiv",
    primaryClass = "hep-th",
    doi = "10.1007/JHEP07(2020)202",
    journal = "JHEP",
    volume = "07",
    pages = "202",
    year = "2020"
}

@article{Hikida:2022ltr,
    author = "Hikida, Yasuaki and Nishioka, Tatsuma and Takayanagi, Tadashi and Taki, Yusuke",
    title = "{CFT duals of three-dimensional de Sitter gravity}",
    eprint = "2203.02852",
    archivePrefix = "arXiv",
    primaryClass = "hep-th",
    reportNumber = "YITP-22-20, IPMU22-0006",
    doi = "10.1007/JHEP05(2022)129",
    journal = "JHEP",
    volume = "05",
    pages = "129",
    year = "2022"
}

@article{Castro:2023dxp,
    author = "Castro, Alejandra and Coman, Ioana and Fliss, Jackson R. and Zukowski, Claire",
    title = "{Keeping matter in the loop in dS$_{3}$ quantum gravity}",
    eprint = "2302.12281",
    archivePrefix = "arXiv",
    primaryClass = "hep-th",
    doi = "10.1007/JHEP07(2023)120",
    journal = "JHEP",
    volume = "07",
    pages = "120",
    year = "2023",
    note = "[Erratum: JHEP 09, 004 (2024)]"
}

@article{Fliss:2023muk,
    author = "Fliss, Jackson",
    title = "{A 3d perspective on de Sitter quantum field theory}",
    doi = "10.22323/1.436.0129",
    journal = "PoS",
    volume = "CORFU2022",
    pages = "129",
    year = "2023"
}

@article{Hertog:2011ky,
    author = "Hertog, Thomas and Hartle, James",
    title = "{Holographic No-Boundary Measure}",
    eprint = "1111.6090",
    archivePrefix = "arXiv",
    primaryClass = "hep-th",
    doi = "10.1007/JHEP05(2012)095",
    journal = "JHEP",
    volume = "05",
    pages = "095",
    year = "2012"
}

@article{Hartle:2012qb,
    author = "Hartle, James B. and Hawking, S. W. and Hertog, Thomas",
    title = "{Accelerated Expansion from Negative $\Lambda$}",
    eprint = "1205.3807",
    archivePrefix = "arXiv",
    primaryClass = "hep-th",
    month = "5",
    year = "2012"
}

@article{Conti:2015ruo,
    author = "Conti, Gabriele and Hertog, Thomas and van der Woerd, Ellen",
    title = "{Holographic Tunneling Wave Function}",
    eprint = "1506.07374",
    archivePrefix = "arXiv",
    primaryClass = "hep-th",
    doi = "10.1007/JHEP12(2015)025",
    journal = "JHEP",
    volume = "12",
    pages = "025",
    year = "2015"
}

@article{Hertog:2015nia,
    author = "Hertog, Thomas and van der Woerd, Ellen",
    title = "{Primordial fluctuations from complex AdS saddle points}",
    eprint = "1509.03291",
    archivePrefix = "arXiv",
    primaryClass = "hep-th",
    doi = "10.1088/1475-7516/2016/02/010",
    journal = "JCAP",
    volume = "02",
    pages = "010",
    year = "2016"
}

@article{Hawking:2017wrd,
    author = "Hawking, S. W. and Hertog, Thomas",
    title = "{A Smooth Exit from Eternal Inflation?}",
    eprint = "1707.07702",
    archivePrefix = "arXiv",
    primaryClass = "hep-th",
    doi = "10.1007/JHEP04(2018)147",
    journal = "JHEP",
    volume = "04",
    pages = "147",
    year = "2018"
}

@article{Hertog:2017ymy,
    author = "Hertog, Thomas and Tartaglino-Mazzucchelli, Gabriele and Van Riet, Thomas and Venken, Victoria",
    title = "{Supersymmetric dS/CFT}",
    eprint = "1709.06024",
    archivePrefix = "arXiv",
    primaryClass = "hep-th",
    doi = "10.1007/JHEP02(2018)024",
    journal = "JHEP",
    volume = "02",
    pages = "024",
    year = "2018"
}

@article{Hertog:2024shf,
    author = "Hertog, Thomas and Pauwels, Jef and Venken, Victoria",
    title = "{Holographic (eternal) inflation}",
    eprint = "2411.18396",
    archivePrefix = "arXiv",
    primaryClass = "hep-th",
    doi = "10.1007/JHEP07(2025)061",
    journal = "JHEP",
    volume = "07",
    pages = "061",
    year = "2025"
}

@article{Buchel:2002kj,
    author = "Buchel, Alex and Langfelder, Peter and Walcher, Johannes",
    title = "{On time dependent backgrounds in supergravity and string theory}",
    eprint = "hep-th/0207214",
    archivePrefix = "arXiv",
    reportNumber = "NSF-ITP-02-64, YITP-SB-02-37",
    doi = "10.1103/PhysRevD.67.024011",
    journal = "Phys. Rev. D",
    volume = "67",
    pages = "024011",
    year = "2003"
}

@article{Hoare:2014pna,
    author = "Hoare, B. and Roiban, R. and Tseytlin, A. A.",
    title = "{On deformations of $AdS_n$ x $S^n$ supercosets}",
    eprint = "1403.5517",
    archivePrefix = "arXiv",
    primaryClass = "hep-th",
    reportNumber = "IMPERIAL-TP-AT-2014-02, HU-EP-14-10",
    doi = "10.1007/JHEP06(2014)002",
    journal = "JHEP",
    volume = "06",
    pages = "002",
    year = "2014"
}

@article{Arutyunov:2014cra,
    author = "Arutyunov, Gleb and van Tongeren, Stijn J.",
    title = "{$\mathrm{AdS}_5 \times \mathrm{S}^5$ mirror model as a string sigma model}",
    eprint = "1406.2304",
    archivePrefix = "arXiv",
    primaryClass = "hep-th",
    reportNumber = "HU-EP-14-21, HU-MATH-14-12, ITP-UU-14-18, SPIN-14-16",
    doi = "10.1103/PhysRevLett.113.261605",
    journal = "Phys. Rev. Lett.",
    volume = "113",
    pages = "261605",
    year = "2014"
}

@article{Cvetic:1996vr,
    author = "Cvetic, Mirjam and Soleng, Harald H.",
    title = "{Supergravity domain walls}",
    eprint = "hep-th/9604090",
    archivePrefix = "arXiv",
    reportNumber = "IASSNS-HEP-96-25, CERN-TH-96-97",
    doi = "10.1016/S0370-1573(96)00035-X",
    journal = "Phys. Rept.",
    volume = "282",
    pages = "159--223",
    year = "1997"
}

@article{Skenderis:2006fb,
    author = "Skenderis, K. and Townsend, P. K.",
    editor = "Sola, Joan",
    title = "{Pseudo-Supersymmetry and the Domain-Wall/Cosmology Correspondence}",
    eprint = "hep-th/0610253",
    archivePrefix = "arXiv",
    reportNumber = "DAMTP-2006-81, ITFA-2006-38",
    doi = "10.1088/1751-8113/40/25/S18",
    journal = "J. Phys. A",
    volume = "40",
    pages = "6733--6742",
    year = "2007"
}

@article{Chakraborty:2025izq,
    author = "Chakraborty, Tuneer and H, Ashik and Raju, Suvrat",
    title = "{Cosmological correlators in gravitationally-constrained de Sitter states}",
    eprint = "2507.15926",
    archivePrefix = "arXiv",
    primaryClass = "hep-th",
    month = "7",
    year = "2025"
}

@article{Chakraborty:2023los,
    author = "Chakraborty, Tuneer and Chakravarty, Joydeep and Godet, Victor and Paul, Priyadarshi and Raju, Suvrat",
    title = "{Holography of information in de Sitter space}",
    eprint = "2303.16316",
    archivePrefix = "arXiv",
    primaryClass = "hep-th",
    doi = "10.1007/JHEP12(2023)120",
    journal = "JHEP",
    volume = "12",
    pages = "120",
    year = "2023"
}

@article{Chakraborty:2023yed,
    author = "Chakraborty, Tuneer and Chakravarty, Joydeep and Godet, Victor and Paul, Priyadarshi and Raju, Suvrat",
    title = "{The Hilbert space of de Sitter quantum gravity}",
    eprint = "2303.16315",
    archivePrefix = "arXiv",
    primaryClass = "hep-th",
    doi = "10.1007/JHEP01(2024)132",
    journal = "JHEP",
    volume = "01",
    pages = "132",
    year = "2024"
}

@article{Liu:1987nz,
    author = "Liu, Jun and Polchinski, Joseph",
    title = "{Renormalization of the Mobius Volume}",
    reportNumber = "UTTG-26-87",
    doi = "10.1016/0370-2693(88)91566-3",
    journal = "Phys. Lett. B",
    volume = "203",
    pages = "39--43",
    year = "1988"
}

@article{Witten:1992qy,
    author = "Witten, Edward",
    title = "{On background independent open string field theory}",
    eprint = "hep-th/9208027",
    archivePrefix = "arXiv",
    reportNumber = "IASSNS-HEP-92-53",
    doi = "10.1103/PhysRevD.46.5467",
    journal = "Phys. Rev. D",
    volume = "46",
    pages = "5467--5473",
    year = "1992"
}

@article{Witten:1992cr,
    author = "Witten, Edward",
    title = "{Some computations in background independent off-shell string theory}",
    eprint = "hep-th/9210065",
    archivePrefix = "arXiv",
    reportNumber = "IASSNS-HEP-92-63",
    doi = "10.1103/PhysRevD.47.3405",
    journal = "Phys. Rev. D",
    volume = "47",
    pages = "3405--3410",
    year = "1993"
}

@article{Tseytlin:2000mt,
    author = "Tseytlin, Arkady A.",
    title = "{Sigma model approach to string theory effective actions with tachyons}",
    eprint = "hep-th/0011033",
    archivePrefix = "arXiv",
    reportNumber = "OHSTPY-HEP-T-00-025",
    doi = "10.1063/1.1376129",
    journal = "J. Math. Phys.",
    volume = "42",
    pages = "2854--2871",
    year = "2001"
}

@article{Witten:2013pra,
    author = "Witten, Edward",
    title = "{The Feynman $i \epsilon$ in String Theory}",
    eprint = "1307.5124",
    archivePrefix = "arXiv",
    primaryClass = "hep-th",
    doi = "10.1007/JHEP04(2015)055",
    journal = "JHEP",
    volume = "04",
    pages = "055",
    year = "2015"
}

@article{Eberhardt:2021ynh,
    author = "Eberhardt, Lorenz and Pal, Sridip",
    title = "{The disk partition function in string theory}",
    eprint = "2105.08726",
    archivePrefix = "arXiv",
    primaryClass = "hep-th",
    doi = "10.1007/JHEP08(2021)026",
    journal = "JHEP",
    volume = "08",
    pages = "026",
    year = "2021"
}

@article{Ahmadain:2022tew,
    author = "Ahmadain, Amr and Wall, Aron C.",
    title = "{Off-shell strings I: S-matrix and action}",
    eprint = "2211.08607",
    archivePrefix = "arXiv",
    primaryClass = "hep-th",
    doi = "10.21468/SciPostPhys.17.1.005",
    journal = "SciPost Phys.",
    volume = "17",
    number = "1",
    pages = "005",
    year = "2024"
}

@article{Tseytlin:1988tv,
    author = "Tseytlin, Arkady A.",
    title = "{Mobius Infinity Subtraction and Effective Action in $\sigma$ Model Approach to Closed String Theory}",
    reportNumber = "Print-88-0440 (LEBEDEV INST)",
    doi = "10.1016/0370-2693(88)90421-2",
    journal = "Phys. Lett. B",
    volume = "208",
    pages = "221--227",
    year = "1988"
}

@article{Maldacena:1997re,
    author = "Maldacena, Juan Martin",
    title = "{The Large $N$ limit of superconformal field theories and supergravity}",
    eprint = "hep-th/9711200",
    archivePrefix = "arXiv",
    reportNumber = "HUTP-97-A097, HUTP-98-A097",
    doi = "10.4310/ATMP.1998.v2.n2.a1",
    journal = "Adv. Theor. Math. Phys.",
    volume = "2",
    pages = "231--252",
    year = "1998"
}

@article{Gubser:1998bc,
    author = "Gubser, S. S. and Klebanov, Igor R. and Polyakov, Alexander M.",
    title = "{Gauge theory correlators from noncritical string theory}",
    eprint = "hep-th/9802109",
    archivePrefix = "arXiv",
    reportNumber = "PUPT-1767",
    doi = "10.1016/S0370-2693(98)00377-3",
    journal = "Phys. Lett. B",
    volume = "428",
    pages = "105--114",
    year = "1998"
}

@article{Witten:1998qj,
    author = "Witten, Edward",
    title = "{Anti de Sitter space and holography}",
    eprint = "hep-th/9802150",
    archivePrefix = "arXiv",
    reportNumber = "IASSNS-HEP-98-15",
    doi = "10.4310/ATMP.1998.v2.n2.a2",
    journal = "Adv. Theor. Math. Phys.",
    volume = "2",
    pages = "253--291",
    year = "1998"
}

@article{Planck:2018vyg,
    author = "Aghanim, N. and others",
    collaboration = "Planck",
    title = "{Planck 2018 results. VI. Cosmological parameters}",
    eprint = "1807.06209",
    archivePrefix = "arXiv",
    primaryClass = "astro-ph.CO",
    doi = "10.1051/0004-6361/201833910",
    journal = "Astron. Astrophys.",
    volume = "641",
    pages = "A6",
    year = "2020",
    note = "[Erratum: Astron.Astrophys. 652, C4 (2021)]"
}

@article{SDSS:2006lmn,
    author = "Tegmark, Max and others",
    collaboration = "SDSS",
    title = "{Cosmological Constraints from the SDSS Luminous Red Galaxies}",
    eprint = "astro-ph/0608632",
    archivePrefix = "arXiv",
    reportNumber = "FERMILAB-PUB-06-511-A-CD",
    doi = "10.1103/PhysRevD.74.123507",
    journal = "Phys. Rev. D",
    volume = "74",
    pages = "123507",
    year = "2006"
}

@article{BOSS:2016wmc,
    author = "Alam, Shadab and others",
    collaboration = "BOSS",
    title = "{The clustering of galaxies in the completed SDSS-III Baryon Oscillation Spectroscopic Survey: cosmological analysis of the DR12 galaxy sample}",
    eprint = "1607.03155",
    archivePrefix = "arXiv",
    primaryClass = "astro-ph.CO",
    doi = "10.1093/mnras/stx721",
    journal = "Mon. Not. Roy. Astron. Soc.",
    volume = "470",
    number = "3",
    pages = "2617--2652",
    year = "2017"
}

@article{SupernovaSearchTeam:1998fmf,
    author = "Riess, Adam G. and others",
    collaboration = "Supernova Search Team",
    title = "{Observational evidence from supernovae for an accelerating universe and a cosmological constant}",
    eprint = "astro-ph/9805201",
    archivePrefix = "arXiv",
    doi = "10.1086/300499",
    journal = "Astron. J.",
    volume = "116",
    pages = "1009--1038",
    year = "1998"
}

@article{SupernovaCosmologyProject:1998vns,
    author = "Perlmutter, S. and others",
    collaboration = "Supernova Cosmology Project",
    title = "{Measurements of $\Omega$ and $\Lambda$ from 42 High Redshift Supernovae}",
    eprint = "astro-ph/9812133",
    archivePrefix = "arXiv",
    reportNumber = "LBNL-41801, LBL-41801",
    doi = "10.1086/307221",
    journal = "Astrophys. J.",
    volume = "517",
    pages = "565--586",
    year = "1999"
}

@article{Guth:1980zm,
    author = "Guth, Alan H.",
    editor = "Fang, Li-Zhi and Ruffini, R.",
    title = "{The Inflationary Universe: A Possible Solution to the Horizon and Flatness Problems}",
    reportNumber = "SLAC-PUB-2576",
    doi = "10.1103/PhysRevD.23.347",
    journal = "Phys. Rev. D",
    volume = "23",
    pages = "347--356",
    year = "1981"
}

@article{Linde:1981mu,
    author = "Linde, Andrei D.",
    editor = "Fang, Li-Zhi and Ruffini, R.",
    title = "{A New Inflationary Universe Scenario: A Possible Solution of the Horizon, Flatness, Homogeneity, Isotropy and Primordial Monopole Problems}",
    reportNumber = "LEBEDEV-81-229",
    doi = "10.1016/0370-2693(82)91219-9",
    journal = "Phys. Lett. B",
    volume = "108",
    pages = "389--393",
    year = "1982"
}

@article{Starobinsky:1980te,
    author = "Starobinsky, Alexei A.",
    editor = "Khalatnikov, I. M. and Mineev, V. P.",
    title = "{A New Type of Isotropic Cosmological Models Without Singularity}",
    doi = "10.1016/0370-2693(80)90670-X",
    journal = "Phys. Lett. B",
    volume = "91",
    pages = "99--102",
    year = "1980"
}

@article{Sundborg:2000wp,
    author = "Sundborg, Bo",
    editor = "Sorokin, Dmitri P.",
    title = "{Stringy gravity, interacting tensionless strings and massless higher spins}",
    eprint = "hep-th/0103247",
    archivePrefix = "arXiv",
    doi = "10.1016/S0920-5632(01)01545-6",
    journal = "Nucl. Phys. B Proc. Suppl.",
    volume = "102",
    pages = "113--119",
    year = "2001"
}

@article{Sezgin:2002rt,
    author = "Sezgin, E. and Sundell, P.",
    title = "{Massless higher spins and holography}",
    eprint = "hep-th/0205131",
    archivePrefix = "arXiv",
    reportNumber = "CTP-TAMU-08-02, UU-01-10",
    doi = "10.1016/S0550-3213(02)00739-3",
    journal = "Nucl. Phys. B",
    volume = "644",
    pages = "303--370",
    year = "2002",
    note = "[Erratum: Nucl.Phys.B 660, 403--403 (2003)]"
}

@article{Klebanov:2002ja,
    author = "Klebanov, I. R. and Polyakov, A. M.",
    title = "{AdS dual of the critical O(N) vector model}",
    eprint = "hep-th/0210114",
    archivePrefix = "arXiv",
    reportNumber = "PUPT-2053",
    doi = "10.1016/S0370-2693(02)02980-5",
    journal = "Phys. Lett. B",
    volume = "550",
    pages = "213--219",
    year = "2002"
}

@article{Giombi:2009wh,
    author = "Giombi, Simone and Yin, Xi",
    title = "{Higher Spin Gauge Theory and Holography: The Three-Point Functions}",
    eprint = "0912.3462",
    archivePrefix = "arXiv",
    primaryClass = "hep-th",
    doi = "10.1007/JHEP09(2010)115",
    journal = "JHEP",
    volume = "09",
    pages = "115",
    year = "2010"
}

@article{Giombi:2012ms,
    author = "Giombi, Simone and Yin, Xi",
    title = "{The Higher Spin/Vector Model Duality}",
    eprint = "1208.4036",
    archivePrefix = "arXiv",
    primaryClass = "hep-th",
    doi = "10.1088/1751-8113/46/21/214003",
    journal = "J. Phys. A",
    volume = "46",
    pages = "214003",
    year = "2013"
}

@article{Anninos:2011ui,
    author = "Anninos, Dionysios and Hartman, Thomas and Strominger, Andrew",
    title = "{Higher Spin Realization of the dS/CFT Correspondence}",
    eprint = "1108.5735",
    archivePrefix = "arXiv",
    primaryClass = "hep-th",
    doi = "10.1088/1361-6382/34/1/015009",
    journal = "Class. Quant. Grav.",
    volume = "34",
    number = "1",
    pages = "015009",
    year = "2017"
}

@article{Anninos:2012ft,
    author = "Anninos, Dionysios and Denef, Frederik and Harlow, Daniel",
    title = "{Wave function of Vasiliev\textquoteright{}s universe: A few slices thereof}",
    eprint = "1207.5517",
    archivePrefix = "arXiv",
    primaryClass = "hep-th",
    reportNumber = "SU-ITP-12-19",
    doi = "10.1103/PhysRevD.88.084049",
    journal = "Phys. Rev. D",
    volume = "88",
    number = "8",
    pages = "084049",
    year = "2013"
}

@article{Anninos:2017eib,
    author = "Anninos, Dionysios and Denef, Frederik and Monten, Ruben and Sun, Zimo",
    title = "{Higher Spin de Sitter Hilbert Space}",
    eprint = "1711.10037",
    archivePrefix = "arXiv",
    primaryClass = "hep-th",
    doi = "10.1007/JHEP10(2019)071",
    journal = "JHEP",
    volume = "10",
    pages = "071",
    year = "2019",
    note = "[Erratum: JHEP 06, 085 (2024)]"
}

@article{Zamolodchikov:2004ce,
    author = "Zamolodchikov, Alexander B.",
    title = "{Expectation value of composite field T anti-T in two-dimensional quantum field theory}",
    eprint = "hep-th/0401146",
    archivePrefix = "arXiv",
    reportNumber = "BONN-TH-2004-02",
    month = "1",
    year = "2004"
}

@article{Smirnov:2016lqw,
    author = "Smirnov, F. A. and Zamolodchikov, A. B.",
    title = "{On space of integrable quantum field theories}",
    eprint = "1608.05499",
    archivePrefix = "arXiv",
    primaryClass = "hep-th",
    doi = "10.1016/j.nuclphysb.2016.12.014",
    journal = "Nucl. Phys. B",
    volume = "915",
    pages = "363--383",
    year = "2017"
}

@article{Aharony:2018vux,
    author = "Aharony, Ofer and Vaknin, Talya",
    title = "{The TT* deformation at large central charge}",
    eprint = "1803.00100",
    archivePrefix = "arXiv",
    primaryClass = "hep-th",
    doi = "10.1007/JHEP05(2018)166",
    journal = "JHEP",
    volume = "05",
    pages = "166",
    year = "2018"
}

@article{Kraus:2018xrn,
    author = "Kraus, Per and Liu, Junyu and Marolf, Donald",
    title = "{Cutoff AdS$_{3}$ versus the $ T\overline{T} $ deformation}",
    eprint = "1801.02714",
    archivePrefix = "arXiv",
    primaryClass = "hep-th",
    reportNumber = "CALT-TH-2018-002",
    doi = "10.1007/JHEP07(2018)027",
    journal = "JHEP",
    volume = "07",
    pages = "027",
    year = "2018"
}

@article{Taylor:2018xcy,
    author = "Taylor, Marika",
    title = "{$T \bar{T}$ deformations in general dimensions}",
    eprint = "1805.10287",
    archivePrefix = "arXiv",
    primaryClass = "hep-th",
    doi = "10.4310/ATMP.2023.v27.n1.a2",
    journal = "Adv. Theor. Math. Phys.",
    volume = "27",
    number = "1",
    pages = "37--63",
    year = "2023"
}

@article{McGough:2016lol,
    author = "McGough, Lauren and Mezei, M{\'a}rk and Verlinde, Herman",
    title = "{Moving the CFT into the bulk with $ T\overline{T} $}",
    eprint = "1611.03470",
    archivePrefix = "arXiv",
    primaryClass = "hep-th",
    doi = "10.1007/JHEP04(2018)010",
    journal = "JHEP",
    volume = "04",
    pages = "010",
    year = "2018"
}

@article{Hartman:2018tkw,
    author = "Hartman, Thomas and Kruthoff, Jorrit and Shaghoulian, Edgar and Tajdini, Amirhossein",
    title = "{Holography at finite cutoff with a $T^2$ deformation}",
    eprint = "1807.11401",
    archivePrefix = "arXiv",
    primaryClass = "hep-th",
    doi = "10.1007/JHEP03(2019)004",
    journal = "JHEP",
    volume = "03",
    pages = "004",
    year = "2019"
}

@article{Shyam:2018sro,
    author = "Shyam, Vasudev",
    title = "{Finite Cutoff AdS$_{5}$ Holography and the Generalized Gradient Flow}",
    eprint = "1808.07760",
    archivePrefix = "arXiv",
    primaryClass = "hep-th",
    doi = "10.1007/JHEP12(2018)086",
    journal = "JHEP",
    volume = "12",
    pages = "086",
    year = "2018"
}

@article{Silverstein:2007ac,
    author = "Silverstein, Eva",
    title = "{Simple de Sitter Solutions}",
    eprint = "0712.1196",
    archivePrefix = "arXiv",
    primaryClass = "hep-th",
    reportNumber = "SLAC-PUB-13016, SITP-07-20",
    doi = "10.1103/PhysRevD.77.106006",
    journal = "Phys. Rev. D",
    volume = "77",
    pages = "106006",
    year = "2008"
}

@article{Silverstein:2008sg,
    author = "Silverstein, Eva and Westphal, Alexander",
    title = "{Monodromy in the CMB: Gravity Waves and String Inflation}",
    eprint = "0803.3085",
    archivePrefix = "arXiv",
    primaryClass = "hep-th",
    reportNumber = "SU-ITP-08-07, SLAC-PUB-13183",
    doi = "10.1103/PhysRevD.78.106003",
    journal = "Phys. Rev. D",
    volume = "78",
    pages = "106003",
    year = "2008"
}

@article{Dong:2010pm,
    author = "Dong, Xi and Horn, Bart and Silverstein, Eva and Torroba, Gonzalo",
    title = "{Micromanaging de Sitter holography}",
    eprint = "1005.5403",
    archivePrefix = "arXiv",
    primaryClass = "hep-th",
    reportNumber = "NSF-KITP-10-068, SU-ITP-10-19, SLAC-PUB-14146",
    doi = "10.1088/0264-9381/27/24/245020",
    journal = "Class. Quant. Grav.",
    volume = "27",
    pages = "245020",
    year = "2010"
}

@article{DeLuca:2021pej,
    author = "De Luca, G. Bruno and Silverstein, Eva and Torroba, Gonzalo",
    title = "{Hyperbolic compactification of M-theory and de Sitter quantum gravity}",
    eprint = "2104.13380",
    archivePrefix = "arXiv",
    primaryClass = "hep-th",
    doi = "10.21468/SciPostPhys.12.3.083",
    journal = "SciPost Phys.",
    volume = "12",
    number = "3",
    pages = "083",
    year = "2022"
}

@article{Gorbenko:2018oov,
    author = "Gorbenko, Victor and Silverstein, Eva and Torroba, Gonzalo",
    title = "{dS/dS and $ T\overline{T} $}",
    eprint = "1811.07965",
    archivePrefix = "arXiv",
    primaryClass = "hep-th",
    doi = "10.1007/JHEP03(2019)085",
    journal = "JHEP",
    volume = "03",
    pages = "085",
    year = "2019"
}

@article{Lewkowycz:2019xse,
    author = "Lewkowycz, Aitor and Liu, Junyu and Silverstein, Eva and Torroba, Gonzalo",
    title = "{$ T\overline{T} $ and EE, with implications for (A)dS subregion encodings}",
    eprint = "1909.13808",
    archivePrefix = "arXiv",
    primaryClass = "hep-th",
    reportNumber = "CALT-TH-2019--031",
    doi = "10.1007/JHEP04(2020)152",
    journal = "JHEP",
    volume = "04",
    pages = "152",
    year = "2020"
}

@article{Shyam:2021ciy,
    author = "Shyam, Vasudev",
    title = "{$ \mathrm{T}\overline{\mathrm{T}} $ + {\ensuremath{\Lambda}}$_{2}$ deformed CFT on the stretched dS$_{3}$ horizon}",
    eprint = "2106.10227",
    archivePrefix = "arXiv",
    primaryClass = "hep-th",
    doi = "10.1007/JHEP04(2022)052",
    journal = "JHEP",
    volume = "04",
    pages = "052",
    year = "2022"
}

@article{Coleman:2021nor,
    author = "Coleman, Evan and Mazenc, Edward A. and Shyam, Vasudev and Silverstein, Eva and Soni, Ronak M. and Torroba, Gonzalo and Yang, Sungyeon",
    title = "{De Sitter microstates from T$ \overline{T} $ + {\ensuremath{\Lambda}}$_{2}$ and the Hawking-Page transition}",
    eprint = "2110.14670",
    archivePrefix = "arXiv",
    primaryClass = "hep-th",
    doi = "10.1007/JHEP07(2022)140",
    journal = "JHEP",
    volume = "07",
    pages = "140",
    year = "2022"
}

@article{Silverstein:2022dfj,
    author = "Silverstein, Eva",
    title = "{Black hole to cosmic horizon microstates in string/M theory: timelike boundaries and internal averaging}",
    eprint = "2212.00588",
    archivePrefix = "arXiv",
    primaryClass = "hep-th",
    doi = "10.1007/JHEP05(2023)160",
    journal = "JHEP",
    volume = "05",
    pages = "160",
    year = "2023"
}

@article{Batra:2024kjl,
    author = "Batra, Gauri and De Luca, G. Bruno and Silverstein, Eva and Torroba, Gonzalo and Yang, Sungyeon",
    title = "{Bulk-local dS$_{3}$ holography: the matter with $ T\overline{T} $ + {\ensuremath{\Lambda}}$_{2}$}",
    eprint = "2403.01040",
    archivePrefix = "arXiv",
    primaryClass = "hep-th",
    doi = "10.1007/JHEP10(2024)072",
    journal = "JHEP",
    volume = "10",
    pages = "072",
    year = "2024"
}

@article{Silverstein:2024xnr,
    author = "Silverstein, Eva and Torroba, Gonzalo",
    title = "{Timelike-bounded dS$_{4}$ holography from a solvable sector of the T$^{2}$ deformation}",
    eprint = "2409.08709",
    archivePrefix = "arXiv",
    primaryClass = "hep-th",
    doi = "10.1007/JHEP03(2025)156",
    journal = "JHEP",
    volume = "03",
    pages = "156",
    year = "2025"
}

@article{Aguilar-Gutierrez:2024nst,
    author = "Aguilar-Gutierrez, Sergio E. and Svesko, Andrew and Visser, Manus R.",
    title = "{$ \textrm{T}\overline{\textrm{T}} $ deformations from AdS$_{2}$ to dS$_{2}$}",
    eprint = "2410.18257",
    archivePrefix = "arXiv",
    primaryClass = "hep-th",
    doi = "10.1007/JHEP01(2025)120",
    journal = "JHEP",
    volume = "01",
    pages = "120",
    year = "2025"
}

@article{Philcox:2025faf,
    author = "Philcox, Oliver H. E. and Silverstein, Eva and Torroba, Gonzalo",
    title = "{Quantum stress-energy at timelike boundaries: testing a new beyond-$\Lambda$CDM parameter with cosmological data}",
    eprint = "2507.00115",
    archivePrefix = "arXiv",
    primaryClass = "astro-ph.CO",
    month = "6",
    year = "2025"
}

@article{Chang:2025ays,
    author = "Chang, Jing-Cheng and He, Yang and Liu, Yu-Xiao and Sun, Yuan",
    title = "{Toward a Unified de Sitter Holography: A Composite $T\bar{T}$ and $T\bar{T}+\Lambda_2$ Flow}",
    eprint = "2511.16098",
    archivePrefix = "arXiv",
    primaryClass = "hep-th",
    month = "11",
    year = "2025"
}

@article{Susskind:2021omt,
    author = "Susskind, Leonard",
    title = "{De Sitter Holography: Fluctuations, Anomalous Symmetry, and Wormholes}",
    eprint = "2106.03964",
    archivePrefix = "arXiv",
    primaryClass = "hep-th",
    doi = "10.3390/universe7120464",
    journal = "Universe",
    volume = "7",
    number = "12",
    pages = "464",
    year = "2021"
}

@article{Rahman:2022jsf,
    author = "Rahman, Adel A.",
    title = "{dS JT Gravity and Double-Scaled SYK}",
    eprint = "2209.09997",
    archivePrefix = "arXiv",
    primaryClass = "hep-th",
    month = "9",
    year = "2022"
}

@article{Susskind:2022bia,
    author = "Susskind, Leonard",
    title = "{De Sitter Space, Double-Scaled SYK, and the Separation of Scales in the Semiclassical Limit}",
    eprint = "2209.09999",
    archivePrefix = "arXiv",
    primaryClass = "hep-th",
    doi = "10.22128/jhap.2024.920.1103",
    journal = "JHAP",
    volume = "5",
    number = "1",
    pages = "1--30",
    year = "2025"
}

@article{Narovlansky:2023lfz,
    author = "Narovlansky, Vladimir and Verlinde, Herman",
    title = "{Double-scaled SYK and de Sitter holography}",
    eprint = "2310.16994",
    archivePrefix = "arXiv",
    primaryClass = "hep-th",
    doi = "10.1007/JHEP05(2025)032",
    journal = "JHEP",
    volume = "05",
    pages = "032",
    year = "2025"
}

@article{Verlinde:2024znh,
    author = "Verlinde, Herman",
    title = "{Double-scaled SYK, chords and de Sitter gravity}",
    eprint = "2402.00635",
    archivePrefix = "arXiv",
    primaryClass = "hep-th",
    doi = "10.1007/JHEP03(2025)076",
    journal = "JHEP",
    volume = "03",
    pages = "076",
    year = "2025"
}

@article{Yuan:2024utc,
    author = "Yuan, Haiming and Ge, Xian-Hui and Kim, Keun-Young",
    title = "{Pole skipping in two-dimensional de Sitter spacetime and double-scaled SYK model}",
    eprint = "2408.12330",
    archivePrefix = "arXiv",
    primaryClass = "hep-th",
    doi = "10.1103/f3cb-kmnc",
    journal = "Phys. Rev. D",
    volume = "112",
    number = "2",
    pages = "026022",
    year = "2025"
}

@article{Verlinde:2024zrh,
    author = "Verlinde, Herman and Zhang, Mengyang",
    title = "{SYK correlators from 2D Liouville-de Sitter gravity}",
    eprint = "2402.02584",
    archivePrefix = "arXiv",
    primaryClass = "hep-th",
    doi = "10.1007/JHEP05(2025)053",
    journal = "JHEP",
    volume = "05",
    pages = "053",
    year = "2025"
}

@article{Anninos:2012qw,
    author = "Anninos, Dionysios",
    title = "{De Sitter Musings}",
    eprint = "1205.3855",
    archivePrefix = "arXiv",
    primaryClass = "hep-th",
    doi = "10.1142/S0217751X1230013X",
    journal = "Int. J. Mod. Phys. A",
    volume = "27",
    pages = "1230013",
    year = "2012"
}

@inproceedings{Flauger:2022hie,
    author = "Flauger, Raphael and Gorbenko, Victor and Joyce, Austin and McAllister, Liam and Shiu, Gary and Silverstein, Eva",
    title = "{Snowmass White Paper: Cosmology at the Theory Frontier}",
    booktitle = "{Snowmass 2021}",
    eprint = "2203.07629",
    archivePrefix = "arXiv",
    primaryClass = "hep-th",
    month = "3",
    year = "2022"
}

@article{Harlow:2022qsq,
    author = "Harlow, Daniel and others",
    title = "{TF1 Snowmass Report: Quantum gravity, string theory, and black holes}",
    eprint = "2210.01737",
    archivePrefix = "arXiv",
    primaryClass = "hep-th",
    month = "10",
    year = "2022"
}

@article{Galante:2023uyf,
    author = "Galante, Damian A.",
    title = "{Modave lectures on de Sitter space {\&} holography}",
    eprint = "2306.10141",
    archivePrefix = "arXiv",
    primaryClass = "hep-th",
    doi = "10.22323/1.435.0003",
    journal = "PoS",
    volume = "Modave2022",
    pages = "003",
    year = "2023"
}

@article{Goodhew:2024eup,
    author = "Goodhew, Harry and Thavanesan, Ayngaran and Wall, Aron C.",
    title = "{The Cosmological CPT Theorem}",
    eprint = "2408.17406",
    archivePrefix = "arXiv",
    primaryClass = "hep-th",
    month = "8",
    year = "2024"
}

@article{Boyle:2018rgh,
    author = "Boyle, Latham and Finn, Kieran and Turok, Neil",
    title = "{The Big Bang, CPT, and neutrino dark matter}",
    eprint = "1803.08930",
    archivePrefix = "arXiv",
    primaryClass = "hep-ph",
    doi = "10.1016/j.aop.2022.168767",
    journal = "Annals Phys.",
    volume = "438",
    pages = "168767",
    year = "2022"
}

@article{Boyle:2018tzc,
    author = "Boyle, Latham and Finn, Kieran and Turok, Neil",
    title = "{CPT-Symmetric Universe}",
    eprint = "1803.08928",
    archivePrefix = "arXiv",
    primaryClass = "hep-ph",
    doi = "10.1103/PhysRevLett.121.251301",
    journal = "Phys. Rev. Lett.",
    volume = "121",
    number = "25",
    pages = "251301",
    year = "2018"
}

@article{Boyle:2021jej,
    author = "Boyle, Latham and Turok, Neil",
    title = "{Two-Sheeted Universe, Analyticity and the Arrow of Time}",
    eprint = "2109.06204",
    archivePrefix = "arXiv",
    primaryClass = "hep-th",
    month = "9",
    year = "2021"
}

@article{Turok:2022fgq,
    author = "Turok, Neil and Boyle, Latham",
    title = "{Gravitational entropy and the flatness, homogeneity and isotropy puzzles}",
    eprint = "2201.07279",
    archivePrefix = "arXiv",
    primaryClass = "hep-th",
    doi = "10.1016/j.physletb.2024.138443",
    journal = "Phys. Lett. B",
    volume = "849",
    pages = "138443",
    year = "2024"
}

@article{Boyle:2022lyw,
    author = "Boyle, Latham and Teuscher, Martin and Turok, Neil",
    title = "{The Big Bang as a Mirror: a Solution of the Strong CP Problem}",
    eprint = "2208.10396",
    archivePrefix = "arXiv",
    primaryClass = "hep-ph",
    month = "8",
    year = "2022"
}

@article{Turok:2023amx,
    author = "Turok, Neil and Boyle, Latham",
    title = "{A Minimal Explanation of the Primordial Cosmological Perturbations}",
    eprint = "2302.00344",
    archivePrefix = "arXiv",
    primaryClass = "hep-ph",
    month = "2",
    year = "2023"
}

@article{Deng:2024uuz,
    author = "Deng, Wei-Ning and Handley, Will",
    title = "{Predicting spatial curvature $\Omega_K$ in globally $CPT$-symmetric universes}",
    eprint = "2407.18225",
    archivePrefix = "arXiv",
    primaryClass = "astro-ph.CO",
    month = "7",
    year = "2024"
}

@article{Hartle:1983ai,
    author = "Hartle, J. B. and Hawking, S. W.",
    editor = "Fang, Li-Zhi and Ruffini, R.",
    title = "{Wave Function of the Universe}",
    reportNumber = "PRINT-83-0937 (CAMBRIDGE)",
    doi = "10.1103/PhysRevD.28.2960",
    journal = "Phys. Rev. D",
    volume = "28",
    pages = "2960--2975",
    year = "1983"
}

@article{Halliwell:1984eu,
    author = "Halliwell, J. J. and Hawking, S. W.",
    editor = "Fang, Li-Zhi and Ruffini, R.",
    title = "{The Origin of Structure in the Universe}",
    reportNumber = "Print-85-0265 (CAMBRIDGE)",
    doi = "10.1103/PhysRevD.31.1777",
    journal = "Phys. Rev. D",
    volume = "31",
    pages = "1777",
    year = "1985"
}

@article{BunchDavies1978,
  author = {Bunch, T. S. and Davies, P. C. W.},
  title = {Quantum field theory in de Sitter space: Renormalization by point-splitting},
  journal = {Proc. Roy. Soc. Lond. A},
  volume = {360},
  year = {1978},
  pages = {117--134}
}

@article{deAlwis:2018sec,
    author = "de Alwis, S. P.",
    title = "{Wave function of the Universe and CMB fluctuations}",
    eprint = "1811.12892",
    archivePrefix = "arXiv",
    primaryClass = "hep-th",
    doi = "10.1103/PhysRevD.100.043544",
    journal = "Phys. Rev. D",
    volume = "100",
    number = "4",
    pages = "043544",
    year = "2019"
}

@article{Khan:2023ljg,
    author = "Khan, Rifath",
    title = "{The Semiclassical Approximation: Its Application to Holography and the Information Paradox}",
    eprint = "2309.08116",
    archivePrefix = "arXiv",
    primaryClass = "gr-qc",
    month = "9",
    year = "2023"
}

@article{Maldacena:2002vr,
    author = "Maldacena, Juan Martin",
    title = "{Non-Gaussian features of primordial fluctuations in single field inflationary models}",
    eprint = "astro-ph/0210603",
    archivePrefix = "arXiv",
    doi = "10.1088/1126-6708/2003/05/013",
    journal = "JHEP",
    volume = "05",
    pages = "013",
    year = "2003"
}

@article{Pimentel:2013gza,
    author = "Pimentel, Guilherme L.",
    title = "{Inflationary Consistency Conditions from a Wavefunctional Perspective}",
    eprint = "1309.1793",
    archivePrefix = "arXiv",
    primaryClass = "hep-th",
    reportNumber = "PUPT-2452",
    doi = "10.1007/JHEP02(2014)124",
    journal = "JHEP",
    volume = "02",
    pages = "124",
    year = "2014"
}

@article{Skenderis:2002wp,
    author = "Skenderis, Kostas",
    editor = "de Wit, B. and Vandoren, S.",
    title = "{Lecture notes on holographic renormalization}",
    eprint = "hep-th/0209067",
    archivePrefix = "arXiv",
    reportNumber = "PUTP-2047",
    doi = "10.1088/0264-9381/19/22/306",
    journal = "Class. Quant. Grav.",
    volume = "19",
    pages = "5849--5876",
    year = "2002"
}

@article{McFadden:2009fg,
    author = "McFadden, Paul and Skenderis, Kostas",
    title = "{Holography for Cosmology}",
    eprint = "0907.5542",
    archivePrefix = "arXiv",
    primaryClass = "hep-th",
    reportNumber = "ITF-22",
    doi = "10.1103/PhysRevD.81.021301",
    journal = "Phys. Rev. D",
    volume = "81",
    pages = "021301",
    year = "2010"
}

@article{McFadden:2010na,
    author = "McFadden, Paul and Skenderis, Kostas",
    editor = "Basilakos, Spyros and Cadoni, Mariano and Cavaglia, Marco and Christodoulakis, Theodosios and Vagenas, Elias C.",
    title = "{The Holographic Universe}",
    eprint = "1001.2007",
    archivePrefix = "arXiv",
    primaryClass = "hep-th",
    reportNumber = "ITFA-10-02",
    doi = "10.1088/1742-6596/222/1/012007",
    journal = "J. Phys. Conf. Ser.",
    volume = "222",
    pages = "012007",
    year = "2010"
}

@inproceedings{McFadden:2010jw,
    author = "McFadden, Paul and Skenderis, Kostas",
    title = "{Observational signatures of holographic models of inflation}",
    booktitle = "{12th Marcel Grossmann Meeting on General Relativity}",
    eprint = "1010.0244",
    archivePrefix = "arXiv",
    primaryClass = "hep-th",
    reportNumber = "ITFA-10-22",
    doi = "10.1142/9789814374552_0468",
    pages = "2315--2323",
    month = "10",
    year = "2010"
}

@article{McFadden:2010vh,
    author = "McFadden, Paul and Skenderis, Kostas",
    title = "{Holographic Non-Gaussianity}",
    eprint = "1011.0452",
    archivePrefix = "arXiv",
    primaryClass = "hep-th",
    reportNumber = "ITFA-10-23",
    doi = "10.1088/1475-7516/2011/05/013",
    journal = "JCAP",
    volume = "05",
    pages = "013",
    year = "2011"
}

@article{McFadden:2011kk,
    author = "McFadden, Paul and Skenderis, Kostas",
    title = "{Cosmological 3-point correlators from holography}",
    eprint = "1104.3894",
    archivePrefix = "arXiv",
    primaryClass = "hep-th",
    reportNumber = "ITFA-11-08",
    doi = "10.1088/1475-7516/2011/06/030",
    journal = "JCAP",
    volume = "06",
    pages = "030",
    year = "2011"
}

@article{Easther:2011wh,
    author = "Easther, Richard and Flauger, Raphael and McFadden, Paul and Skenderis, Kostas",
    title = "{Constraining holographic inflation with WMAP}",
    eprint = "1104.2040",
    archivePrefix = "arXiv",
    primaryClass = "astro-ph.CO",
    doi = "10.1088/1475-7516/2011/09/030",
    journal = "JCAP",
    volume = "09",
    pages = "030",
    year = "2011"
}

@article{Bzowski:2011ab,
    author = "Bzowski, Adam and McFadden, Paul and Skenderis, Kostas",
    title = "{Holographic predictions for cosmological 3-point functions}",
    eprint = "1112.1967",
    archivePrefix = "arXiv",
    primaryClass = "hep-th",
    doi = "10.1007/JHEP03(2012)091",
    journal = "JHEP",
    volume = "03",
    pages = "091",
    year = "2012"
}

@article{Bzowski:2012ih,
    author = "Bzowski, Adam and McFadden, Paul and Skenderis, Kostas",
    title = "{Holography for inflation using conformal perturbation theory}",
    eprint = "1211.4550",
    archivePrefix = "arXiv",
    primaryClass = "hep-th",
    doi = "10.1007/JHEP04(2013)047",
    journal = "JHEP",
    volume = "04",
    pages = "047",
    year = "2013"
}

@article{Bzowski:2013sza,
    author = "Bzowski, Adam and McFadden, Paul and Skenderis, Kostas",
    title = "{Implications of conformal invariance in momentum space}",
    eprint = "1304.7760",
    archivePrefix = "arXiv",
    primaryClass = "hep-th",
    doi = "10.1007/JHEP03(2014)111",
    journal = "JHEP",
    volume = "03",
    pages = "111",
    year = "2014"
}

@article{Afshordi:2016dvb,
    author = "Afshordi, Niayesh and Coriano, Claudio and Delle Rose, Luigi and Gould, Elizabeth and Skenderis, Kostas",
    title = "{From Planck data to Planck era: Observational tests of Holographic Cosmology}",
    eprint = "1607.04878",
    archivePrefix = "arXiv",
    primaryClass = "astro-ph.CO",
    doi = "10.1103/PhysRevLett.118.041301",
    journal = "Phys. Rev. Lett.",
    volume = "118",
    number = "4",
    pages = "041301",
    year = "2017"
}

@article{Afshordi:2017ihr,
    author = "Afshordi, Niayesh and Gould, Elizabeth and Skenderis, Kostas",
    title = "{Constraining holographic cosmology using Planck data}",
    eprint = "1703.05385",
    archivePrefix = "arXiv",
    primaryClass = "astro-ph.CO",
    doi = "10.1103/PhysRevD.95.123505",
    journal = "Phys. Rev. D",
    volume = "95",
    number = "12",
    pages = "123505",
    year = "2017"
}

@article{Nastase:2019rsn,
    author = "Nastase, Horatiu and Skenderis, Kostas",
    title = "{Holography for the very early Universe and the classic puzzles of Hot Big Bang cosmology}",
    eprint = "1904.05821",
    archivePrefix = "arXiv",
    primaryClass = "hep-th",
    doi = "10.1103/PhysRevD.101.021901",
    journal = "Phys. Rev. D",
    volume = "101",
    number = "2",
    pages = "021901",
    year = "2020"
}

@article{Bzowski:2019kwd,
    author = "Bzowski, Adam and McFadden, Paul and Skenderis, Kostas",
    title = "{Conformal $n$-point functions in momentum space}",
    eprint = "1910.10162",
    archivePrefix = "arXiv",
    primaryClass = "hep-th",
    doi = "10.1103/PhysRevLett.124.131602",
    journal = "Phys. Rev. Lett.",
    volume = "124",
    number = "13",
    pages = "131602",
    year = "2020"
}

@article{Penin:2021sry,
    author = "Pen{\'\i}n, Jos{\'e} Manuel and Skenderis, Kostas and Withers, Benjamin",
    title = "{Massive holographic QFTs in de Sitter}",
    eprint = "2112.14639",
    archivePrefix = "arXiv",
    primaryClass = "hep-th",
    doi = "10.21468/SciPostPhys.12.6.182",
    journal = "SciPost Phys.",
    volume = "12",
    number = "6",
    pages = "182",
    year = "2022"
}

@article{Bzowski:2023nef,
    author = "Bzowski, Adam and McFadden, Paul and Skenderis, Kostas",
    title = "{Renormalisation of IR divergences and holography in de Sitter}",
    eprint = "2312.17316",
    archivePrefix = "arXiv",
    primaryClass = "hep-th",
    doi = "10.1007/JHEP05(2024)053",
    journal = "JHEP",
    volume = "05",
    pages = "053",
    year = "2024"
}

@article{McFadden:2013ria,
    author = "McFadden, Paul",
    title = "{On the power spectrum of inflationary cosmologies dual to a deformed CFT}",
    eprint = "1308.0331",
    archivePrefix = "arXiv",
    primaryClass = "hep-th",
    doi = "10.1007/JHEP10(2013)071",
    journal = "JHEP",
    volume = "10",
    pages = "071",
    year = "2013"
}

@article{WFCtoCorrelators1,
    author = "Anninos, Dionysios and Anous, Tarek and Freedman, Daniel Z. and Konstantinidis, George",
    title = "{Late-time Structure of the Bunch-Davies De Sitter Wavefunction}",
    eprint = "1406.5490",
    archivePrefix = "arXiv",
    primaryClass = "hep-th",
    reportNumber = "MIT-CTP-4561",
    doi = "10.1088/1475-7516/2015/11/048",
    journal = "JCAP",
    volume = "11",
    pages = "048",
    year = "2015"
}

@article{WFCtoCorrelators2,
    author = "Goon, Garrett and Hinterbichler, Kurt and Joyce, Austin and Trodden, Mark",
    title = "{Shapes of gravity: Tensor non-Gaussianity and massive spin-2 fields}",
    eprint = "1812.07571",
    archivePrefix = "arXiv",
    primaryClass = "hep-th",
    doi = "10.1007/JHEP10(2019)182",
    journal = "JHEP",
    volume = "10",
    pages = "182",
    year = "2019"
}

@article{Abolhasani:2022twf,
    author = "Abolhasani, Aliakbar and Goodhew, Harry",
    title = "{Derivative interactions during inflation: a systematic approach}",
    eprint = "2201.05117",
    archivePrefix = "arXiv",
    primaryClass = "hep-th",
    doi = "10.1088/1475-7516/2022/06/032",
    journal = "JCAP",
    volume = "06",
    number = "06",
    pages = "032",
    year = "2022"
}

@misc{BaumannJoyce:2023Lecs,
  author = "Baumann, Daniel and Joyce, Austin",
  title = {Lectures on Cosmological Correlations},
  year = {2023},
  publisher = {GitHub},
  journal = {GitHub repository},
  howpublished = {\url{https://github.com/ddbaumann/cosmo-correlators}},
}

@article{Cespedes:2023aal,
    author = "C\'espedes, Sebasti\'an and Davis, Anne-Christine and Wang, Dong-Gang",
    title = "{On the IR divergences in de Sitter space: loops, resummation and the semi-classical wavefunction}",
    eprint = "2311.17990",
    archivePrefix = "arXiv",
    primaryClass = "hep-th",
    doi = "10.1007/JHEP04(2024)004",
    journal = "JHEP",
    volume = "04",
    pages = "004",
    year = "2024"
}

@article{COT,
    author = "Goodhew, Harry and Jazayeri, Sadra and Pajer, Enrico",
    title = "{The Cosmological Optical Theorem}",
    eprint = "2009.02898",
    archivePrefix = "arXiv",
    primaryClass = "hep-th",
    month = "9",
    year = "2020"
}

@article{Cespedes:2020xqq,
    author = "C\'espedes, Sebasti\'an and Davis, Anne-Christine and Melville, Scott",
    title = "{On the time evolution of cosmological correlators}",
    eprint = "2009.07874",
    archivePrefix = "arXiv",
    primaryClass = "hep-th",
    doi = "10.1007/JHEP02(2021)012",
    journal = "JHEP",
    volume = "02",
    pages = "012",
    year = "2021"
}

@article{Melville:2021lst,
    author = "Melville, Scott and Pajer, Enrico",
    title = "{Cosmological Cutting Rules}",
    eprint = "2103.09832",
    archivePrefix = "arXiv",
    primaryClass = "hep-th",
    doi = "10.1007/JHEP05(2021)249",
    journal = "JHEP",
    volume = "05",
    pages = "249",
    year = "2021"
}

@article{Goodhew:2021oqg,
    author = "Goodhew, Harry and Jazayeri, Sadra and Gordon Lee, Mang Hei and Pajer, Enrico",
    title = "{Cutting Cosmological Correlators}",
    eprint = "2104.06587",
    archivePrefix = "arXiv",
    primaryClass = "hep-th",
    month = "4",
    year = "2021"
}

@article{Baumann:2021fxj,
    author = "Baumann, Daniel and Chen, Wei-Ming and Duaso Pueyo, Carlos and Joyce, Austin and Lee, Hayden and Pimentel, Guilherme L.",
    title = "{Linking the singularities of cosmological correlators}",
    eprint = "2106.05294",
    archivePrefix = "arXiv",
    primaryClass = "hep-th",
    doi = "10.1007/JHEP09(2022)010",
    journal = "JHEP",
    volume = "09",
    pages = "010",
    year = "2022"
}

@article{Marolf:2012kh,
    author = "Marolf, Donald and Morrison, Ian A. and Srednicki, Mark",
    title = "{Perturbative S-matrix for massive scalar fields in global de Sitter space}",
    eprint = "1209.6039",
    archivePrefix = "arXiv",
    primaryClass = "hep-th",
    doi = "10.1088/0264-9381/30/15/155023",
    journal = "Class. Quant. Grav.",
    volume = "30",
    pages = "155023",
    year = "2013"
}

@article{Melville:2023kgd,
    author = "Melville, Scott and Pimentel, Guilherme L.",
    title = "{de Sitter S matrix for the masses}",
    eprint = "2309.07092",
    archivePrefix = "arXiv",
    primaryClass = "hep-th",
    doi = "10.1103/PhysRevD.110.103530",
    journal = "Phys. Rev. D",
    volume = "110",
    number = "10",
    pages = "103530",
    year = "2024"
}

@article{Melville:2024ove,
    author = "Melville, Scott and Pimentel, Guilherme L.",
    title = "{A de Sitter S-matrix from amputated cosmological correlators}",
    eprint = "2404.05712",
    archivePrefix = "arXiv",
    primaryClass = "hep-th",
    doi = "10.1007/JHEP08(2024)211",
    journal = "JHEP",
    volume = "08",
    pages = "211",
    year = "2024"
}

@article{Stefanyszyn:2023qov,
    author = "Stefanyszyn, David and Tong, Xi and Zhu, Yuhang",
    title = "{Cosmological correlators through the looking glass: reality, parity, and factorisation}",
    eprint = "2309.07769",
    archivePrefix = "arXiv",
    primaryClass = "hep-th",
    doi = "10.1007/JHEP05(2024)196",
    journal = "JHEP",
    volume = "05",
    pages = "196",
    year = "2024"
}

@article{Stefanyszyn:2024msm,
    author = "Stefanyszyn, David and Tong, Xi and Zhu, Yuhang",
    title = "{There and Back Again: Mapping and Factorizing Cosmological Observables}",
    eprint = "2406.00099",
    archivePrefix = "arXiv",
    primaryClass = "hep-th",
    doi = "10.1103/PhysRevLett.133.221501",
    journal = "Phys. Rev. Lett.",
    volume = "133",
    number = "22",
    pages = "221501",
    year = "2024"
}

@article{Stefanyszyn:2025yhq,
    author = "Stefanyszyn, David and Tong, Xi and Zhu, Yuhang",
    title = "{A Match Made in Heaven: Linking Observables in Inflationary Cosmology}",
    eprint = "2505.16071",
    archivePrefix = "arXiv",
    primaryClass = "hep-th",
    month = "5",
    year = "2025"
}

@article{Liu:2019fag,
    author = "Liu, Tao and Tong, Xi and Wang, Yi and Xianyu, Zhong-Zhi",
    title = "{Probing P and CP Violations on the Cosmological Collider}",
    eprint = "1909.01819",
    archivePrefix = "arXiv",
    primaryClass = "hep-ph",
    doi = "10.1007/JHEP04(2020)189",
    journal = "JHEP",
    volume = "04",
    pages = "189",
    year = "2020"
}

@article{Cabass:2022rhr,
    author = "Cabass, Giovanni and Jazayeri, Sadra and Pajer, Enrico and Stefanyszyn, David",
    title = "{Parity violation in the scalar trispectrum: no-go theorems and yes-go examples}",
    eprint = "2210.02907",
    archivePrefix = "arXiv",
    primaryClass = "hep-th",
    doi = "10.1007/JHEP02(2023)021",
    journal = "JHEP",
    volume = "02",
    pages = "021",
    year = "2023"
}

@article{Thavanesan:2025kyc,
    author = "Thavanesan, Ayngaran",
    title = "{No-go Theorem for Cosmological Parity Violation}",
    eprint = "2501.06383",
    archivePrefix = "arXiv",
    primaryClass = "hep-th",
    month = "1",
    year = "2025"
}

@article{Anous:2020nxu,
    author = "Anous, Tarek and Skulte, Jim",
    title = "{An invitation to the principal series}",
    eprint = "2007.04975",
    archivePrefix = "arXiv",
    primaryClass = "hep-th",
    doi = "10.21468/SciPostPhys.9.3.028",
    journal = "SciPost Phys.",
    volume = "9",
    number = "3",
    pages = "028",
    year = "2020"
}

@article{Anninos:2023lin,
    author = {Anninos, Dionysios and Anous, Tarek and Pethybridge, Ben and \c{S}eng\"or, Gizem},
    title = "{The discreet charm of the discrete series in dS$_{2}$}",
    eprint = "2307.15832",
    archivePrefix = "arXiv",
    primaryClass = "hep-th",
    doi = "10.1088/1751-8121/ad14ad",
    journal = "J. Phys. A",
    volume = "57",
    number = "2",
    pages = "025401",
    year = "2024"
}

@article{Joung:2006gj,
    author = "Joung, E. and Mourad, J. and Parentani, R.",
    title = "{Group theoretical approach to quantum fields in de Sitter space. I. The Principle series}",
    eprint = "hep-th/0606119",
    archivePrefix = "arXiv",
    doi = "10.1088/1126-6708/2006/08/082",
    journal = "JHEP",
    volume = "08",
    pages = "082",
    year = "2006"
}

@article{Sengor:2019mbz,
    author = {Seng\"or, Gizem and Skordis, Constantinos},
    title = "{Unitarity at the Late time Boundary of de Sitter}",
    eprint = "1912.09885",
    archivePrefix = "arXiv",
    primaryClass = "hep-th",
    doi = "10.1007/JHEP06(2020)041",
    journal = "JHEP",
    volume = "06",
    pages = "041",
    year = "2020"
}

@article{Sengor:2021zlc,
    author = "Sengor, Gizem and Skordis, Constantinos",
    title = "{Scalar two-point functions at the late-time boundary of de Sitter}",
    eprint = "2110.01635",
    archivePrefix = "arXiv",
    primaryClass = "hep-th",
    doi = "10.1007/JHEP02(2024)076",
    journal = "JHEP",
    volume = "02",
    pages = "076",
    year = "2024"
}

@article{Sengor:2022hfx,
    author = "Sengor, Gizem and Skordis, Constantinos",
    editor = "Dobrev, Vladimir",
    title = "{Principal and Complementary Series Representations at the Late-Time Boundary of de Sitter}",
    eprint = "2205.11550",
    archivePrefix = "arXiv",
    primaryClass = "hep-th",
    doi = "10.1007/978-981-19-4751-3_21",
    journal = "Springer Proc. Math. Stat.",
    volume = "396",
    pages = "269--276",
    year = "2022"
}

@article{Sengor:2022lyv,
    author = {Seng\"or, Gizem},
    title = "{The de Sitter group and its presence at the late-time boundary}",
    eprint = "2206.04719",
    archivePrefix = "arXiv",
    primaryClass = "hep-th",
    doi = "10.22323/1.406.0356",
    journal = "PoS",
    volume = "CORFU2021",
    pages = "356",
    year = "2022"
}

@article{Sengor:2022kji,
    author = {\c{S}eng\"or, Gizem},
    title = "{Particles of a de Sitter Universe}",
    eprint = "2212.10626",
    archivePrefix = "arXiv",
    primaryClass = "hep-th",
    doi = "10.3390/universe9020059",
    journal = "Universe",
    volume = "9",
    number = "2",
    pages = "59",
    year = "2023"
}

@inproceedings{Sengor:2023buj,
    author = {\c{S}eng\"or, Gizem},
    title = "{Searching for discrete series representations at the late-time boundary of de Sitter}",
    booktitle = "{15th International Workshop on Lie Theory and Its Applications in Physics}",
    eprint = "2312.00363",
    archivePrefix = "arXiv",
    primaryClass = "hep-th",
    month = "12",
    year = "2023"
}

@article{Dey:2024zjx,
    author = "Dey, Indranil and Nanda, Kanhu Kishore and Roy, Akashdeep and Trivedi, Sandip P.",
    title = "{Aspects of dS/CFT Holography}",
    eprint = "2407.02417",
    archivePrefix = "arXiv",
    primaryClass = "hep-th",
    month = "7",
    year = "2024"
}

@article{Lee:2023jby,
    author = "Lee, Mang Hei Gordon and McCulloch, Ciaran and Pajer, Enrico",
    title = "{Leading loops in cosmological correlators}",
    eprint = "2305.11228",
    archivePrefix = "arXiv",
    primaryClass = "hep-th",
    doi = "10.1007/JHEP11(2023)038",
    journal = "JHEP",
    volume = "11",
    pages = "038",
    year = "2023"
}

@article{Senatore:2009cf,
    author = "Senatore, Leonardo and Zaldarriaga, Matias",
    title = "{On Loops in Inflation}",
    eprint = "0912.2734",
    archivePrefix = "arXiv",
    primaryClass = "hep-th",
    doi = "10.1007/JHEP12(2010)008",
    journal = "JHEP",
    volume = "12",
    pages = "008",
    year = "2010"
}

@article{Thavanesan:2025tha,
    author = "Thavanesan, Ayngaran",
    title = "{Going through phases of UV and IR divergences in Cosmology}",
    eprint = "25xx.xxxxx",
    primaryClass = "hep-th",
    month = "x",
    year = "2025"
}

@article{Strominger:2001pn,
    author = "Strominger, Andrew",
    title = "{The dS / CFT correspondence}",
    eprint = "hep-th/0106113",
    archivePrefix = "arXiv",
    doi = "10.1088/1126-6708/2001/10/034",
    journal = "JHEP",
    volume = "10",
    pages = "034",
    year = "2001"
}

@article{Ng:2012xp,
    author = "Ng, Gim Seng and Strominger, Andrew",
    title = "{State/Operator Correspondence in Higher-Spin dS/CFT}",
    eprint = "1204.1057",
    archivePrefix = "arXiv",
    primaryClass = "hep-th",
    doi = "10.1088/0264-9381/30/10/104002",
    journal = "Class. Quant. Grav.",
    volume = "30",
    pages = "104002",
    year = "2013"
}

@article{Das:2012dt,
    author = "Das, Diptarka and Das, Sumit R. and Jevicki, Antal and Ye, Qibin",
    title = "{Bi-local Construction of Sp(2N)/dS Higher Spin Correspondence}",
    eprint = "1205.5776",
    archivePrefix = "arXiv",
    primaryClass = "hep-th",
    reportNumber = "UK-12-04, BROWN-HET-1635",
    doi = "10.1007/JHEP01(2013)107",
    journal = "JHEP",
    volume = "01",
    pages = "107",
    year = "2013"
}

@article{LeClair:2006kb,
    author = "LeClair, Andre",
    title = "{Quantum critical spin liquids, the 3D Ising model, and conformal field theory in 2+1 dimensions}",
    eprint = "cond-mat/0610639",
    archivePrefix = "arXiv",
    month = "10",
    year = "2006"
}

@article{LeClair:2007iy,
    author = "LeClair, Andre and Neubert, Matthias",
    title = "{Semi-Lorentz invariance, unitarity, and critical exponents of symplectic fermion models}",
    eprint = "0705.4657",
    archivePrefix = "arXiv",
    primaryClass = "hep-th",
    reportNumber = "MZ-TH-07-09",
    doi = "10.1088/1126-6708/2007/10/027",
    journal = "JHEP",
    volume = "10",
    pages = "027",
    year = "2007"
}

@article{Robinson:2009xm,
    author = "Robinson, Dean J. and Kapit, Eliot and LeClair, Andre",
    title = "{Lorentz Symmetric Quantum Field Theory for Symplectic Fermions}",
    eprint = "0903.2399",
    archivePrefix = "arXiv",
    primaryClass = "hep-th",
    doi = "10.1063/1.3248256",
    journal = "J. Math. Phys.",
    volume = "50",
    pages = "112301",
    year = "2009"
}

@article{Freidel:2008sh,
    author = "Freidel, Laurent",
    title = "{Reconstructing AdS/CFT}",
    eprint = "0804.0632",
    archivePrefix = "arXiv",
    primaryClass = "hep-th",
    month = "4",
    year = "2008"
}

@article{Cotler:2019nbi,
    author = "Cotler, Jordan and Jensen, Kristan and Maloney, Alexander",
    title = "{Low-dimensional de Sitter quantum gravity}",
    eprint = "1905.03780",
    archivePrefix = "arXiv",
    primaryClass = "hep-th",
    doi = "10.1007/JHEP06(2020)048",
    journal = "JHEP",
    volume = "06",
    pages = "048",
    year = "2020"
}

@article{Araujo-Regado:2025elv,
    author = "Araujo-Regado, Goncalo and Thavanesan, Ayngaran and Wall, Aron C.",
    title = "{Holographic Cosmology at Finite Time}",
    eprint = "2511.04511",
    archivePrefix = "arXiv",
    primaryClass = "hep-th",
    month = "11",
    year = "2025"
}

@article{Collier:2024kmo,
    author = {Collier, Scott and Eberhardt, Lorenz and M\"uhlmann, Beatrix and Rodriguez, Victor A.},
    title = "{The complex Liouville string}",
    eprint = "2409.17246",
    archivePrefix = "arXiv",
    primaryClass = "hep-th",
    month = "9",
    year = "2024"
}

@article{Collier:2024kwt,
    author = {Collier, Scott and Eberhardt, Lorenz and M\"uhlmann, Beatrix and Rodriguez, Victor A.},
    title = "{The complex Liouville string: the worldsheet}",
    eprint = "2409.18759",
    archivePrefix = "arXiv",
    primaryClass = "hep-th",
    month = "9",
    year = "2024"
}

@article{Collier:2024lys,
    author = {Collier, Scott and Eberhardt, Lorenz and M\"uhlmann, Beatrix and Rodriguez, Victor A.},
    title = "{The complex Liouville string: the matrix integral}",
    eprint = "2410.07345",
    archivePrefix = "arXiv",
    primaryClass = "hep-th",
    month = "10",
    year = "2024"
}

@article{Collier:2025pbm,
    author = {Collier, Scott and Eberhardt, Lorenz and M\"uhlmann, Beatrix},
    title = "{The complex Liouville string: the gravitational path integral}",
    eprint = "2501.10265",
    archivePrefix = "arXiv",
    primaryClass = "hep-th",
    month = "1",
    year = "2025"
}

@article{Collier:2025lux,
    author = {Collier, Scott and Eberhardt, Lorenz and M\"uhlmann, Beatrix},
    title = "{A microscopic realization of dS$_3$}",
    eprint = "2501.01486",
    archivePrefix = "arXiv",
    primaryClass = "hep-th",
    doi = "10.21468/SciPostPhys.18.4.131",
    journal = "SciPost Phys.",
    volume = "18",
    pages = "131",
    year = "2025"
}

@article{Araujo-Regado:2022gvw,
    author = "Araujo-Regado, Goncalo and Khan, Rifath and Wall, Aron C.",
    title = "{Cauchy slice holography: a new AdS/CFT dictionary}",
    eprint = "2204.00591",
    archivePrefix = "arXiv",
    primaryClass = "hep-th",
    doi = "10.1007/JHEP03(2023)026",
    journal = "JHEP",
    volume = "03",
    pages = "026",
    year = "2023"
}

@article{Araujo-Regado:2022jpj,
    author = "Araujo-Regado, Goncalo",
    title = "{Holographic Cosmology on Closed Slices in 2+1 Dimensions}",
    eprint = "2212.03219",
    archivePrefix = "arXiv",
    primaryClass = "hep-th",
    month = "12",
    year = "2022"
}

@article{Soni:2024aop,
    author = "Soni, Ronak M. and Wall, Aron C.",
    title = "{A New Covariant Entropy Bound from Cauchy Slice Holography}",
    eprint = "2407.16769",
    archivePrefix = "arXiv",
    primaryClass = "hep-th",
    month = "7",
    year = "2024"
}

@article{Shyam:2025ttb ,
    author = "Shyam, Vasudev and Silverstein, Eva and Soni, Ronak M. and Thavanesan, Ayngaran and Torroba, Gonzalo",
    title = "{dS/CFT from T$\overline{T}$+$\Lambda_d$}",
    eprint = "25xx.xxxxx",
    primaryClass = "hep-th",
    month = "x",
    year = "2025"
}

@article{Hogervorst:2021uvp,
    author = "Hogervorst, Matthijs and Penedones, Jo\~ao and Vaziri, Kamran Salehi",
    title = "{Towards the non-perturbative cosmological bootstrap}",
    eprint = "2107.13871",
    archivePrefix = "arXiv",
    primaryClass = "hep-th",
    month = "7",
    year = "2021"
}

@article{DiPietro:2021sjt,
    author = "Di Pietro, Lorenzo and Gorbenko, Victor and Komatsu, Shota",
    title = "{Analyticity and Unitarity for Cosmological Correlators}",
    eprint = "2108.01695",
    archivePrefix = "arXiv",
    primaryClass = "hep-th",
    month = "8",
    year = "2021"
}

@phdthesis{SalehiVaziri:2022sdr,
    author = "Salehi Vaziri, Kamran",
    title = "{Nonperturbative Quantum Field Theory in Curved Spacetime}",
    doi = "10.5075/epfl-thesis-9263",
    school = "Ecole Polytechnique, Lausanne",
    year = "2022"
}

@article{Penedones:2023uqc,
    author = "Penedones, Joao and Salehi Vaziri, Kamran and Sun, Zimo",
    title = "{Hilbert space of quantum field theory in de Sitter spacetime}",
    eprint = "2301.04146",
    archivePrefix = "arXiv",
    primaryClass = "hep-th",
    doi = "10.1103/PhysRevD.111.045001",
    journal = "Phys. Rev. D",
    volume = "111",
    number = "4",
    pages = "045001",
    year = "2025"
}

@article{Loparco:2023rug,
    author = "Loparco, Manuel and Penedones, Joao and Salehi Vaziri, Kamran and Sun, Zimo",
    title = {{The K{\"a}ll{\'e}n-Lehmann representation in de Sitter spacetime}},
    eprint = "2306.00090",
    archivePrefix = "arXiv",
    primaryClass = "hep-th",
    doi = "10.1007/JHEP12(2023)159",
    journal = "JHEP",
    volume = "12",
    pages = "159",
    year = "2023"
}

@article{Green:2023ids,
    author = "Green, Daniel and Huang, Yiwen and Shen, Chia-Hsien and Baumann, Daniel",
    title = "{Positivity from Cosmological Correlators}",
    eprint = "2310.02490",
    archivePrefix = "arXiv",
    primaryClass = "hep-th",
    doi = "10.1007/JHEP04(2024)034",
    journal = "JHEP",
    volume = "04",
    pages = "034",
    year = "2024"
}

@article{Loparco:2023akg,
    author = "Loparco, Manuel and Qiao, Jiaxin and Sun, Zimo",
    title = "{A radial variable for de Sitter two-point functions}",
    eprint = "2310.15944",
    archivePrefix = "arXiv",
    primaryClass = "hep-th",
    doi = "10.21468/SciPostPhys.18.5.164",
    journal = "SciPost Phys.",
    volume = "18",
    number = "5",
    pages = "164",
    year = "2025"
}

@article{Loparco:2024ibp,
    author = "Loparco, Manuel",
    title = "{RG flows in de Sitter: C-functions and sum rules}",
    eprint = "2404.03739",
    archivePrefix = "arXiv",
    primaryClass = "hep-th",
    doi = "10.21468/SciPostPhys.17.3.079",
    journal = "SciPost Phys.",
    volume = "17",
    number = "3",
    pages = "079",
    year = "2024"
}

@article{Loparco:2025azm,
    author = "Loparco, Manuel and Penedones, Joao and Ulrich, Yannis",
    title = "{What is a photon in de Sitter spacetime?}",
    eprint = "2505.00761",
    archivePrefix = "arXiv",
    primaryClass = "hep-th",
    month = "5",
    year = "2025"
}

@article{Chakraborty:2025mhh,
    author = "Chakraborty, Priyesh and Cohen, Timothy and Green, Daniel and Huang, Yiwen",
    title = "{A Compact Story of Positivity in de Sitter}",
    eprint = "2508.08359",
    archivePrefix = "arXiv",
    primaryClass = "hep-th",
    reportNumber = "CERN-TH-2025-154, CERN-TH-2025-154",
    month = "8",
    year = "2025"
}

@article{Cotler:2024xzz,
    author = "Cotler, Jordan and Jensen, Kristan",
    title = "{Non-perturbative de Sitter Jackiw-Teitelboim gravity}",
    eprint = "2401.01925",
    archivePrefix = "arXiv",
    primaryClass = "hep-th",
    doi = "10.1007/JHEP12(2024)016",
    journal = "JHEP",
    volume = "12",
    pages = "016",
    year = "2024"
}

@incollection{Deligne1990TannakianCategories,
  author       = {Pierre Deligne and James S. Milne},
  title        = {Tannakian Categories},
  booktitle    = {Hodge Cycles, Motives, and Shimura Varieties},
  series       = {Lecture Notes in Mathematics},
  volume       = {900},
  editor       = {P. Deligne et al.},
  publisher    = {Springer–Verlag},
  address      = {Berlin},
  year         = {1982},
  pages        = {101--228},
}

@article{Deligne2002TensorCategories,
  author       = {Pierre Deligne},
  title        = {Cat{\'e}gories tensorielles},
  journal      = {Moscow Mathematical Journal},
  volume       = {2},
  number       = {2},
  year         = {2002},
  pages        = {227--248},
  doi          = {10.17323/1609-4514-2002-2-2-227-248},
}

@incollection{Deligne2007RepsSt,
  author       = {Pierre Deligne},
  title        = {La cat\'egorie des repr\'esentations du groupe sym\'etrique $S_t$, lorsque $t$ n’est pas un entier naturel},
  booktitle    = {Algebraic Groups and Homogeneous Spaces},
  publisher    = {Narosa Publishing House},
  year         = {2007},
  pages        = {209--273},
  note         = {Proceedings of the International Colloquium on Algebraic Groups and Homogeneous Spaces, TIFR, Mumbai, January 2004},
}

@article{ComesOstrik2011BlocksRepSt,
  author       = {Jonathan Comes and Victor Ostrik},
  title        = {On blocks of Deligne’s category $\mathrm{Rep}(S_t)$},
  journal      = {Advances in Mathematics},
  volume       = {226},
  number       = {2},
  year         = {2011},
  pages        = {1331--1377},
  doi          = {10.1016/j.aim.2010.08.010},
}

@article{Knop2007TensorEnvelopes,
  author       = {Friedrich Knop},
  title        = {Tensor envelopes of regular categories},
  journal      = {Advances in Mathematics},
  volume       = {214},
  number       = {2},
  year         = {2007},
  pages        = {571--617},
  doi          = {10.1016/j.aim.2007.03.007},
  eprint       = {math/0610552},
  archivePrefix = {arXiv},
  primaryClass  = {math.CT},
  url          = {https://arxiv.org/abs/math/0610552}
}

@article{Etingof2014RepComplexI,
  author       = {Pavel Etingof},
  title        = {Representation theory in complex rank, I},
  journal      = {Transformation Groups},
  volume       = {19},
  number       = {2},
  year         = {2014},
  pages        = {359--381},
  eprint       = {1401.6321},
  archivePrefix = {arXiv},
  primaryClass  = {math.RT},
  url          = {https://arxiv.org/abs/1401.6321}
}

@article{Etingof2014RepComplexII,
  author       = {Pavel Etingof},
  title        = {Representation theory in complex rank, II},
  year         = {2014},
  eprint       = {1407.0373},
  archivePrefix = {arXiv},
  primaryClass  = {math.RT},
  note         = {See also ``I'' (arXiv:1401.6321) for background and further references},
  url          = {https://arxiv.org/abs/1407.0373}
}

@article{EntovaHinichSerganova2015DeligneGLmnn,
  author       = {Inna Entova-Aizenbud and Vladimir Hinich and Vera Serganova},
  title        = {Deligne categories and the limit of categories $\mathrm{Rep}(\mathrm{GL}(m|n))$},
  journal      = {International Mathematics Research Notices},
  volume       = {2020},
  number       = {15},
  pages        = {4602--4666},
  year         = {2020},
  doi          = {10.1093/imrn/rny144},
  eprint       = {1511.07699},
  archivePrefix = {arXiv},
  primaryClass  = {math.RT},
  url          = {https://arxiv.org/abs/1511.07699}
}

@article{EntovaSerganova2018Periplectic,
  author       = {Inna Entova-Aizenbud and Vera Serganova},
  title        = {Deligne categories and the periplectic Lie superalgebra},
  year         = {2018},
  eprint       = {1807.09478},
  archivePrefix = {arXiv},
  primaryClass  = {math.RT},
  url          = {https://arxiv.org/abs/1807.09478}
}

@article{Harman2016DeligneLimits,
  author       = {Nate Harman},
  title        = {Deligne categories as limits in rank and characteristic},
  year         = {2016},
  eprint       = {1601.03426},
  archivePrefix = {arXiv},
  primaryClass  = {math.RT},
  url          = {https://arxiv.org/abs/1601.03426}
}

@article{BarterEntovaHeidersdorf2017InfiniteSymm,
  author       = {Daniel Barter and Inna Entova-Aizenbud and Thorsten Heidersdorf},
  title        = {Deligne categories and representations of the infinite symmetric group},
  year         = {2017},
  eprint       = {1706.03645},
  archivePrefix = {arXiv},
  primaryClass  = {math.RT},
  url          = {https://arxiv.org/abs/1706.03645}
}

@article{Meir2021Interpolations,
  author       = {Eran Meir},
  title        = {Interpolations of monoidal categories and algebraic structures by invariant theory},
  journal      = {Selecta Mathematica, New Series},
  year         = {2021},
  volume       = {27},
  number       = {58},
  pages        = {Paper No. 58, 44 pp.},
  eprint       = {2105.04622},
  archivePrefix = {arXiv},
  primaryClass  = {math.CT},
  url          = {https://arxiv.org/abs/2105.04622}
}

@article{KhovanovOstrikKononov2020TQFT,
  author       = {Mikhail Khovanov and Victor Ostrik and Yakov Kononov},
  title        = {Two-dimensional topological theories, rational functions and their tensor envelopes},
  year         = {2020},
  eprint       = {2011.14758},
  archivePrefix = {arXiv},
  primaryClass  = {math.QA},
  url          = {https://arxiv.org/abs/2011.14758}
}

@misc{DeligneNotesOnline,
  author       = {Pierre Deligne},
  title        = {Deligne's papers and notes on tensor categories and interpolation},
  howpublished = {\url{https://publications.ias.edu/deligne/}},
  note         = {Collection of papers and PDFs hosted by the IAS Deligne page}
}

@article{Maldacena:2000mw,
    author = "Maldacena, Juan Martin and Nunez, Carlos",
    editor = "Duff, Michael J. and Liu, J. T. and Lu, J.",
    title = "{Supergravity description of field theories on curved manifolds and a no go theorem}",
    eprint = "hep-th/0007018",
    archivePrefix = "arXiv",
    doi = "10.1142/S0217751X01003937",
    journal = "Int. J. Mod. Phys. A",
    volume = "16",
    pages = "822--855",
    year = "2001"
}

@book{Polchinski:1998rq,
    author = "Polchinski, J.",
    title = "{String theory. Vol. 1: An introduction to the bosonic string}",
    doi = "10.1017/CBO9780511816079",
    isbn = "978-0-511-25227-3, 978-0-521-67227-6, 978-0-521-63303-1",
    publisher = "Cambridge University Press",
    series = "Cambridge Monographs on Mathematical Physics",
    month = "12",
    year = "2007"
}

@article{Anninos:2019nib,
    author = "Anninos, D. and De Luca, V. and Franciolini, G. and Kehagias, A. and Riotto, A.",
    title = "{Cosmological Shapes of Higher-Spin Gravity}",
    eprint = "1902.01251",
    archivePrefix = "arXiv",
    primaryClass = "hep-th",
    doi = "10.1088/1475-7516/2019/04/045",
    journal = "JCAP",
    volume = "04",
    pages = "045",
    year = "2019"
}

@article{Giddings:2001yu,
    author = "Giddings, Steven B. and Kachru, Shamit and Polchinski, Joseph",
    title = "{Hierarchies from fluxes in string compactifications}",
    eprint = "hep-th/0105097",
    archivePrefix = "arXiv",
    reportNumber = "SLAC-PUB-8807, NSF-ITP-01-37, SU-ITP-01-16",
    doi = "10.1103/PhysRevD.66.106006",
    journal = "Phys. Rev. D",
    volume = "66",
    pages = "106006",
    year = "2002"
}

@article{Kachru:2003aw,
    author = "Kachru, Shamit and Kallosh, Renata and Linde, Andrei D. and Trivedi, Sandip P.",
    title = "{De Sitter vacua in string theory}",
    eprint = "hep-th/0301240",
    archivePrefix = "arXiv",
    reportNumber = "SLAC-PUB-9630, SU-ITP-03-01, TIFR-TH-03-03",
    doi = "10.1103/PhysRevD.68.046005",
    journal = "Phys. Rev. D",
    volume = "68",
    pages = "046005",
    year = "2003"
}

@article{Danielsson:2018ztv,
    author = "Danielsson, Ulf H. and Van Riet, Thomas",
    title = "{What if string theory has no de Sitter vacua?}",
    eprint = "1804.01120",
    archivePrefix = "arXiv",
    primaryClass = "hep-th",
    reportNumber = "UUITP-09/18, UUITP-09-18",
    doi = "10.1142/S0218271818300070",
    journal = "Int. J. Mod. Phys. D",
    volume = "27",
    number = "12",
    pages = "1830007",
    year = "2018"
}

@article{Obied:2018sgi,
    author = "Obied, Georges and Ooguri, Hirosi and Spodyneiko, Lev and Vafa, Cumrun",
    title = "{De Sitter Space and the Swampland}",
    eprint = "1806.08362",
    archivePrefix = "arXiv",
    primaryClass = "hep-th",
    reportNumber = "CALT-TH-2018-020, IPMU18-0100",
    month = "6",
    year = "2018"
}

\end{document}